\documentclass[twocolumn]{revtex4}
\usepackage{ulem}
\usepackage{mathrsfs}
\usepackage{hyperref}
\usepackage{graphicx,color}

\begin{document}

\title{A First Step Towards Effectively Nonperturbative Scattering Amplitudes in the Perturbative Regime}

\author{Neil Christensen}
\thanks{nchris3@ilstu.edu}
\author{Joshua Henderson}
\author{Santiago Pinto}
\author{Cory Russ}
\affiliation{Department of Physics, Illinois State University, Normal, IL 61790-4560 USA}

\begin{abstract}
We propose an effectively nonperturbative approach to calculating scattering amplitudes in the perturbative regime.  We do this in a discretized momentum space by using the QSE method to calculate all the contributions (to all orders in perturbation theory) to the scattering eigenstates that are above a precision cutoff.  We then calculate the scattering amplitude by directly taking the inner product between these eigenstates.  In the current work we have analyzed this procedure for a $\lambda\phi^4$ theory in one spatial dimension and compared our results with perturbation theory obtaining favorable results suggestive that further research in this direction might be worthwhile.  In particular, we show that the efficiency of our method scales much better than second- and higher-order perturbation theory as the momentum lattice spacing decreases and as the eigenstate energy increases.
\end{abstract}

\maketitle

The textbook technique for calculating scattering amplitudes is an approximation scheme known as the Feynman diagram method.  It is a perturbative approach that expands the scattering amplitude in powers of the coupling constant, with each higher order in the perturbative series being significantly, exponentially more difficult than the previous one.  As a result, typically only the tree-level, sometimes the one-loop, and occasionally higher-loop Feynman diagrams are calculated, corresponding with first, second and higher orders in the perturbative expansion.  This leads to a precision of calculation that is fundamentally limited by the perturbative order of the calculation.  The only way to improve the precision (assuming it is above the precision of the computer algorithm that computes it) is to calculate a higher order in the perturbative series, which can be extremely difficult and even impossible.   Although spectacular progress has been and continues to be made in perturbative calculations, for example \cite{Parke:1986gb,Britto:2005fq,ArkaniHamed:2008gz,Feng:2011np,Elvang:2013cua,Dixon:2013uaa,Benincasa:2013faa,Arkani-Hamed:2013jha}, we believe developing new approaches to scattering amplitudes could be useful.

In this paper, we propose a different approximation scheme for calculating scattering amplitudes.  One whose fundamental approximation is not due to an order in a perturbative series, but is rather due to a cutoff on contributions to the final result.  That is, we propose to calculate all contributions to the scattering amplitude that are above the cutoff (no matter the order of perturbation theory) but ignore all contributions below the cutoff (again no matter the perturbative order).  Of course, this is not trivial, and in order to implement it, we must make a major modification of our theory that introduces another level of approximation.  We must latticize momentum space.  Of course, we do this with the hope that we will eventually be able to extrapolate to the continuum limit.  As expected, this approach has its own limitations as we attempt to increase its precision.  It becomes significantly, exponentially more difficult the lower the cutoff is taken and the smaller the momentum spacing becomes.  Nevertheless, we believe this new method complements well the current perturbative approach.

Our proposed method is the following.  We first propose to directly calculate all contributions (to all orders of perturbation theory), that are above a cutoff, to scattering eigenstates of the Hamiltonian in a discrete momentum space.  We do this by using the quasi-sparse eigenvector (QSE) method described in \cite{Lee:2000ac}.  This is a cyclic algorithm that randomly searches the Hilbert space in the vicinity of the eigenstate for basis states contributing above the cutoff.  Briefly, the way it does this is that, during each cycle, it randomly chooses new basis states that are connected by one application of the Hamiltonian to basis states already found to be above the cutoff in previous cycles.  It then tests these new basis states by diagonalizing the Hamiltonian with respect to the new basis states as well as the basis states already found to be above the cutoff.  Further details are given in App.~\ref{app:QSE method}.  Once we have the scattering eigenstates, we propose to calculate the scattering amplitude by directly taking the inner product between the scattering eigenstates.  Finally, we propose to calculate the inner product at multiple small values of the momentum lattice spacing and extrapolate the results to the continuum.  Before giving more details, we note that although our method is effectively nonperturbative, it only works in the perturbative regime where the important basis states are connected to the main scattering basis state by a small number of applications of the Hamiltonian.  

Accomplishing our full proposal for a nontrivial theory is our long-term goal.  There will be many challenges, both expected and unexpected.  However, we have made a beginning in the present work and summarize what we have accomplished and what we have not.  We have written numerical code implementing both the QSE method and perturbation theory up to third order for comparison, for $\lambda\phi^4$ theory.  We have used this code to calculate three eigenstates: the vacuum, a two-particle eigenstate and a four-particle eigenstate, with the last two being ``high" above the vacuum and representative of scattering in and out states.  Moreover, we have compared our results with perturbation theory and with what we expect on physical grounds and found agreement (with some important limitations described in detail in the text). We have further calculated the inner product between the two scattering eigenstates and found a null result in agreement with both perturbation theory and physical expectation.  We were not able to calculate a nonzero inner product in the present work since all our eigenstates are nondegenerate.  Achieving a nonzero inner product is a major goal for a near-future work.  Further details will be given in the text.  Finally, we studied the dependence of our results on the momentum lattice spacing and found that our inner product was stable as we decrease the momentum spacing.  On the other hand, we find that a better understanding of the density of basis states is required for an extrapolation to the continuum limit.  Again, this will be a major topic of future research.  Beyond this, we also studied the dependence of our results on the coupling constant, the energy of the scattering eigenstates and, very importantly, we analyzed the efficiency of the QSE code relative to higher-order perturbation theory.  We show that the QSE method is faster than second-order perturbation theory (often orders of magnitude faster) for the entire range of parameters we studied.  Furthermore, we show that the QSE code scales better than second-order perturbation theory both as the momentum lattice spacing decreases and as the eigenstate energy increases.

To the best of our knowledge, nobody else is attempting to directly calculate scattering eigenstates of the Hamiltonian that are high above the vacuum and use them to calculate the scattering amplitude.  Our first attempt in this direction used a truncation of the Hilbert space \cite{Christensen:2016naf} via a cutoff on the energy.  However, we have since realized the futility of such an approach (see App.~\ref{sec:Hsize}) and have come to appreciate the power of the QSE method which we use in the present work.  We are not the only ones using the QSE method.  However, their purpose is different and they are not interested in the scattering eigenstates as we are \cite{Lee:2000gm,Salwen:2000gn,Lee:2000xna,Lee:2000yt,Lee:2000yu,Borasoy:2001pb,Salwen:2002dx}.  We also note that there is much important work being done on diagonalizing the Hamiltonian by using an energy cutoff to truncate the Hilbert space.  For example, in $\lambda\phi^4$ theory, \cite{Salwen:1999pw,Martinovic:2002bv,Wagner:1,Wagner:2,Hogervorst:2014rta,Rychkov:2014eea,Rychkov:2015vap,Elias-Miro:2015bqk,Bajnok:2015bgw,Katz:2016hxp,Burkardt:2016ffk,Anand:2017yij}.  Again, their emphasis is different.  They are not interested in scattering eigenstates far above the vacuum and their associated S-Matrix elements.

The outline of this paper is as follows.  In Sec.~\ref{sec:results}, we describe our three eigenstates of the Hamiltonian and their comparison with first-order perturbation theory.  In Sec.~\ref{sec:Higher Order}, we compare with second- and third-order perturbation theory.  In Sec.~\ref{sec:lambda}, we describe the dependence of our results on the coupling constant of the theory.    In Sec.~\ref{sec:dp}, we describe how our results depend on the momentum lattice spacing.  In Sec.~\ref{sec:Hsize}, we explore the dependence of the QSE code on the size of the reduced Hilbert space (see App.~\ref{app:QSE method}).  We also brielfy look at the dependence on the scattering eigenstate energy.  In Sec.~\ref{sec:conclusions}, we summarize our results and conclude.  In App.~\ref{sec:truncated}, we describe why diagonalizing the Hamiltonian through brute force methods in a truncated Hilbert space is hopeless for scattering eigenstates.  In App.~\ref{app:QSE method}, we review the QSE method and give some details of our numerical code. In App.~\ref{app:renormalization of the mass}, we describe the Hamiltonian of our $\lambda\phi^4$ theory.  In App.~\ref{app:perturbative solution}, we review perturbation theory.

\section{\label{sec:results}Results and Comparison with First-Order Perturbation Theory}
\begin{figure*}[!]
\begin{center}
\includegraphics[scale=0.89]{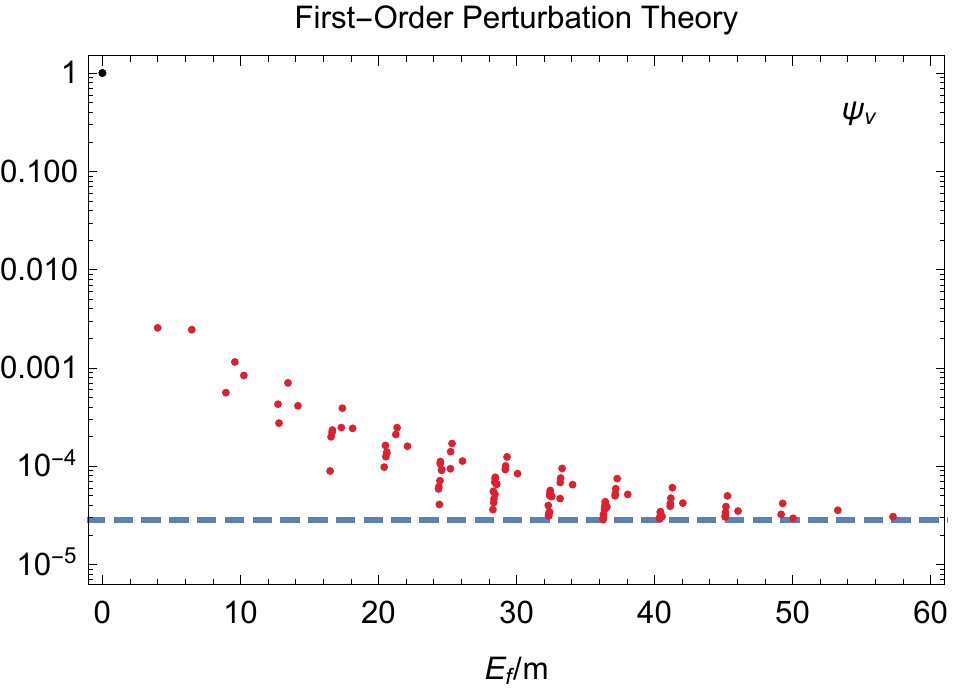}
\hfill
\includegraphics[scale=0.89]{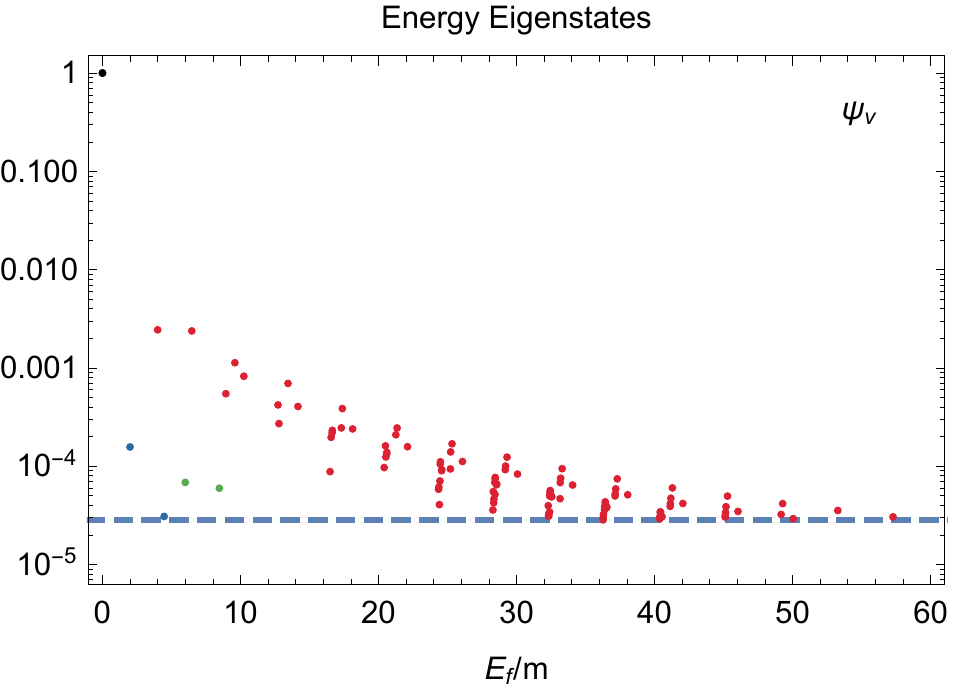}\\
\includegraphics[scale=0.89]{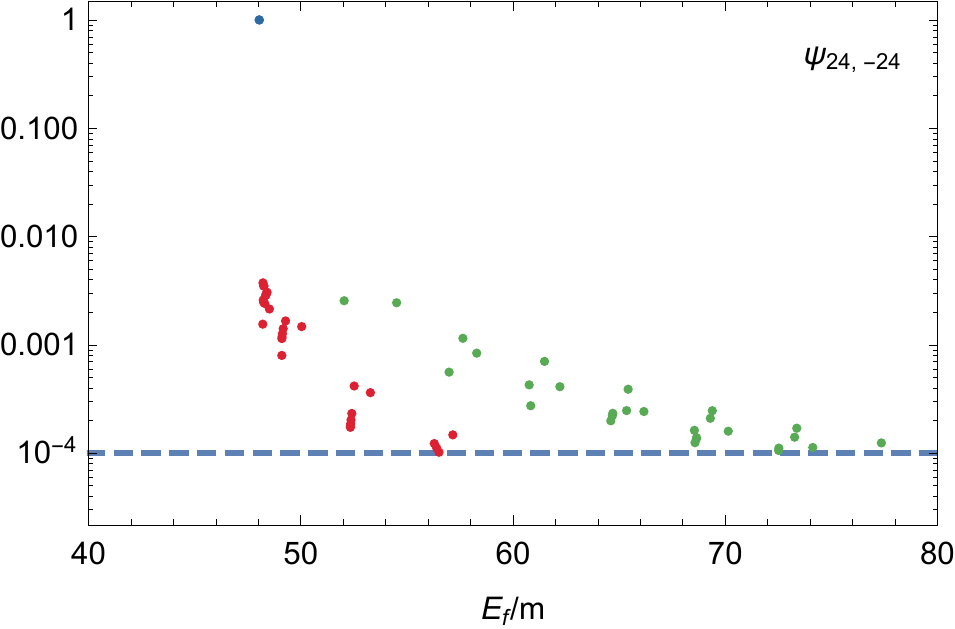}
\hfill
\includegraphics[scale=0.89]{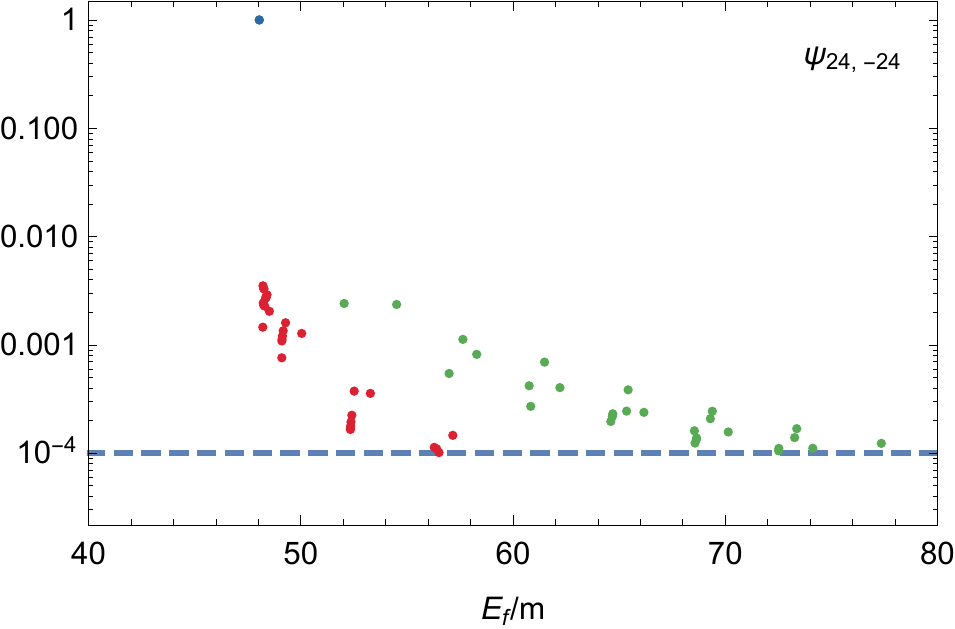}\\
\includegraphics[scale=0.89]{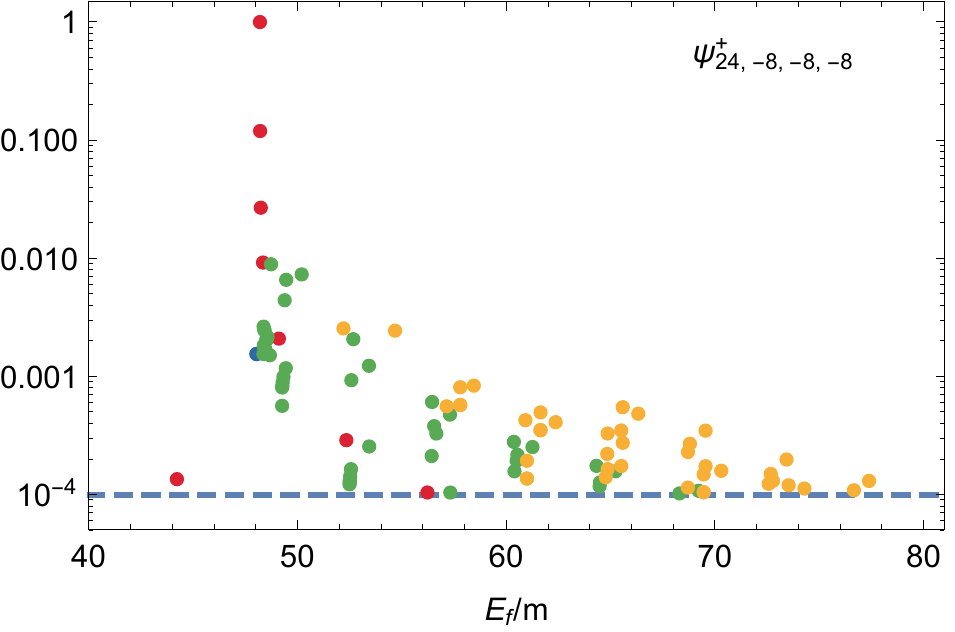}
\hfill
\includegraphics[scale=0.89]{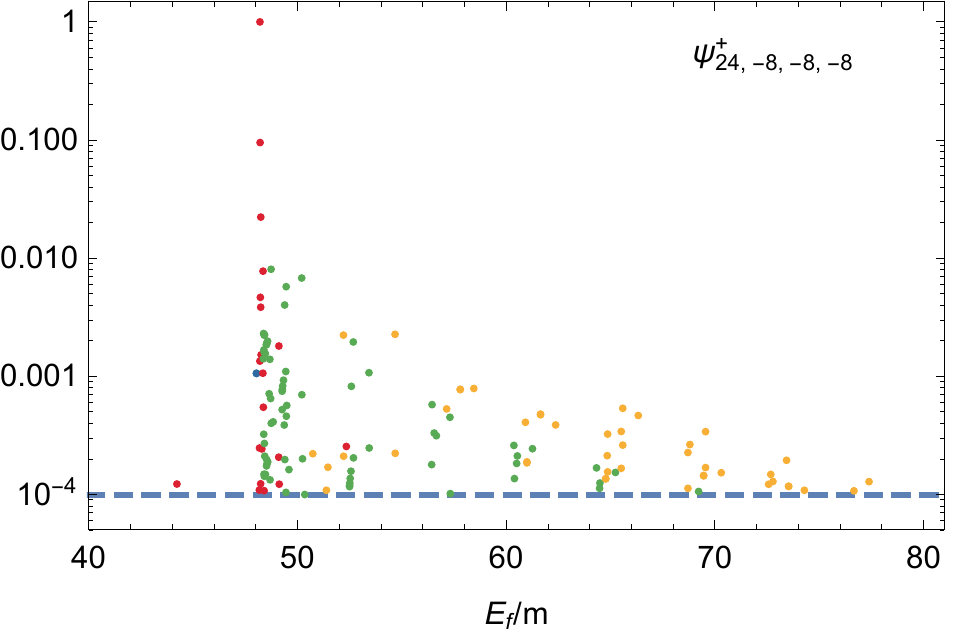}
\end{center}
\caption{\label{fig:Energy Eigenstates}Eigenstates of the Hamiltonian given in Eq.~(\ref{eq:Discrete Hamiltonian}) with $\Delta p=2m$ and $\lambda=0.1m^2$, where $m$ is the mass parameter.  The top row is of the vacuum $\Psi_v$, the middle row is the two-particle eigenstate $\Psi_{24,-24}$, and the bottom row is the four-particle eigenstate $\Psi^+_{24,-8,-8,-8}$.  The black, blue, red, green, yellow and orange dots represent 0-, 2-, 4-, 6-, 8- and 10-particle basis states, respectively.  The horizontal axis gives the free energy of the basis states divided by the mass while the vertical axis gives the absolute value of the coefficient of the basis state.  The dashed blue line gives the cutoff used in the calculation.}
\end{figure*}
In order to make this concrete, we have focused on three eigenstates of the Hamiltonian [Eq.~(\ref{eq:Discrete Hamiltonian})].  The first is the vacuum, which is the state of lowest energy.  We call this eigenstate $\Psi_v$.  It can be seen in the first row of Fig.~\ref{fig:Energy Eigenstates}.  (We will describe this figure in detail throughout this section.)  The second is the two-particle eigenstate where the two particles have equal but opposite momentum.  We chose the magnitude of each momentum to be $24m$, where $m$ is the mass parameter.  We label this eigenstate $\Psi_{24,-24}$.  For our numerical calculations, we have taken $m=1$.  The reason we chose this state was that it was a scattering-type eigenstate far above the vacuum where the Hilbert space is too large to directly diagonalize on a personal computer, yet it was still low enough that second- (and even third-) order perturbation theory could still be achieved for comparison.  This two-particle state is shown in the second row of Fig.~\ref{fig:Energy Eigenstates}.   The third state we chose is the parity-even four-particle eigenstate $\Psi^+_{24,-8,-8,-8}=(\Psi_{24,-8,-8,-8}+\Psi_{8,8,8,-24})/\sqrt{2}$.  It is a linear combination of two eigenstates related by a parity transformation.   The first has one particle of momentum $24m$ and three particles of momentum $-8m$ and the second has the same momenta  but with the opposite sign.  We chose this eigenstate because it was close in energy to our two-particle eigenstate, and was thus a potential final state of a scattering event.  In other words, we were interested in calculating an S-Matrix element between a two-particle eigenstate and a four-particle eigenstate and this seemed like a good candidate for that calculation.  We display this four-particle eigenstate in the third row of Fig.~\ref{fig:Energy Eigenstates}.

The left column of Fig.~\ref{fig:Energy Eigenstates} is the result of a first-order perturbation-theory calculation (see App.~\ref{app:perturbative solution} for a brief review) while the right column is the result of ten iterations of our cyclic QSE code (described in App.~\ref{app:QSE method}).  Although they are very similar, there are small, but significant, differences.  Our code fills in basis states missed by first-order perturbation theory.  We will show (see Sec.~\ref{sec:Higher Order}) that these points are included in second- and third-order perturbation theory, but first, we would like to describe these plots for each eigenstate and the differences between the left and right columns (the results of first-order perturbation theory and our QSE code, respectively.)  The vertical axis of these plots is the absolute value of the coefficient [e.g. see Eq.~(\ref{eq:pert coeff})] of the basis states while the horizontal axis gives the free-particle energy of the basis states normalized by the mass.  The numerical calculations done to produce these plots used a momentum spacing of $\Delta p=2m$.  This is a rather large momentum spacing and requires explanation since smaller values of $\Delta p\ll m$ give better physical results as described in \cite{Christensen:2016naf}.  Although this is true, perturbation theory becomes impossible for these scattering states at such small momentum spacings and a critical objective of the current paper is to compare this method with perturbation theory for these scattering states.  The final parameter in our calculation is the coupling constant $\lambda$.  We took a value of $\lambda=0.1m^2$ for the plots in Fig.~\ref{fig:Energy Eigenstates}.  We will consider other values of $\lambda$ in Sec.~\ref{sec:lambda}.

We begin by describing the vacuum, which is the top row of Fig.~\ref{fig:Energy Eigenstates}.  This is the state of lowest energy.  We found its energy to be 0 at first order in perturbation theory and -0.000119m in our cyclic QSE code.  This is inline with higher orders in perturbation theory as we will describe in the next section.  The color coding of the points is as follows: black, blue, red, green and yellow dots represent 0-, 2-, 4-, 6- and 8-particle basis states.   We see that the vacuum is dominated by a 0-particle basis state, with a coefficient of very nearly 1 (all of the eigenstates described in this paper are normalized so that the sum of the squares equals 1).  If the coupling constant $\lambda$ were equal to zero, the vacuum would be identically equal to the 0-particle basis state and the contribution from all other basis states would vanish.  As the coupling constant is turned on and grows in size, the contribution from the other basis states grows along with it.  The next most important basis states to the vacuum are 4-particle basis state seen in red.  The left-most red point is the basis state with 4 free particles at rest.  It contributes to the vacuum with a coefficient of approximately -0.00255 at first order in perturbation theory (in the left plot) and -0.00244 in our QSE code (in the right plot).  (We have normalized the overall phase so that the coefficient of the 0-particle basis state is positive).  The basis state directly to its right has 2 free particles at rest and 2 free particles with momenta of $\pm 2m$.  It's coefficient is also approximately -0.00244 at first-order in perturbation theory (in the left plot) and -0.00237 in our QSE code (in the right plot).  We will describe the differences between these coefficients from perturbation theory and our code in greater detail in Sec.~\ref{sec:Higher Order}.  Other 4-particle basis states of greater complexity can be seen exponentially falling off in importance to the right in these plots.  Another feature of these plots is that the points come in clusters with a horizontal spacing between the groups.  The reason for this is the rather large value of $\Delta p$ that we chose.  In Sec.~\ref{sec:dp}, we will show plots with smaller values of $\Delta p$, where this feature will be less pronounced or disappear altogether.  

The largest difference between the left plots and the right plots is the points that are present in the cyclic QSE code (on the right) but not in first-order perturbation theory (on the left).  For the vacuum, there are no 2-particle basis states (blue points) at first order in perturbation theory.  This may seem strange but can be easily understood.  At first order in perturbation theory, the coefficient of these 2-particle basis states is proportional to $\langle b_{2p}|V|b_{0p}\rangle$ where $|b_{2p}\rangle$ is the 2-particle basis state we are interested in and $|b_{0p}\rangle$ is the 0-particle basis state [see Eq.~(\ref{eq:app:pert:cj1})].  $V$ is the potential of our Hamiltonian and is the second term of Eq.~(\ref{eq:Discrete Hamiltonian}).  In order to give a nonzero result, the potential $V$ would have to add 2 free particles to the 0-particle basis state.   The only term in the Hamiltonian that could do this is the fourth term of the potential, the term with $a^\dagger_{p_1}a^\dagger_{p_2}a^\dagger_{p_3}a_{-p_4}$.  However, we see that this term has an annihilation operator at the right-most position and, therefore, annihilates the 0-particle basis state on the right, giving zero for this coefficient.  As a result, the 2-particle basis states do not contribute to the vacuum until at least second order in perturbation theory.  Our QSE code has no problem discovering these points and found two above the cutoff.  The left-most blue point is the basis state with two free particles at rest and has a coefficient of 0.000156 while the one directly below and to the right of it has two free particles with momenta $\pm 2m$ and has a coefficient of 0.000031.  Other 2-particle basis states contribute below the cutoff.  Finally, we also see contributions from 6-particle basis states (green points) in the plot on the right.  These were also missed by first-order perturbation theory for a similar reason.  Their coefficients are proportional to $\langle b_{6p}|V|b_{0p}\rangle$ and $\langle b_{8p}|V|b_{0p}\rangle$, respectively, where $|b_{6p}\rangle$ is a 6-particle basis state and $|b_{8p}\rangle$ is an 8-particle basis state.  For these to be nonzero, the potential would have to contain operators that created 6 and 8 particles, respectively.  However, the potential in Eq.~(\ref{eq:Discrete Hamiltonian}) does not have any operators of this form.  All of these basis states (missing in first-order perturbation theory but present in our QSE code) receive contributions at second order in perturbation theory.  Their coefficients at second order, for example, are proportional to $\langle b_j|V|\psi^1\rangle$ [see Eq.~(\ref{eq:app:pert:cj2})] where $|b_j\rangle$ is any of the basis states that contributes only at second order and $\psi^1$ is the first-order perturbative eigenstate.  For the vacuum, the left-most green point is the basis state with six free particles at rest.  It's coefficient is 0.000068.  The green point just to the right has four free particles at rest and two free particles with momenta $\pm 2m$ with a coefficient of 0.000059.  Other green points have exponentially smaller contributions as the free energy of the basis states increases to the right.  

The second row of Fig.~\ref{fig:Energy Eigenstates} shows the eigenstate with two particles of momenta $\pm24m$.  We found its energy to be 48.0417m at first order in perturbation theory and 48.0416m in our cyclic QSE code.  As expected, it is dominated by the basis state with two free particles of momenta $\pm24m$, which has a coefficient of 0.99994 at first order in perturbation theory (on the left) and the same 0.99994 in our cyclic QSE code (on the right).  If the coupling constant $\lambda$ were set equal to zero, this coefficients would become 1 and all others would vanish.  The 0-particle basis state does not contribute to this state at first order in perturbation theory.  The reason is that it's contribution is proportional to $\langle|V|24m,-24m\rangle$.  This would only be nonzero if $V$ contained an operator that removed two free particles.  The only operator in Eq.~(\ref{eq:Discrete Hamiltonian}) that does this is $a^\dagger_{p_1}a_{-p_2}a_{-p_3}a_{-p_4}$ which completely annihilates $|24m,-24m\rangle$, and so gives zero.  (This is the hermitian conjugate of the reason the 2-particle basis state did not appear in the vacuum at first order.) It does contribute at higher order, however, and our cyclic QSE code discovers it but finds its contribution to be below the cutoff.  Therefore, it does not appear in either plot.  The most important 4-free-particle basis states, that are above the cutoff, are found by first-order perturbation theory and appear in both plots.  The most important one is $|24m,-6m,-8m,-10m\rangle_+$ [where throughout this paper $|p_1,p_2,\cdots\rangle_+=\left(|p_1,p_2,\cdots\rangle+|-p_1,-p_2,\cdots\rangle\right)/\sqrt{2}$], whose coefficient is -0.00373 at first order in perturbation theory and -0.00351 in our cyclic QSE code.  After this is the basis state $|24m,-4m,-6m,-14m\rangle_+$ with a coefficient of -0.00348 at first order in perturbation theory and -0.00328 in our cyclic QSE code.  Other 4-particle basis states have coefficients that fall off exponentially as their free energy increases or are below the cutoff entirely.  This is true for the other eigenstates as well.  The 6-particle basis states begin with the basis state $|24m,-24m,0,0,0,0\rangle$ which contributes with a coefficient of -0.00255 at first order in perturbation theory and -0.00241 in our cyclic QSE code.  Following this is the basis state $|24m,-24m,1m,0,0,-1m\rangle$ with a coefficient of -0.00244 at first order in perturbation theory and -0.00236 in our cyclic QSE code.  Other 6-particle basis states contribute at exponentially smaller levels as their free energy increases as can be seen in the figure. 

The third row of Fig.~\ref{fig:Energy Eigenstates} shows the parity-even 4-particle eigenstate $\Psi^+_{24,-8,-8,-8}$.  We found its energy to be 48.2095m at first order in perturbation theory and 48.2090m in our cyclic QSE code.  As expected, its most significant contribution is from the parity-even basis state $|24m,-8m,-8m,-8m\rangle_+$ with a coefficient of 0.99239 at first order in perturbation theory and 0.99508 in our QSE code.  The next two most important basis states are also 4-particle basis states.  The first is the basis state $|24m,-6m,-8m,-10m\rangle_+$ with a coefficient of 0.11869 at first order in perturbation theory and 0.09477 in our QSE code.  After this is the basis state $|24m,-4m,-8m,-12m\rangle_+$ with a coefficient of 0.02662 at first order in perturbation theory and 0.02217 in our QSE code.  There are several more 4-particle basis states contributing at lower values.  Some appear only in the right plot.  An example of this is the basis state $|24m,-6m,-6m,-12m\rangle_+$.  The reason this does not appear at first order in perturbation theory is because its coefficient is proportional to $\langle24m,-6m,-6m,-12m|V|24m,-8m,-8m,-8m\rangle$.  For this to be nonzero, $V$ would have to have an operator that annihilates 3 free particles and creates 3 new free particles, but there is no such operator in $V$.  The 0-particle basis state does not contribute above the cutoff.  There is one 2-particle basis state above the cutoff.  It is $|24m,-24m\rangle$.  It has a coefficient of 0.00154 at first order in perturbation theory and 0.00106 in our  QSE code.  We will see in the next section, however, that these coefficients are inline with higher orders of perturbation theory.  This basis state is the main basis state for the state $\Psi_{24,-24}$.  Its large coefficient is the reason we chose this state for our study.  Other 2-particle basis states fall below the cutoff.  The largest 6-particle basis state is $|8m,8m,4m,2m,2m,-24m\rangle_+$ with a coefficient of 0.00884 at first order in perturbation theory and 0.00802 in our  QSE code.  
The next largest is $|8m,8m,8m,0,0,-24m\rangle_+$ with a coefficient of 0.00726 at first order in perturbation theory and 0.00676 in our QSE code.  Other 6-particle basis states give smaller contributions.  The most important 8-particle basis state is $|8m,8m,8m,2m,0,0,-2m,-24m\rangle_+$ with a coefficient of 0.00242 at first order in perturbation theory and 0.00226 in our QSE code.  After this comes the basis state $|8m,8m,8m,0,0,0,0,-24m\rangle_+$ which contributes with a coefficient of 0.00253 at first order in perturbation theory and 0.00222 in our  QSE code.  A few other 8-particle basis states contribute to the eigenstate with smaller coefficients.  Our code did not find any basis states with 10 or more free particles above the cutoff.  As we will discuss in the next section, neither did second- or third-order perturbation theory.

\begin{figure}[!]
\begin{center}
\includegraphics[scale=0.89]{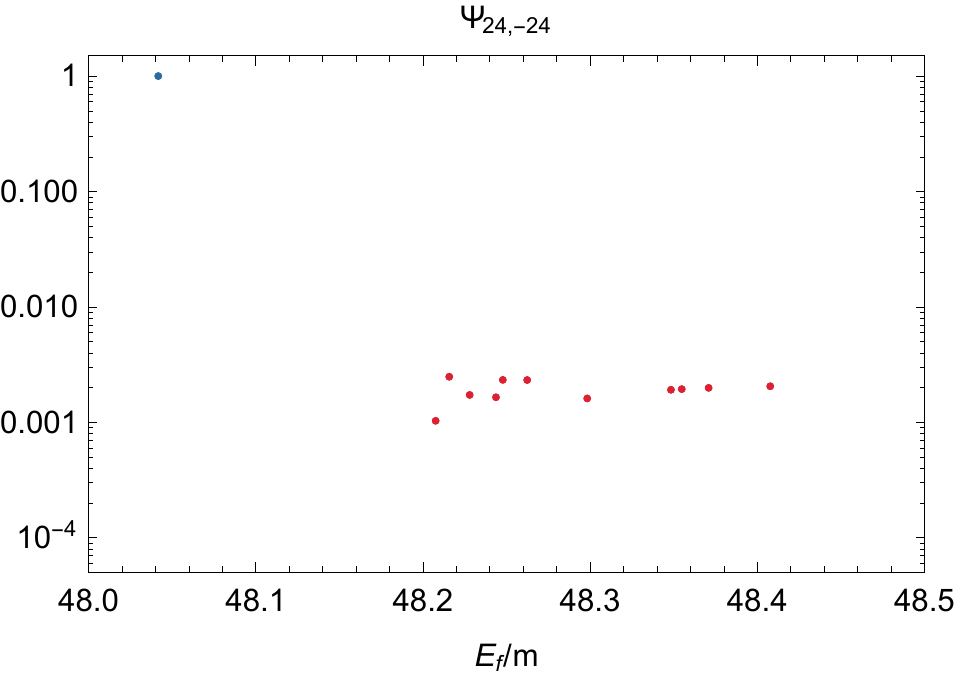}
\end{center}
\caption{\label{image:Psi-24--24}A magnification of the right middle plot of Fig.~\ref{fig:Energy Eigenstates}.}
\end{figure}

We also calculated the inner product between $\Psi_{24,-24}$ and $\Psi^+_{24,-8,-8,-8}$.  We remind the reader that the eigenvalues ($48.0416m$ and $48.2909m$ respectively) are not exactly the same and so we expect the result to be zero.  This is a symptom of the very large $\Delta p$ that we used for this calculation.  If we magnify the plot of $\Psi_{24,-24}$ around the basis state $|24m,-24m\rangle$, as in Fig.~\ref{image:Psi-24--24},
we see that there is a gap in basis states around $|24m,-24m\rangle$.  In fact, the nearest basis state in this plot is $|24m,-8m,-8m,-8m\rangle_+$ at the left-most edge of the red points.  There are other 4-particle basis states that are closer in free energy to $|24m,-24m\rangle$.  For example, the basis state $|12m,12m,-12m,-12m\rangle$ has a free energy of 48.17m.  However, its contribution to $\Psi_{24,-24}$ is below the cutoff, and therefore does not appear in this plot.  Because of this, the eigenstate $\Psi_{12,12,-12,-12}$ which is dominated by this basis state has an inner product with $\Psi_{24,-24}$ of $7\times10^{-7}$ at first order in perturbation theory and $5\times10^{-7}$ in our cyclic QSE code, both below the precision of this calculation.  In other words, their inner product is consistent with zero.  We chose, instead, to focus on the eigenstate $\Psi^+_{24,-8,-8,-8}$ because its main basis state $|24m,-8m,-8m,-8m\rangle_+$ had a larger contribution at leading order in perturbation theory to $\Psi_{24,-24}$ and would be a better test of the eigenstates.  In fact, we found that the inner product between $\Psi^+_{24,-8,-8,-8}$ and $\Psi_{24,-24}$ was 
0.00565 at first order in perturbation theory and $5.7\times10^{-6}$ in our  QSE code.  At first order in perturbation theory, we find that the result is above the precision of the calculation, which is approximately $10^{-4}$.  (The reason this is the approximate precision of the calculation is because we did not include basis states below this cutoff.)   The reason first-order perturbation theory gave a nonzero result for this inner product is that it missed several important basis states, that are above the cutoff as can be seen in the comparison between the left and right plots of Fig.~\ref{fig:Energy Eigenstates} and discussed in this section.  On the other hand, our cyclic QSE code is able to find all the basis states contributing above this cutoff, and therefore, the calculation of the inner product using our QSE code is in agreement with zero at the precision of the calculation.  These important basis states are filled in by higher orders in perturbation theory and the inner product is in agreement with zero at the precision of the calculation at higher order, as we will describe in the next section.

\section{\label{sec:Higher Order}Comparison with Higher Orders of Perturbation Theory}
\begin{figure*}[!]
\begin{center}
\includegraphics[scale=0.85]{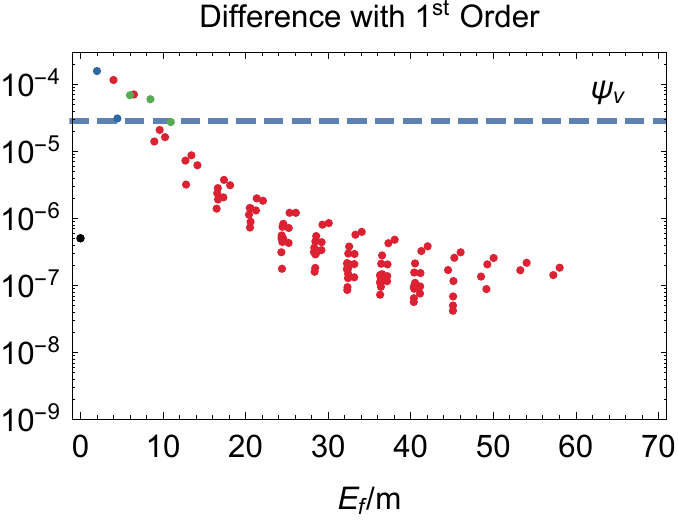}
\includegraphics[scale=0.85]{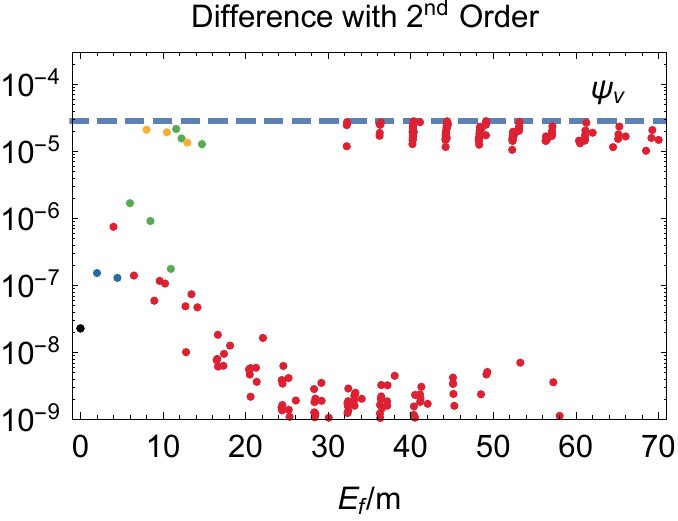}\\
\includegraphics[scale=0.85]{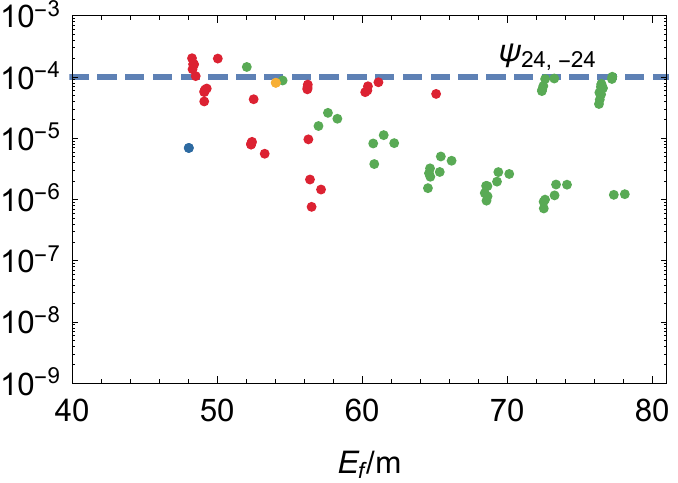}
\includegraphics[scale=0.85]{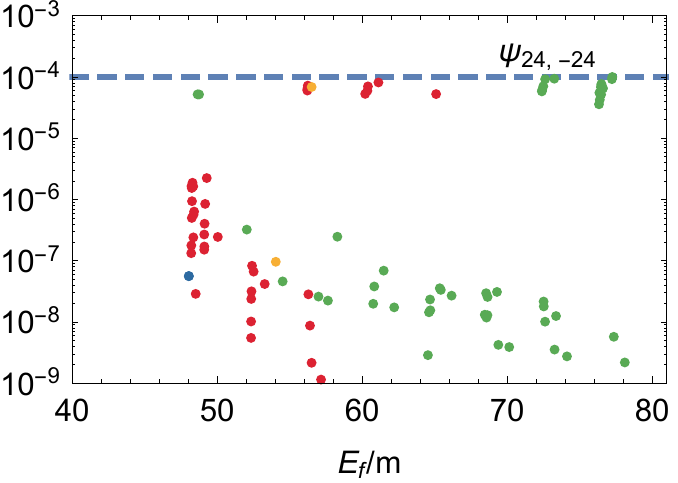}\\
\includegraphics[scale=0.85]{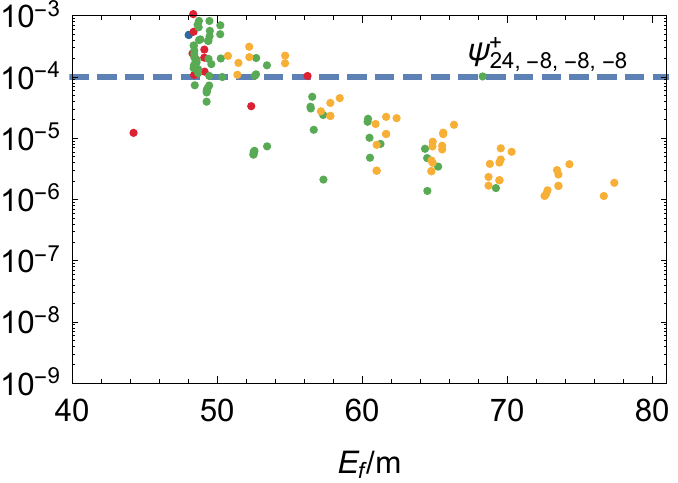}
\includegraphics[scale=0.85]{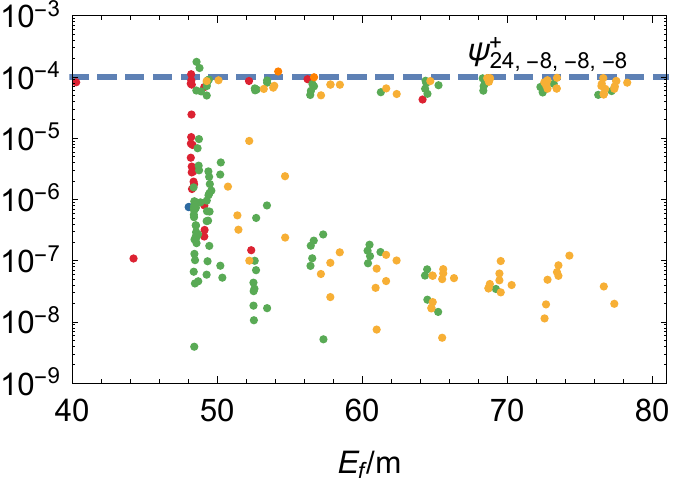}
\end{center}
\caption{\label{fig:Diffs}Plots of the difference between our cyclic QSE code and first-order (left column) and second-order (right column) perturbation theory.  The color coding, the dashed line and the horizontal axis are the same as in Fig.~\ref{fig:Energy Eigenstates}.  The vertical axis measures the absolute value of the difference between the coefficients for each basis state.  }
\end{figure*}
We also compared the results of our cyclic QSE code with second- and third-order perturbation theory.  We review perturbation theory in App.~\ref{app:perturbative solution} including a subtlety regarding nearly degenerate basis states.  However, because of its importance for our calculation, we will also briefly discuss it here.  The textbook formulas for time-independent perturbation theory (e.g. \cite{Merzbacher}) are valid when no basis states are degenerate or nearly degenerate with the main basis state.  At first order, we have no problem with this issue and that is why we have not brought it up before this point.  However, beginning at second order, this is an issue for the two- and four-particle eigenstates, $\Psi_{24,-24}$ and $\Psi^+_{24,-8,-8,-8}$.  We get very good agreement between naive perturbation theory and our  QSE code for basis states whose free energies are not nearly degenerate with the main basis state, but very bad agreement for the basis states that are nearly degenerate.  The reason for this is a breakdown of perturbation theory for the degenerate and nearly degenerate basis states.  The textbook way to handle this at first order is to explicitly diagonalize the degenerate and nearly degenerate basis states before doing perturbation theory.  This gives a slightly better than naive perturbation theory for the degenerate and nearly degenerate basis states but the perturbative result for the other basis states.  This works well at first order, but our problem does not appear until second order, where the diagonalization step would require basis states that connect the nearly degenerate basis states (via the Hamiltonian), and these intermediate basis states are not necessarily nearby in free energy.  Our approach is to find all the basis states given by the naive first- and second- (and third-) order perturbation theory, that are above the cutoff with the naive formulas.  This already includes both the nearly degenerate basis states and the basis states that connect them (that are above the cutoff).  We then simply diagonalize the Hamiltonian with respect to this entire set.  We have done this at both second and third order and find that, for our eigenstates, the results are identical since the basis states above the cutoff are the same for second and third order, and therefore, their diagonalization is the same.  Since the results of diagonalized third-order perturbation theory are  identical to those of diagonalized second-order perturbation theory, we only show plots for second order.  For the rest of this paper, we will assume that whenever we talk about perturbation theory beyond first order, we are discussing the diagonalized versions of perturbation theory.

The columns of Fig.~\ref{fig:Diffs} are as follows.  The left column shows the absolute value of the difference between our QSE-code results and first-order perturbation theory.  The right column shows the absolute difference with second-order perturbation theory.  We used the same parameters as in the previous section, $\lambda=0.1m^2$ and $\Delta p=2m$.  Once again, the reason we chose such a large value of $\Delta p$ in these first two sections was to enable comparison with perturbation theory all the way to the third order.  We will discuss timing further in Sec.~\ref{sec:dp}, but for now we note that with these parameters, third-order perturbation theory took approximately seven and a half hours and grows exponentially as $\Delta p$ becomes smaller.  The three rows represent the same eigenstates as in Fig.~\ref{fig:Energy Eigenstates} and are the vacuum $\Psi_v$ at the top, the two-particle eigenstate $\Psi_{24,-24}$ in the middle and the four-particle eigenstate $\Psi^+_{24,-8,-8,-8}$ at the bottom.  The dashed blue lines give the cutoff below which we did not keep basis states and forms an approximate precision for our calculation.  This line is the same as in Fig.~\ref{fig:Energy Eigenstates}.  We can not expect to do any better than this line and the distance below the line is not significant.  What is more important are the points above the line.  These are the basis states which are present in either perturbation theory or our QSE code but missing in the other or, if the basis state is present, both in perturbation theory and our QSE code, it is the difference between the coefficients.

We begin by focusing on the vacuum, the top row of Fig.~\ref{fig:Diffs}.  We obtained an energy of -0.000124m at second-order which is in agreement with the result from our cyclic code.  In the left plot, there are six points above the cutoff line. Two (in blue) are 2-particle basis states, two (red) are 4-particle basis states and two (green) are 6-particle basis states. The 2-(blue) and 6-(green) particle basis states are missed by first-order perturbation theory as described in Sec.~\ref{sec:results}.  Because of this, their difference with our code is the full size of their coefficient, which is above the cutoff. The two red points are found by first-order perturbation theory but are given coefficients that are slightly higher than the value achieved by our cyclic QSE code. If all we had were first-order perturbation theory, it would be impossible to tell whether our QSE code was more or less correct than perturbation theory. This is the reason we also implemented second- and third-order perturbation theory. In the right column of the first row, we see the difference with second-order perturbation theory. We immediately see that the points that were above the cutoff line are now at or below it. The 2- (blue) and 6- (green) particle basis states that were missed by first-order perturbation theory were filled in by second-order perturbation theory. Moreover, the coefficients are in agreement with those of our QSE code within the precision of our calculation. The coefficients of the four-particle basis states (in red) are also now in agreement with our QSE code.   We find the same agreement with third-order perturbation theory.  We note that where our cyclic QSE code disagrees with first-order perturbation theory, it agrees with higher orders of perturbation theory.  This suggests that our QSE code is correctly constructing the vacuum.

We now move on to the two-particle eigenstate $\Psi_{24,-24}$, which is shown in the second row of Fig.~\ref{fig:Diffs}.  The left column shows the difference with first-order perturbation theory.  We see that there are seven 4-particle basis states (in red) and one 6-particle basis state (in green) above the cut off.  All of these basis states were included in first-order perturbation theory but their coefficients are slightly higher than in our cyclic QSE code.  In the right column, on the other hand, we see that second-order perturbation theory is in agreement with our  QSE code results.  We also find the same agreement with third-order perturbation theory.  Again, the agreement with higher orders of perturbation theory suggests that our QSE code is working properly.

\begin{figure*}[!]
\begin{center}
\includegraphics[scale=0.8]{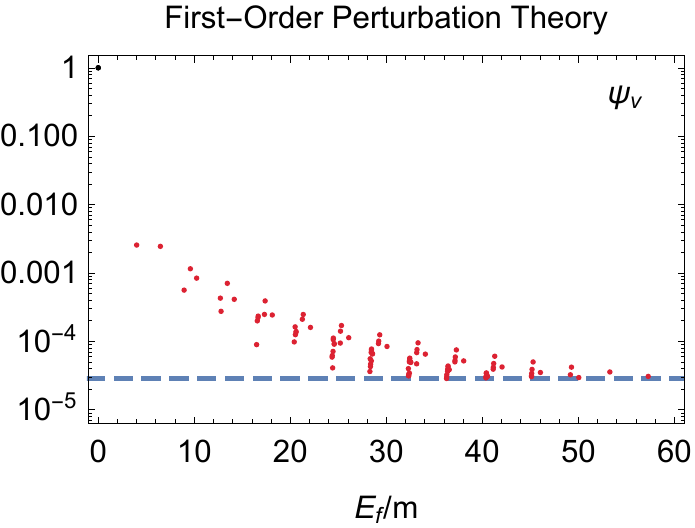}
\includegraphics[scale=0.8]{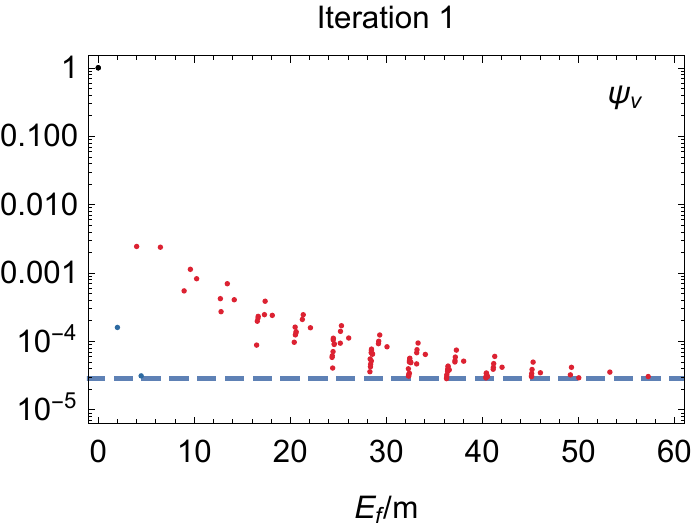}
\includegraphics[scale=0.8]{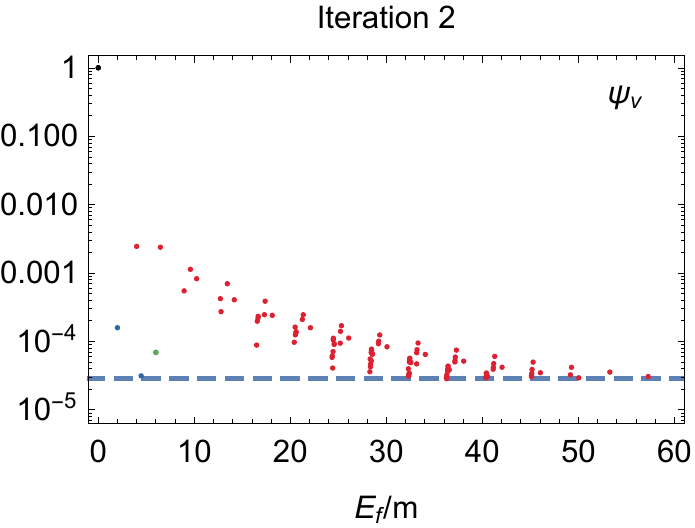}\\
\includegraphics[scale=0.8]{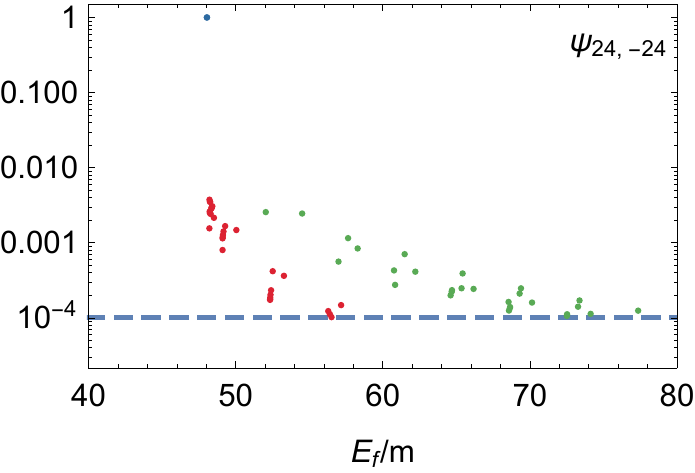}
\includegraphics[scale=0.8]{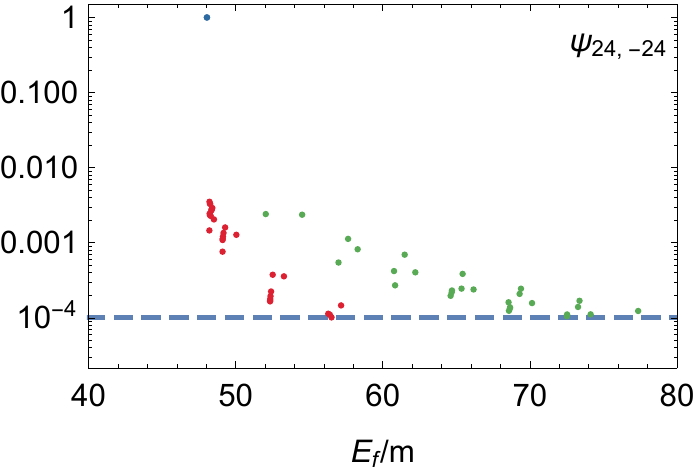}
\includegraphics[scale=0.8]{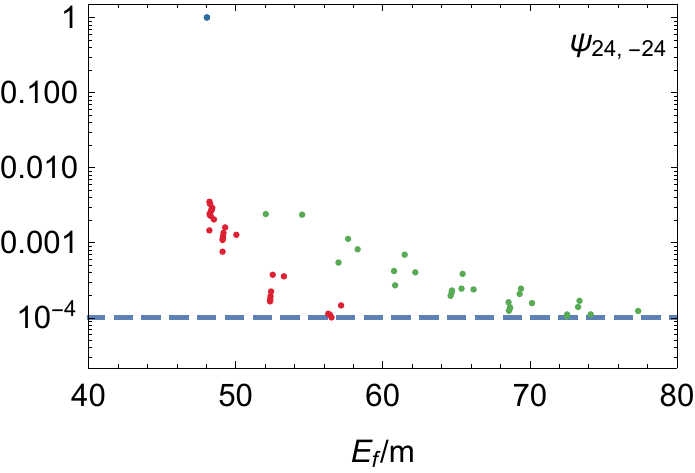}\\
\includegraphics[scale=0.8]{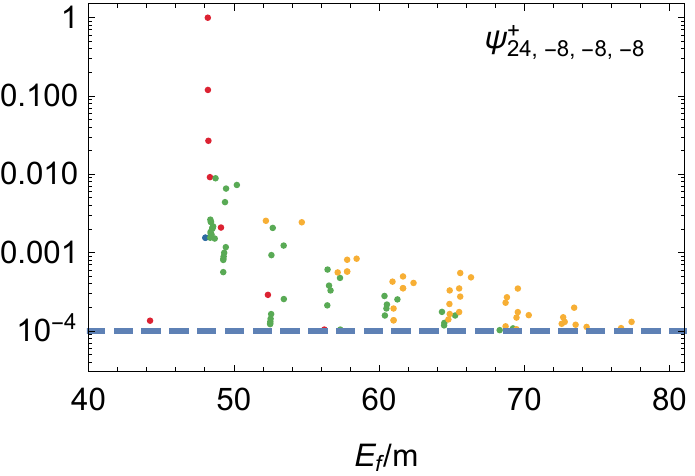}
\includegraphics[scale=0.8]{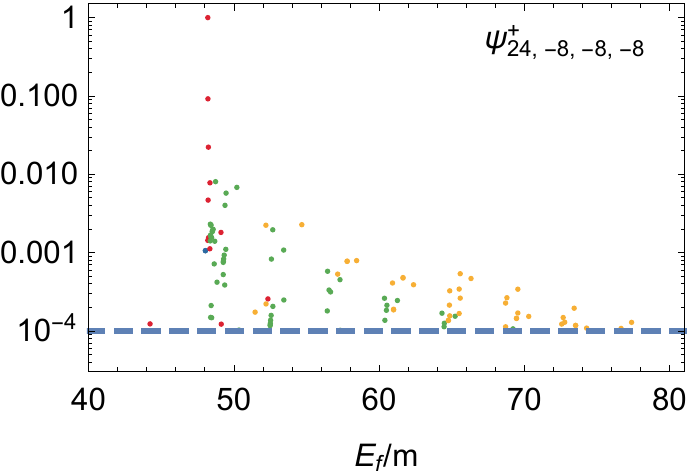}
\includegraphics[scale=0.8]{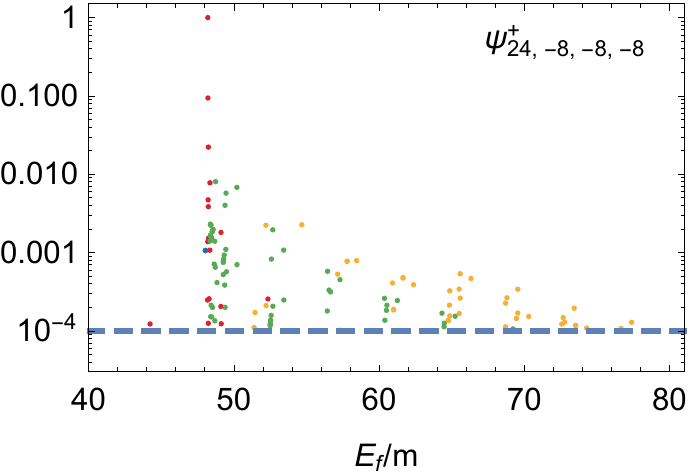}
\end{center}
\caption{\label{fig:Iterations}Plots of the zeroth, first and second iteration of our cyclic QSE code.  The left column is the result of the first order of perturbation theory and is exactly the same as the left column of Fig.~\ref{fig:Energy Eigenstates}.  The middle column is the result of the first iteration of our QSE code and the right column is the result of the second iteration of our QSE code.  The rows, the color coding, the dashed line and the axes are the same as in Fig.~\ref{fig:Energy Eigenstates}. }
\end{figure*}

The four-particle eigenstate $\Psi^+_{12,-4,-4,-4}$, shown in the final row of Fig.~\ref{fig:Diffs}, is similar to the previous two cases, but with a new twist.  At first order of perturbation theory, shown in the left column, there are one 2- (in blue), ten 4- (in red), thirty-seven 6- (in green) and seven  8- (in yellow) particle basis states above the cutoff.  All of these, with the exception of thirteen 6- and two 8-particle basis states, are missed by first-order perturbation theory but found by our cyclic QSE code and therefore their difference is above the cutoff.  The other fifteen, the thirteen 6- and two 8-particle basis states not missed, were found by both first-order perturbation theory and our QSE code but had slightly different coefficients at first order in perturbation theory.  All of the basis states missed by first-order perturbation theory are found at second and third order and all their coefficients are in agreement with our QSE code within the precision of our calculation.  The coefficients of the thirteen 6- and two 8-particle basis states are also brought into agreement with our QSE code at second and third order.  However, there are four basis states that are missed by both our QSE code and first-order perturbation theory but are found at second and third order.  Two of these are 6-particle basis states (green points), one is a 4-particle basis state (red point) and one is a 10-particle basis state (orange point).  They can be seen above the blue dashed line in the right plot.  The highest of these is the basis state $|24,-4,-4,-4,-4,-8\rangle_+$.  To understand the reason that our cyclic QSE code missed these points, we remind the reader of two points described in App.~\ref{app:QSE method}.  The first is that our code uses a random procedure to discover the important basis states and, therefore, it is always possible that basis states are missed.  However, it is also important to note that the way our random procedure chooses new points emphasizes basis states with higher coefficients more strongly than basis states with lower coefficients.  Therefore, it is much more likely that a basis state near the precision cutoff is missed than a basis state high above the cutoff.  Indeed, the coefficient of this basis state is $1.8\times10^{-4}$ (as determined by second- and third-order perturbation theory), which is just barely above the cutoff, which is approximately $1\times10^{-4}$.  Whether these points are discovered by our cyclic QSE code or not is very sensitive to the size of the coupling constant $\lambda$ and the number of new basis states randomly added to the reduced Hilbert space each cycle.  We will describe these effects further in Secs.~\ref{sec:lambda} and \ref{sec:Hsize}, respectively.

In the previous paragraphs, we have described the differences between perturbation theory and the tenth and final iteration of our cyclic QSE code.  We would now like to explore how our cyclic QSE code builds up its final solution as the iterations progress.  In Fig.~\ref{fig:Iterations}, we show the first two iterations of our cyclic QSE code for each of the three states under investigation.  In the first column, we display the results of first-order perturbation theory for comparison.  This is the initial configuration that our QSE code works with.  It is the same as the first column of Fig.~\ref{fig:Energy Eigenstates}.  We include it for the convenience of the reader.  In the second and third columns, we show, respectively, the result of the first and second iterations of our cyclic QSE code.  As the iterations progress, our results quickly approach second- and third-order perturbation theory, often filling in most of the missing points within the first few iterations.  Focusing on the vacuum (the first row), we remind the reader that first-order perturbation theory misses two 2- (blue) and two 6- (green) particle basis states.  We can see these points in the top left plot of Fig.~\ref{fig:Diffs}.
\begin{figure}[!]
\begin{center}
\includegraphics[scale=0.89]{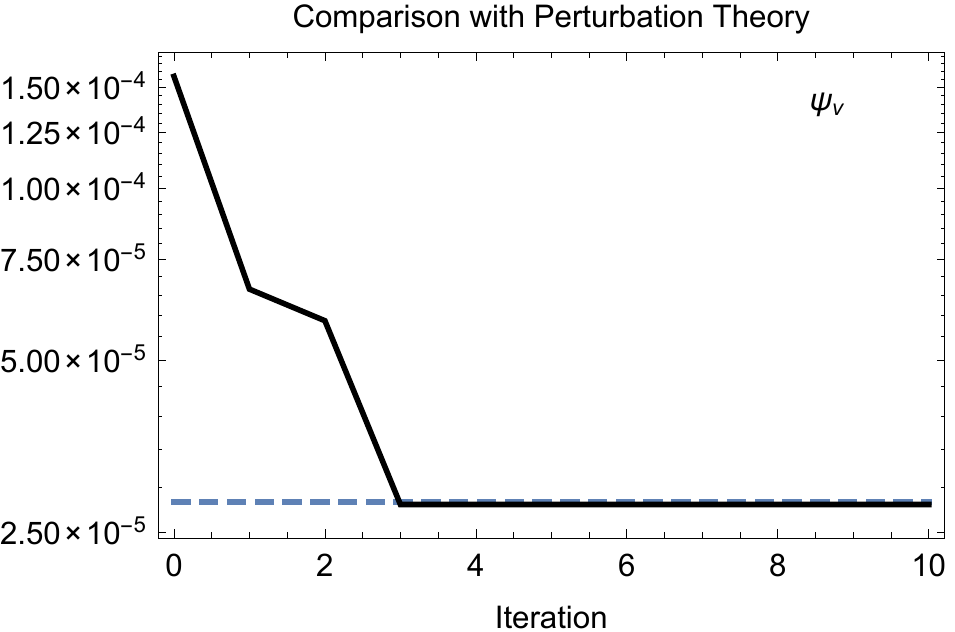}\\
\includegraphics[scale=0.89]{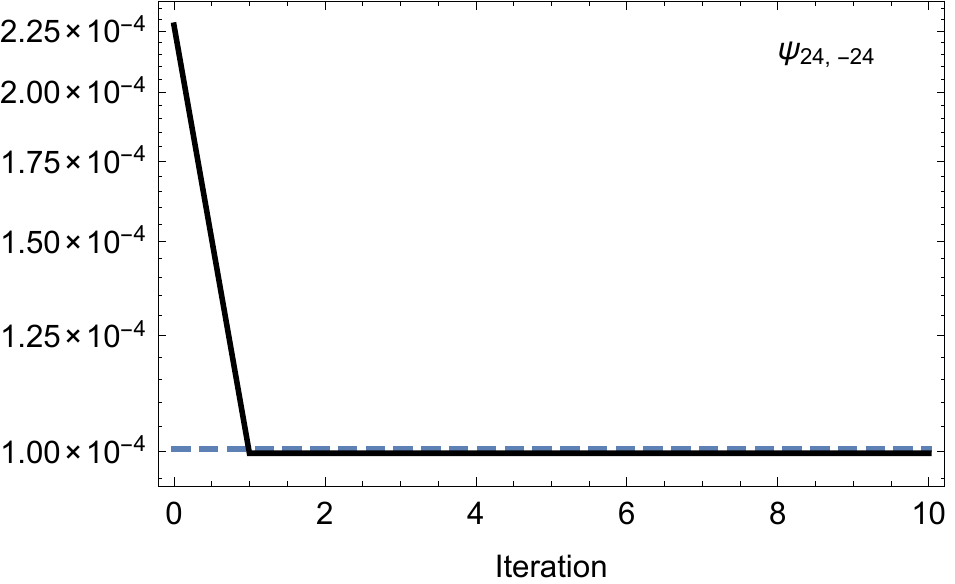}\\
\includegraphics[scale=0.89]{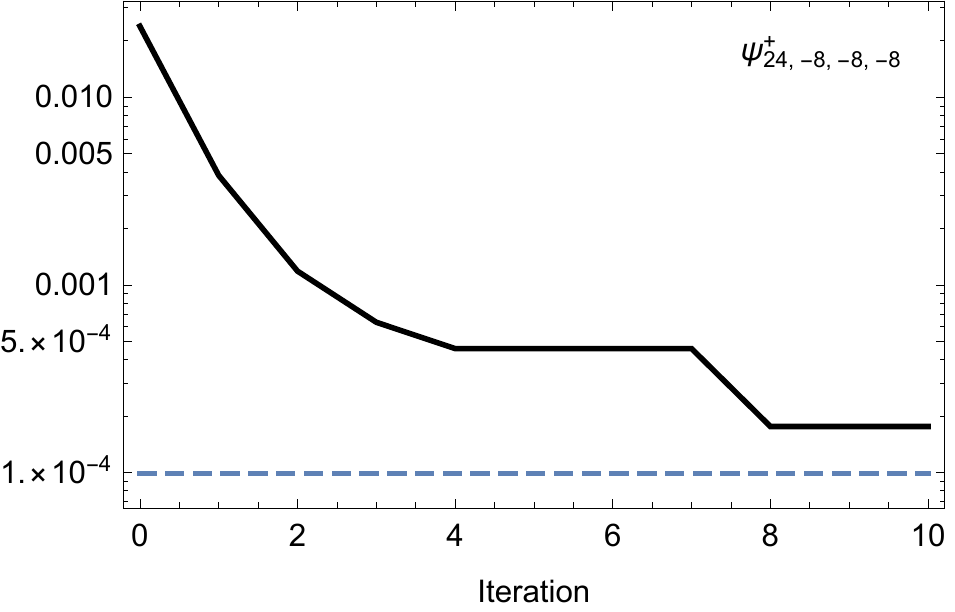}
\end{center}
\caption{\label{fig:DiffVsIt}Plots of the largest differences between our cyclic QSE code and second-order perturbation theory for each iteration.  The blue dashed line is the cutoff used in our calculations.}
\end{figure}
  As we can see in the middle plot of the first row of Fig.~\ref{fig:Iterations}, the first iteration of our cyclic QSE code has already found both 2-particle basis states (in blue).   By the second iteration, shown in the right plot, one of the 6-particle basis states (in green) has also been found.  The other missing 6-particle basis state is found in the third iteration.  In the top row of Fig.~\ref{fig:DiffVsIt}, we show the largest difference between our cyclic QSE code and second-order perturbation theory as a function of the iteration.  We see that after the third iteration the largest difference becomes equal with the precision of our calculation and thus our cyclic QSE code is in agreement with second- and third-order perturbation theory. From this point on, all of the important basis states have been found and the cyclic QSE code is randomly generating less important basis states below the cutoff.  %that are causing small fluctuations below the cutoff.    

The two-particle eigenstate $\Psi_{24,-24}$ (shown in the second row of Fig.~\ref{fig:Iterations}) has the unique property  that it initially begins with all of the important basis states.  Consequently, our cyclic QSE code does not add any new basis states above the cutoff.  However, the coefficients coming from first-order perturbation theory are not yet correct.  At the first iteration of our cyclic QSE code, although no new basis states are found above the cutoff, our code directly diagonalizes the Hamiltonian with respect to these basis states (as well as some random basis states below the cutoff).  This is already sufficient to bring this state into agreement with second- and third-order perturbation theory, as we can see in the second row of Fig.~\ref{fig:DiffVsIt}.  After the first iteration, our cyclic QSE code continues to randomly add basis states below the cutoff.  %, which cause the small random fluctuations below the dashed blue line as seen in the second row of Fig.~\ref{fig:DiffVsIt}.  
Because no new important basis states are added, and the change to the coefficients is so small, the three plots in the second row of Fig.~\ref{fig:Iterations} are indistinguishable by eye.

Turning now to the four-particle eigenstate $\Psi^+_{24,-8,-8,-8}$, the initial iterations are shown in the third row of Fig.~\ref{fig:Iterations}.  As described earlier in this section, one 2- (in blue), ten 4- (in red), thirty-seven 6- (in green) and seven  8- (in yellow) particle basis states are missing at first order in perturbation theory, and consequently, in the left-most plot.  We can see in the middle plot that three of the missing 4-particle basis states, five of the missing 6-particle basis states, and two of the missing 8-particle basis states have been filled in by the first iteration of our cyclic QSE code.  In the second iteration, it further adds four 4-particle basis states, six 6-particle basis states, and one 8-particle basis state, as can be seen in the right-most plot.  The rest of the missing basis states that are found by our cyclic QSE code appear in the following order: six basis states are found in the third iteration, eight basis states are found in the fourth iteration, zero basis states are found in the fifth through seventh iteration, three basis states are found in the eighth iteration and no basis states are found in the ninth and tenth iterations.  In addition to finding missing basis states, the diagonalization of the Hamiltonian improves the coefficients of the basis states as in the previous cases.  Since our cyclic QSE code does not find the final four basis state (two green, one red and one orange point above the dashed line in the bottom right plot of Fig~\ref{fig:Diffs}), the largest difference between our cyclic QSE code and second- and third-order perturbation theory never goes below the cutoff line, although it gets very close, as seen in the bottom plot of Fig.~\ref{fig:DiffVsIt}.  The height of the black line above the dashed blue line is precisely due to the cyclic QSE code missing these basis states.  In Sec.\ref{sec:Hsize},  we will describe how this basis state can be found by our cyclic QSE code by increasing the number of new basis states randomly chosen each cycle.

\section{\label{sec:lambda}Dependence on $\mathbf{\lambda}$}
\begin{figure*}[!]
\begin{center}
\includegraphics[scale=0.83]{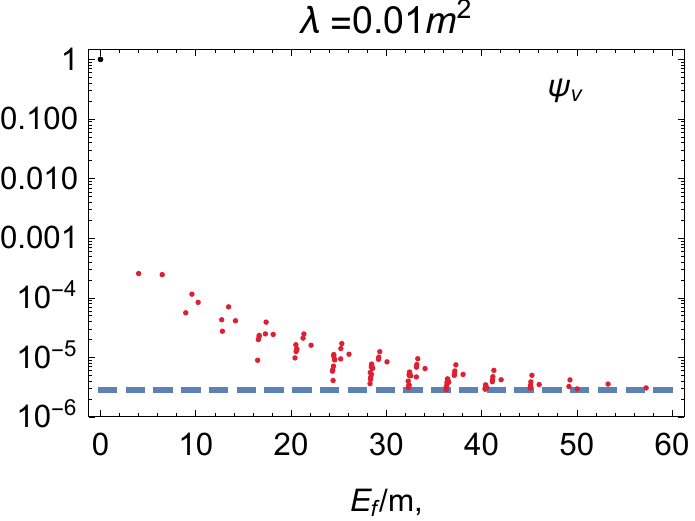}
\includegraphics[scale=0.83]{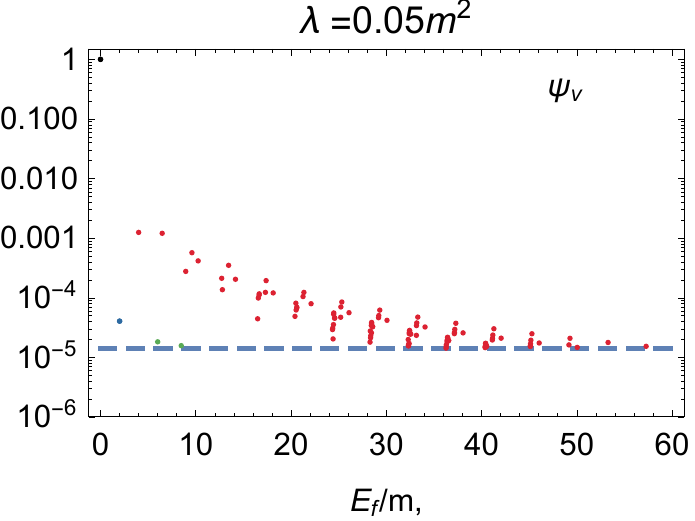}
\includegraphics[scale=0.83]{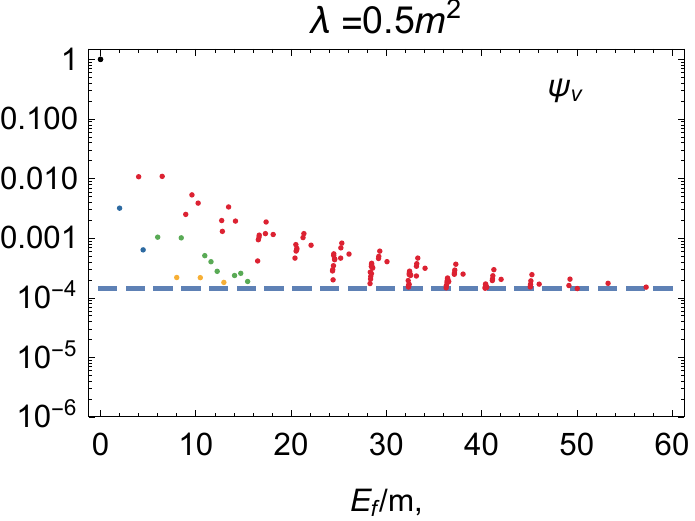}\\
\includegraphics[scale=0.83]{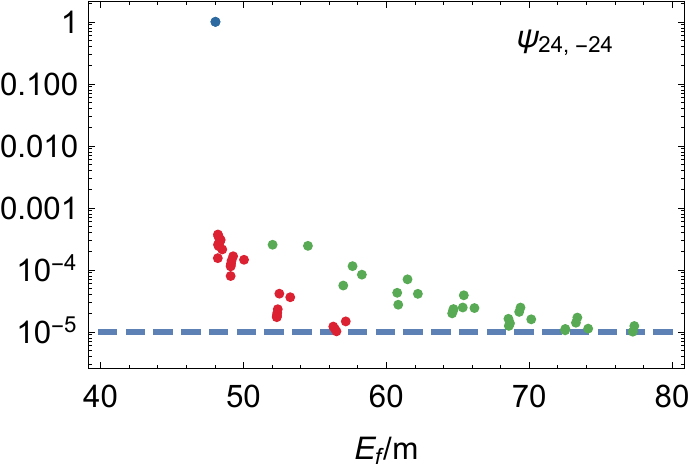}
\includegraphics[scale=0.83]{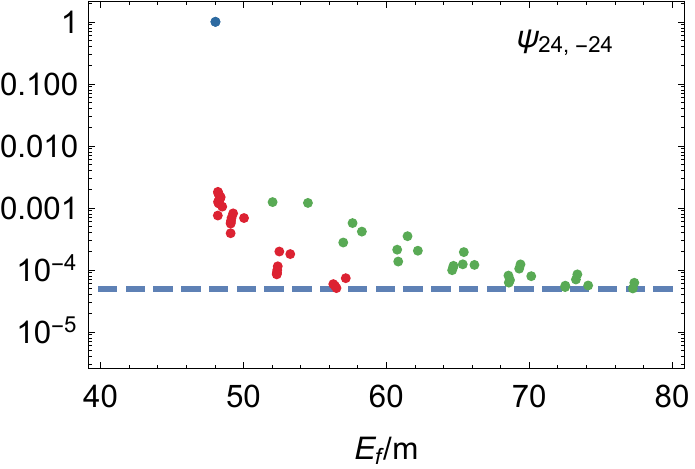}
\includegraphics[scale=0.83]{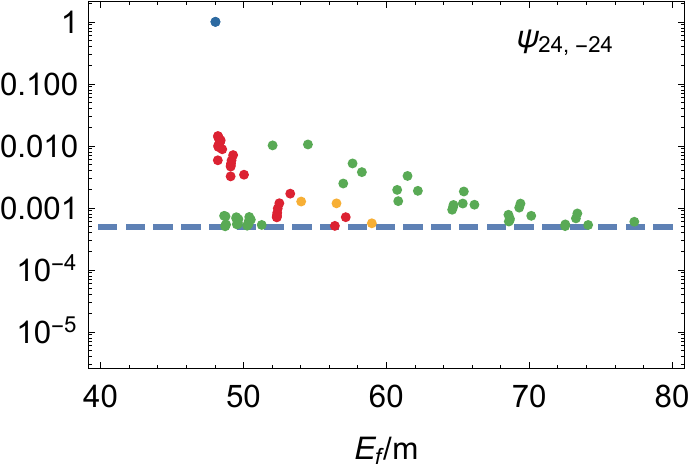}\\
\includegraphics[scale=0.83]{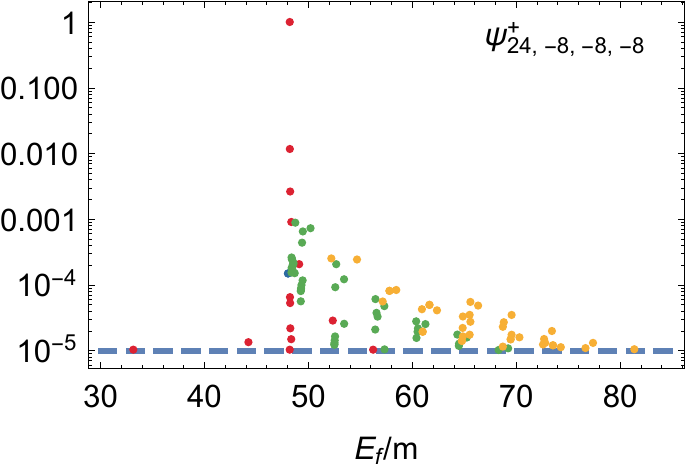}
\includegraphics[scale=0.83]{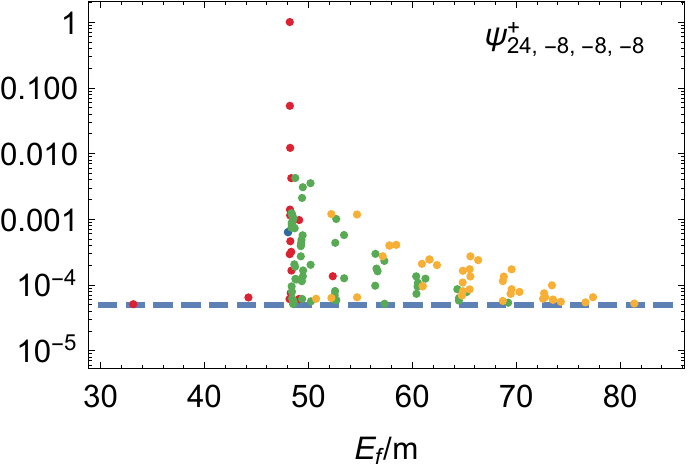}
\includegraphics[scale=0.83]{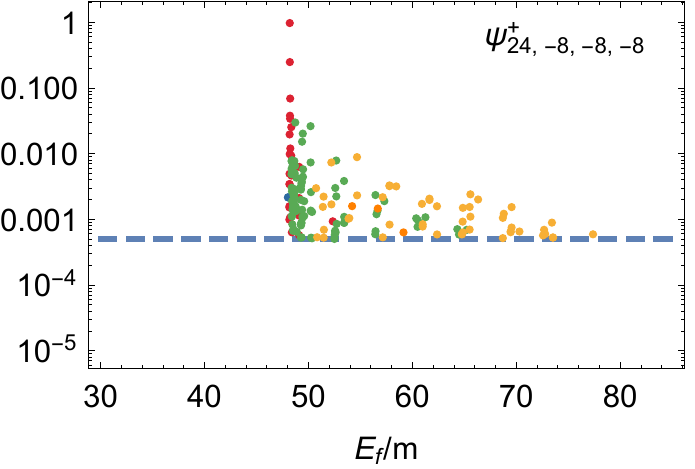}
\end{center}
\caption{\label{fig:lambdas}Plots of the eigenstates of the Hamiltonian for a range of values of $\lambda$.  The left column is for $\lambda=0.01m^2$, the middle column is for $\lambda=0.05m^2$ and the right column is for $\lambda=0.5m^2$.  Plots for $\lambda=0.1m^2$ can be seen in the right column of Fig.~\ref{fig:Energy Eigenstates}.  The rows, the color coding, the dashed line, the vertical axis and the horizontal axis are the same as in Fig.~\ref{fig:Energy Eigenstates}.  }
\end{figure*}

In this section, we explore the dependence of the eigenstates on $\lambda$, the coupling constant.  When $\lambda\to0$, the theory becomes free.  In this limit, the eigenstates become free states with only one basis state (which we call the main basis state) whose coefficient is 1.  When $\lambda$ is turned on, other basis states begin to contribute to the eigenstate and grow with $\lambda$ while the coefficient of the main basis state very slowly diminishes, but remains close to 1 while the theory remains perturbative.  If a basis state is connected to the main basis state by one application of the Hamiltonian, its coefficient grows linearly with $\lambda$.  This is because it appears at first order in perturbation theory.  Similarly, if a basis state is connected to the main basis state by two applications of the Hamiltonian, its coefficient grows quadratically with $\lambda$, as expected since it appears at second order in perturbation theory [see Eqs.~(\ref{eq:app:pert:cjn})-(\ref{eq:app:pert:psi total})].  We have calculated our eigenstates for a range of values of $\lambda$ from $0.01m^2$ to $1m^2$, beginning in the deeply perturbative region and ending in the nearly non-perturbative regime.  We show these eigenstates for $\lambda=0.01m^2, 0.05m^2$ and $0.5m^2$ in Fig.~\ref{fig:lambdas} while the eigenstate for $\lambda=0.1m^2$ can be seen in Fig.~\ref{fig:Energy Eigenstates}.  We can see that the coefficients grow as expected in these plots, as we will describe in greater detail below.  Moreover, because the importance of the basis states grow with increasing $\lambda$, a greater number of basis states contribute above a fixed precision cutoff.  In order to focus on the same region of the eigenstate, we adjust the cutoff for each value of $\lambda$.  The basis states contributing at first order in perturbation theory give us a natural pattern to follow.  We scale the precision of the cutoff linearly with $\lambda$ and present a table of our cutoffs in Table~\ref{tab:cutoffs}.  As in previous plots, these cutoffs are displayed as dashed blue lines in the figure.  The same $\Delta p=2m$ was used in this section as in the previous sections.  

\begin{table}[!]
\begin{center}
\begin{tabular}{|l|lll|}
\hline
$\lambda$ & V & 2 & 4\\
\hline\hline 
$0.01m^2$ & $2.8\times10^{-6}$ & $9.8\times10^{-6}$ & $9.8\times10^{-6}$\\
$0.05m^2$ & $1.4\times10^{-5}$ & $4.9\times10^{-5}$ & $4.9\times10^{-5}$\\
$0.1m^2$ & $2.8\times10^{-5}$ & $9.8\times10^{-5}$ & $9.8\times10^{-5}$\\
$0.5m^2$ & $1.4\times10^{-4}$ & $4.9\times10^{-4}$ & $4.9\times10^{-4}$\\
$1m^2$ & $2.8\times10^{-4}$ & $9.8\times10^{-4}$ & $9.8\times10^{-4}$\\
\hline
\end{tabular}
\end{center}
\caption{\label{tab:cutoffs}A table of the coefficient cutoffs used in this section.}
\end{table}

\begin{figure}[!]
\begin{center}
\includegraphics[scale=0.89]{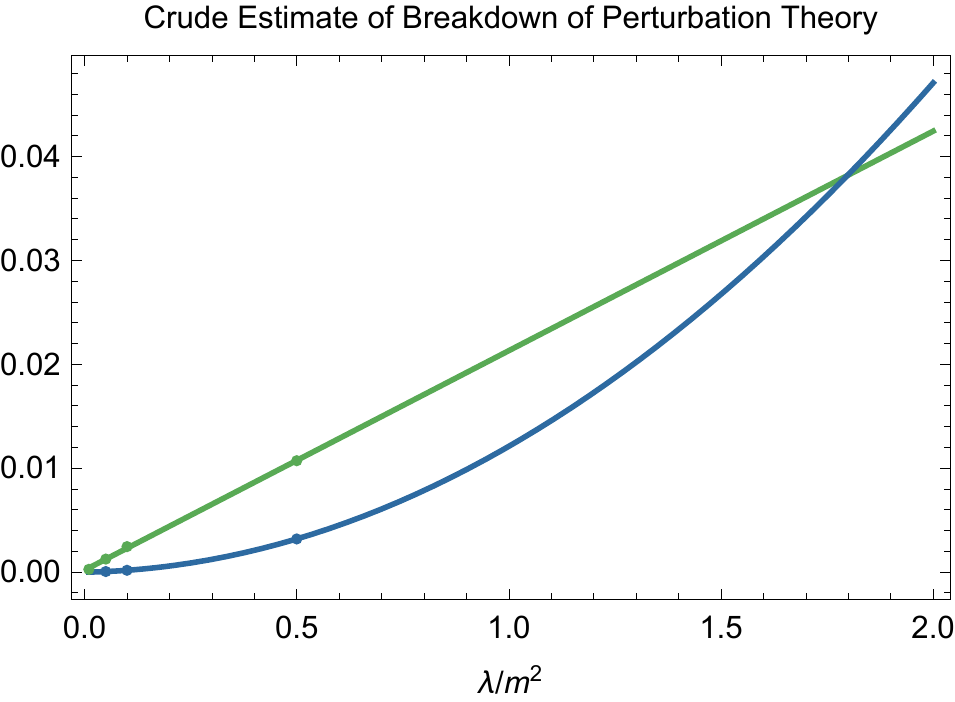}
\end{center}
\caption{\label{fig:PertBreak}For the vacuum, a plot of the coefficients of $|0,0,0,0\rangle$ (green dots) along with their best-fit straight line (green) and the coefficients of $|0,0\rangle$ (blue dots) along with their best-fit parabola (blue).  Their crossing point is a crude estimate of the breakdown of perturbation theory.}
\end{figure}

The vacuum is shown in the first row of Fig.~\ref{fig:lambdas}.  As described in Sec.~\ref{sec:results}, the four-particle basis states (in red) enter at first order in perturbation theory.  As just described, they grow proportional to $\lambda$.  For example, the two most important 4-particle basis states are $|0,0,0,0\rangle$ and $|-2, 0, 0, 2\rangle$.  They have coefficients -0.000254 and -0.000244 (for $\lambda=0.01m^2$), -0.00125 and -0.00120 (for $\lambda=0.05m^2$), -0.00244 and -0.00237 (for $\lambda=0.1m^2$),  -0.0107 and -0.0109 (for $\lambda=0.5m^2$), and -0.0194 and -0.0202 (for $\lambda=1m^2$), respectively.  As we can see, the scaling behavior is very nearly linear, especially for smaller $\lambda$.  The small deviation from linearity for larger $\lambda$ comes from two sources.  The first is that as the coefficients grow, the normalization changes. The second is that although the leading contribution to these coefficients is at first order in perturbation theory, there is also a subleading contribution at second order.  This is most strongly manifested for larger values of $\lambda$ as expected.  On the other hand, the 2-particle basis state $|0,0\rangle$ has its coefficient below the cutoff at $\lambda=0.01m^2$. Its coefficient is 0.00004 at $\lambda=0.05m^2$, 0.00016 at $\lambda=0.1m^2$,  0.0032 at $\lambda=0.5m^2$, and 0.0106 at $\lambda=1m^2$.  Again, the growth of the coefficient is very nearly quadratic in $\lambda$ deviating mostly towards larger $\lambda$, where subleading corrections at third order in perturbation theory become important.  We can use this information to estimate when perturbation theory breaks down by evaluating when the contribution from second order is as great as that from first order.  If we fit the growth of the coefficient of $|0,0,0,0\rangle$ to a straight line and the coefficient of $|0,0\rangle$ to a parabola, we find a crossing point of $\lambda\sim1.8m^2$ as shown in Fig.~\ref{fig:PertBreak}.   
However, we think this estimate is rather crude and that it needs further investigation.  The two most important 6-particle basis states just barely appear above the cutoff at $\lambda=0.05m^2$, both with coefficient 0.00002.  Their coefficients also grow quadratically as $\lambda$ increases.  The largest is $|0,0,0,0,0,0\rangle$ and its coefficient takes the values 0.00007 at $\lambda=0.1m^2$, 0.0010 at $\lambda=0.5m^2$ and 0.0028 at $\lambda=1m^2$.  The eight-particle basis states do not appear above the cutoff until $\lambda=0.5m^2$.  The largest of these is $|0,0,0,0,0,0,0,0\rangle$.  Its coefficient takes the values 0.0002 at $\lambda=0.5m^2$ and 0.0004 at $\lambda=1m^2$.  We expect the deviation from quadratic growth is due to subleading corrections from third order at these large values of $\lambda$ since they are connected to the vacuum by two powers of the Hamiltonian.  Basis states with 10 or more free particles contribute to the vacuum at third- or higher order perturbation theory.  They are all below the cutoff for the values of $\lambda$ that we tested.

The two-particle eigenstate $\Psi_{24,-24}$ is shown in the second row of Fig.~\ref{fig:lambdas}.  The 4-particle (red) and 6-particle (green) basis states shown in the left plot contribute at first order.  They grow linearly with $\lambda$ until higher-order corrections become important.  For example, the largest 4-particle basis state is $|-24,6,8,10\rangle_+$.  Its coefficient is 0.000371 at $\lambda=0.01m^2$, 0.00181 at $\lambda=0.05m^2$, 0.00351 at $\lambda=0.1m^2$, 0.0144 at $\lambda=0.5m^2$ and 0.0237 at $\lambda=1m^2$.  The largest 6-particle basis state is $|-24,0,0,0,0,24\rangle$.  Its coefficient is -0.000254 at $\lambda=0.01m^2$, -0.00124 at $\lambda=0.05m^2$, -0.00241 at $\lambda=0.1m^2$, -0.0102 at $\lambda=0.5m^2$ and -0.0178 at $\lambda=1m^2$.  Other 2-particle basis states contribute at first order but are below the cutoff.  The 0-particle basis state (the free vacuum) contributes at second order in perturbation theory and does not appear above the cutoff for the range of coupling constants that we tested.  Some further 6-particle basis states appear above the cutoff at $\lambda=0.5m^2$ (green points near the blue dashed line with a free energy of approximately 50$m$.)  One of these is $|-24,2,2,4,6,10\rangle_+$.  It has a coefficient of 0.0007 at $\lambda=0.5m^2$ and 0.0017 at $\lambda=1m^2$, getting its dominant contribution at second order.  Since we only have the coefficient at large values of $\lambda$, subleading corrections are already important.  We also see three 8-particle basis states (yellow points) appear above the cutoff when $\lambda=0.5m^2$.  The largest is $|-24,0,0,0,0,0,0,24\rangle$ and has a coefficient of 0.0013 at $\lambda=0.5m^2$ and 0.0033 at $\lambda=1m^2$, which appears to be slightly less than quadratic growth due to subleading effects.  We do not see 10-particle basis states above the cutoff for any of our tested values of $\lambda$.  The same is true for basis states with 12 or more free particles, which contribute at higher than second order.   Because the basis states contributing at second order do not appear above the cutoff until large values of $\lambda$, where subleading contributions are important, we do not use them to estimate the breakdown of perturbation theory.

The four-particle eigenstate $\Psi^+_{24,-8,-8,-8}$ is shown in the third row of Fig.~\ref{fig:lambdas}.  Most of the other 4-particle basis states (red points) contribute at first order in perturbation theory.  For example, the most important 4-particle basis state (after the main basis state) is $|-10,-8,-6,24\rangle_+$.  Its coefficients are  0.011661, 0.05291, 0.09476, 0.2545 and 0.3064 at $\lambda=0.01m^2, 0.05m^2, 0.1m^2, 0.5m^2$ and $1m^2$, respectively.  However, a few of the 4-particle basis states contribute only at second order.  For example, the basis state $|-10,-10,-4,24\rangle_+$ has coefficients 0.000053, 0.00114, 0.00383, 0.0337 and 0.0429 at $\lambda=0.01m^2, 0.05m^2, 0.1m^2, 0.5m^2$ and $1m^2$, respectively.  The reason that this basis state require the action of the Hamiltonian twice is that at least three of the momenta are different.  So, it has to remove at least three momenta and add at least three new momenta.  This requires the application of at least three annihilation operators and at least three creation operators which can only be achieved in two applications of the Hamiltonian.  [See Eq.~(\ref{eq:Discrete Hamiltonian}) where the maximum number of creation and annihilation operators is four.]  Using these coefficients, we might attempt to estimate the breakdown of perturbation theory again and see how it correlates with what we found using the vacuum.  However, there are two challenges.  The first is that the coefficients are already quite large and would naively reach unity levels at $\lambda\sim1.1m^2$ and $\lambda\sim1.8m^2$ before they even cross each other.  Also, as we will shortly see, our results for the four-particle eigenstate at large $\lambda$ appear to be much less trust worthy than for the vacuum.  For this reason, in this paragraph, we give the coefficients coming from diagonalized second-order perturbation theory.   Moving on to other contributing basis states.   The 0-particle basis state also contributes at first order but is below the cutoff for all the values of $\lambda$ that we tested.   There is a single 2-particle basis state above the cutoff.  It is the basis state $|24,-24\rangle$.    Its coefficients are  0.000149, 0.00063, 0.00105, 0.0019 and 0.0012 at $\lambda=0.01m^2, 0.05m^2, 0.1m^2, 0.5m^2$ and $1m^2$, respectively.  We can see that it grows nearly linearly for small $\lambda$.  This makes sense because it can be reached from the main basis state by one application of the Hamiltonian.  Some 6-particle basis states (green points), contribute at first order.  For example, the highest such point is $|-24,2,2,4,8,8\rangle_+$ with coefficients 0.000881, 0.00422, 0.00803, 0.0300 and 0.0479 at $\lambda=0.01m^2, 0.05m^2, 0.1m^2, 0.5m^2$ and $1m^2$.  Others contribute at second order.  For example, the basis state $|-10,-8,-6,0,0,24\rangle_+$ appears above the cutoff at $\lambda=0.05m^2$ and has coefficients 0.00020, 0.00070, 0.0073 and 0.0142 at $\lambda=0.05m^2, 0.1m^2, 0.5m^2$ and $1m^2$.  We can see that it grows nearly quadratically for small $\lambda$.  The same story applies to the 8-particle basis states (yellow points).  Some contribute at first order while others at second order.  The basis state $|-24,0,0,0,0,8,8,8\rangle_+$ has coefficients 0.000252, 0.00119, 0.00223, 0.0073 and 0.0101 at $\lambda=0.01m^2, 0.05m^2, 0.1m^2, 0.5m^2$ and $1m^2$ and contributes at first order while the basis state $|-10,-8,-6,0,0,0,0,24\rangle_+$ has coefficients 0.000063, 0.00021, 0.0017 and 0.0023 at $\lambda=0.05m^2, 0.1m^2, 0.5m^2$ and $1m^2$ and contributes at second order.  Finally, the 10-particle basis state can only begin contributing at second order since they require the addition of six new particles and therefore need at least two applications of the Hamiltonian.  They first appear above the cutoff at $\lambda=0.1m^2$ with a few more appearing at larger $\lambda$.  The first one is the basis state $|-24,0,0,0,0,0,0,8,8,8\rangle_+$ which has coefficients 0.00001, 0.0016 and 0.0036 at $\lambda=0.1m^2, 0.5m^2$ and $1m^2$.

\begin{figure}[!]
\begin{center}
\includegraphics[scale=0.89]{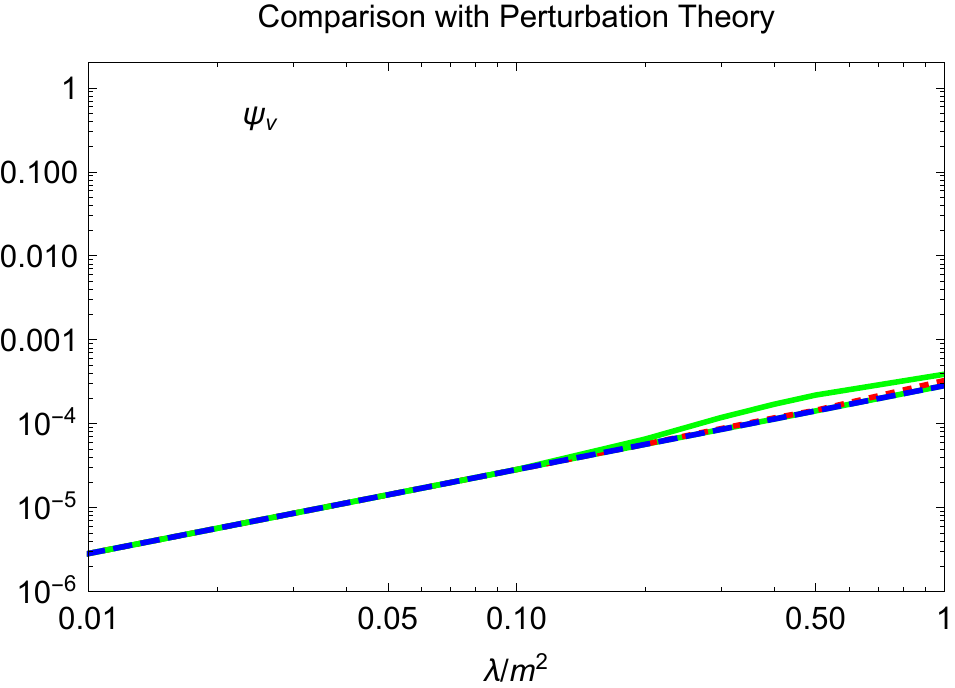}\\
\includegraphics[scale=0.89]{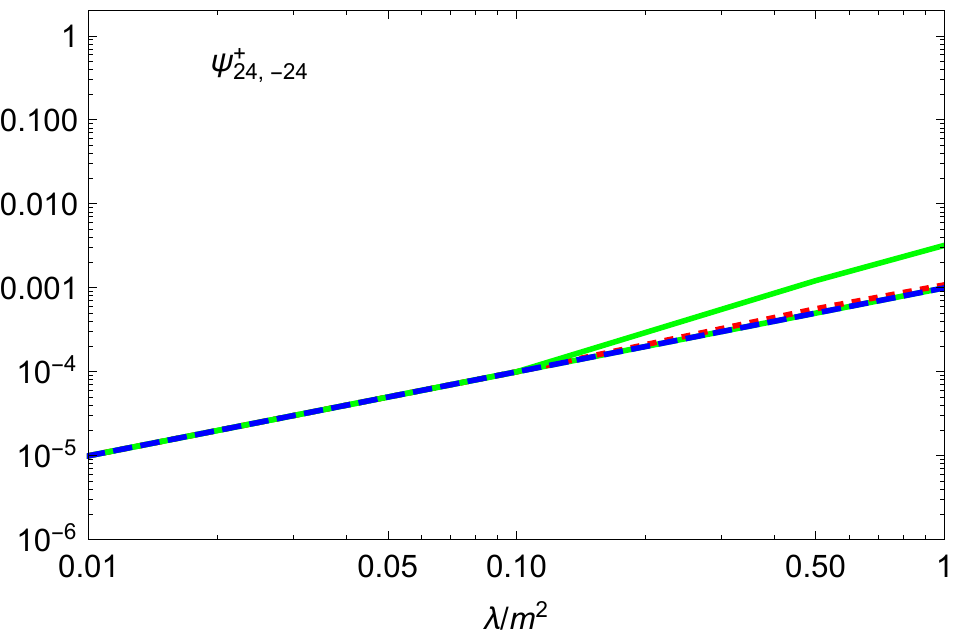}\\
\includegraphics[scale=0.89]{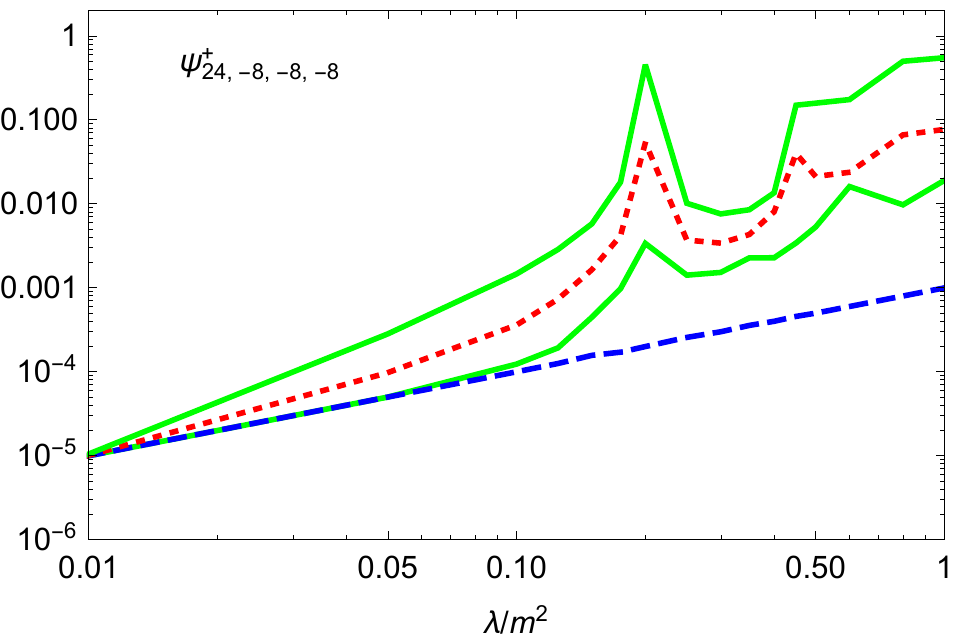}
\end{center}
\caption{\label{fig:DiffVsLambda}Plots of the largest differences between our cyclic QSE code and second-order perturbation for a range of the coupling constant $\lambda$.  The blue dashed line is the cutoff used in our calculations.  The green lines give the maximum and minimum largest differences after 100 independent runs of the code.  The red dotted line gives the mean largest difference.}
\end{figure}
We have also compared the results of our code with perturbation theory for the range of $\lambda$ that we have tested here.  We show plots of the largest absolute difference with second-order perturbation theory in Fig.~\ref{fig:DiffVsLambda}.  Because of the random nature of our cyclic QSE code, each time it is run, we will potentially obtain different results.  It is important to analyze this randomness and develop distributions giving the statistics of how frequently it is successful.  We have chosen to do that here as it depends on the value of the coupling constant $\lambda$.  For each value of $\lambda$ and each eigenstate, we ran our cyclic QSE code 100 independent times.  We then compared each run with second-order perturbation theory and found the largest difference.  We used the largest difference for each run to determine the statistics.  In Fig.~\ref{fig:DiffVsLambda}, we show the cutoff we used in dashed blue.  The solid green curves are for the maximum largest difference and minimum largest difference encountered in the 100 independent runs while the red dotted line is the average largest difference.  The top and middle plots are for the vacuum and the two-particle eigenstate, respectively.  We can see that our cyclic QSE code is extremely successful at accurately constructing these eigenstates.  In fact, if we are able to calculate these two-particle eigenstates in two spatial dimensions with as much success, we should be able to successfully calculate elastic $2\to2$ scattering amplitudes with this method.  

On the other hand, we can see that the four-particle eigenstate, shown at the bottom of Fig.~\ref{fig:DiffVsLambda}, is much more difficult.  Our cyclic QSE code seems to do an acceptable job for $\lambda\lesssim0.1m^2$.  But, it does not appear to be trustworthy above this value.  In fact, what is happening is that our cyclic QSE code is sometimes not finding important basis states.  In particular, at the very peak at $\lambda=0.2m^2$, the maximum largest difference is due to the QSE code not finding the basis state $|18,6,-8,-16\rangle_+$ which has a coefficient 0.45123.  At $\lambda=0.3m^2$, on the other hand, the maximum largest difference is caused by missing the basis state $|24,-6,-8,-10\rangle_+$ which has a coefficient 0.01065.  At $\lambda=0.5m^2$, it is the basis state $|18,6,-10,-14\rangle_+$ with a coefficient of 0.1570.  Whether or not it finds these basis states is very sensitive to what other basis states are present and their coefficients, as well as the size of the reduced Hilbert space (see Sec.~\ref{sec:Hsize}).  It may be possible to fix this deficiency by a more clever random choice of new basis states at each cycle.  We spent considerable time tuning this process but believe there is further room for improvement.  Clearly, if this method is to be used to calculate $2\to4$ scattering amplitudes, this will have to be improved if these larger values of $\lambda$ are important.

\begin{figure}[!]
\begin{center}
\includegraphics[scale=0.89]{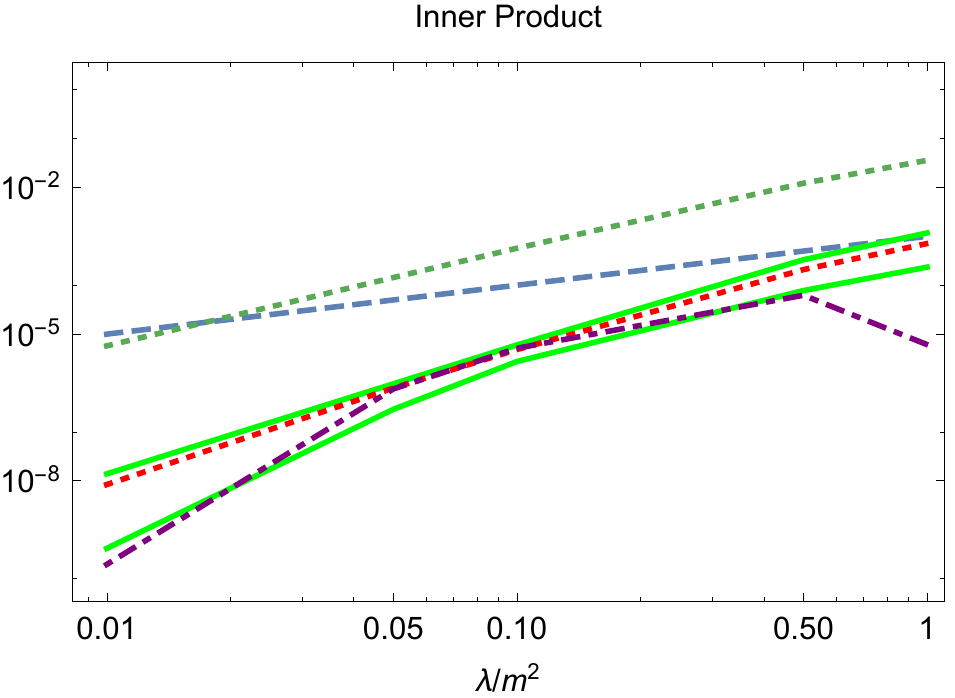}
\end{center}
\caption{\label{fig:IPvsLambda}A plot of the inner product of the two-particle eigenstate $\Psi_{24,-24}$ and the four-particle eigenstate $\Psi^+_{24,-8,-8,-8}$ for a range of coupling constants $\lambda$.  The blue dashed line represents the cutoff we used for these states, the dotted green line gives the inner product at first order, the dot-dashed orange line gives the inner product at second order, the solid green lines give the largest and smallest inner products given by our code after 100 trials while the dotted red line gives the mean value given by our code.  }
\end{figure}
To end this section, we calculate the inner product between the eigenstates $\Psi_{24,-24}$ and $\Psi^+_{24,-8,-8,-8}$ for each value of $\lambda$ and show the results in Fig.~\ref{fig:IPvsLambda}, where the dashed blue line gives the approximate precision of this calculation, which is the cutoff we used for the basis states.  As we discussed in Sec.~\ref{sec:results}, we expect this inner product to be zero because the energies of these eigenstates are slightly different so a nonzero result would indicate non-conservation of energy.  Our result from first-order perturbation theory is shown as the dotted green line.  We can see that it is above the precision of the calculation for most of the range of $\lambda$, indicating a nonzero result.  We are not claiming, by this result, that perturbation theory itself violates energy conservation.  The problem with our first-order result is that it is missing important basis states and the coefficients are not yet sufficiently accurate.  That combination leads to a violation of energy conservation.  On the other hand, the second-order result is shown in dot-dashed purple and is below the precision of the calculation for the entire range of $\lambda$.  The result of our cyclic QSE code is also at or below the precision of the calculation for the entire range of $\lambda$ showing that it is also in agreement with a zero result and successfully conserves energy.  In fact, we calculated the inner product for the entire set of 100 trials for each value of $\lambda$.  We show the smallest and largest values of the inner product we got from our cyclic QSE code as the two solid green lines, while the average inner product is displayed as a dotted red line.  The entire range of our results is at or below the precision of the calculation.  It might, naturally, be wondered how our inner product did so well at large $\lambda$ when the disagreement between our QSE code and second-order perturbation theory for the eigenstate $\Psi^+_{24,-8,-8,-8}$ was so large (see the bottom plot of Fig.~\ref{fig:DiffVsLambda}).  The reason is that the basis states that were missed by our cyclic QSE code when constructing $\Psi^+_{24,-8,-8,-8}$ were below the cutoff for the eigenstate $\Psi_{24,-24}$, so the contribution of these basis states was also below the cutoff.  That is to say, these basis states don't contribute significantly to this inner product.

\section{\label{sec:dp}Dependence on $\mathbf{\Delta p}$}
\begin{figure*}[!]
\begin{center}
\includegraphics[scale=0.83]{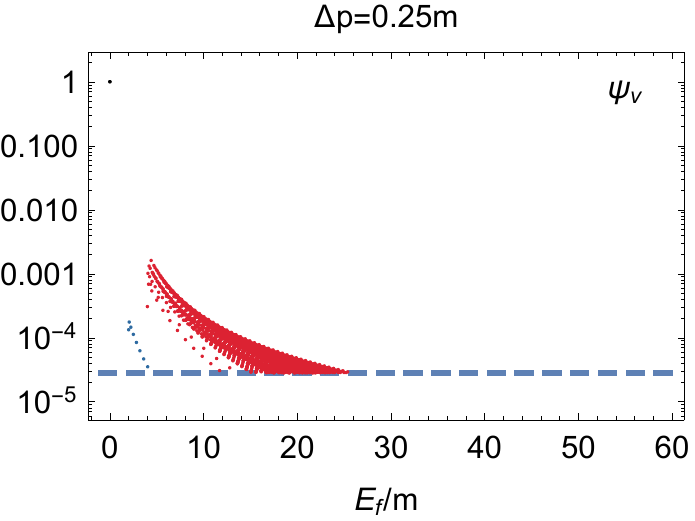}
\includegraphics[scale=0.83]{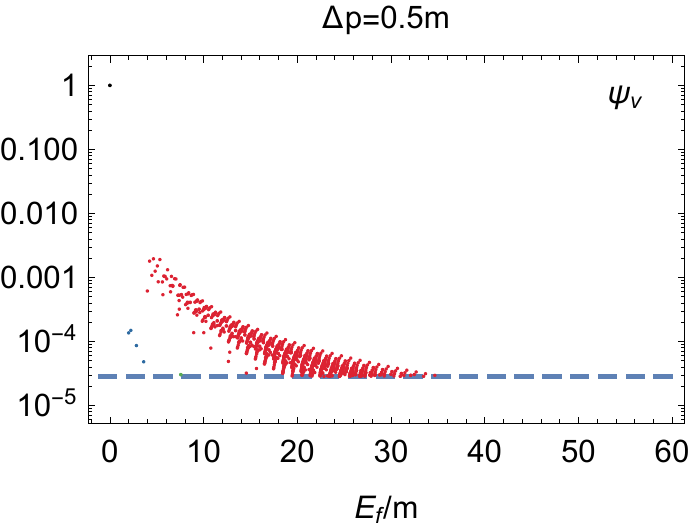}
\includegraphics[scale=0.83]{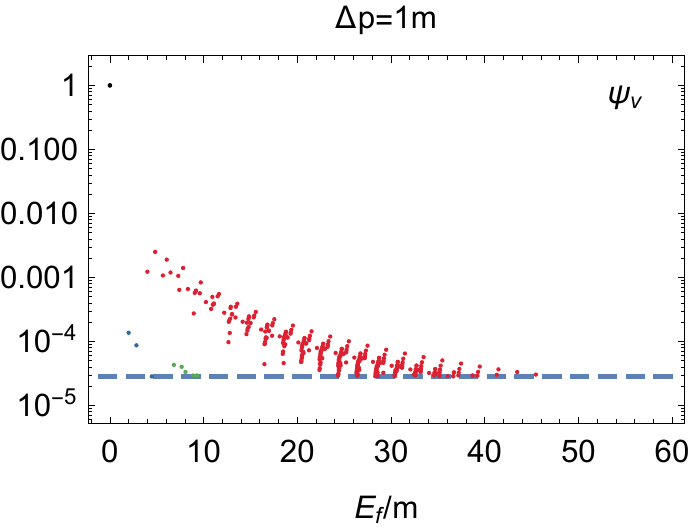}
\\
\includegraphics[scale=0.83]{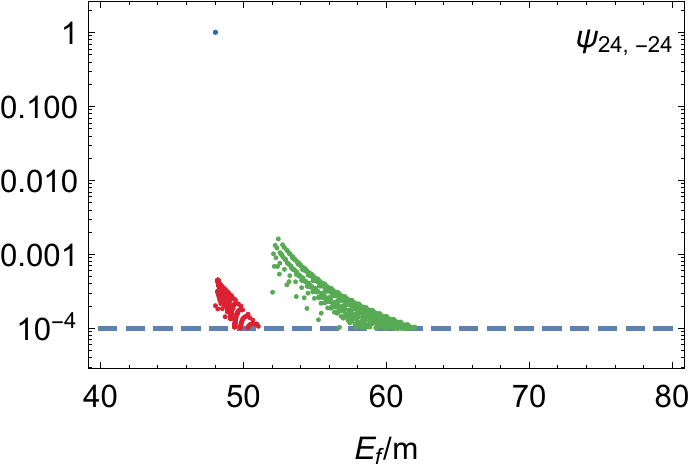}
\includegraphics[scale=0.83]{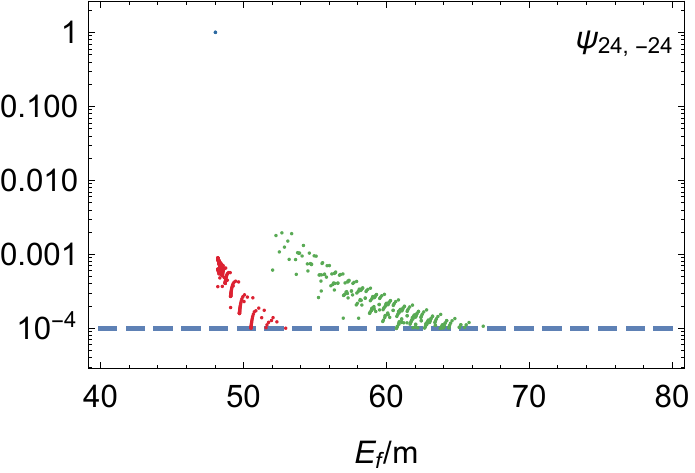}
\includegraphics[scale=0.83]{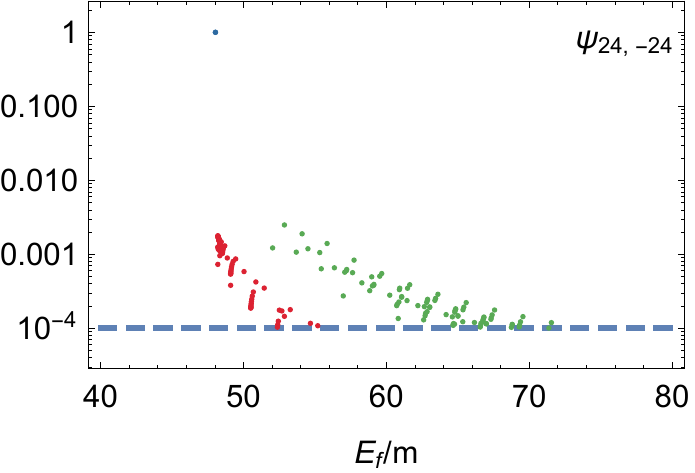}
\\
\includegraphics[scale=0.83]{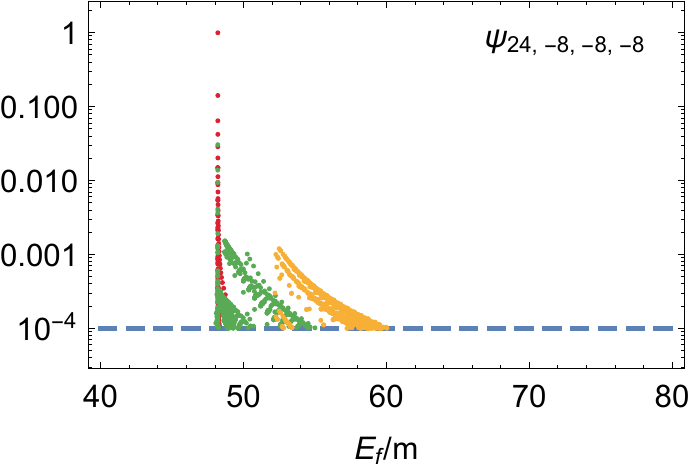}
\includegraphics[scale=0.83]{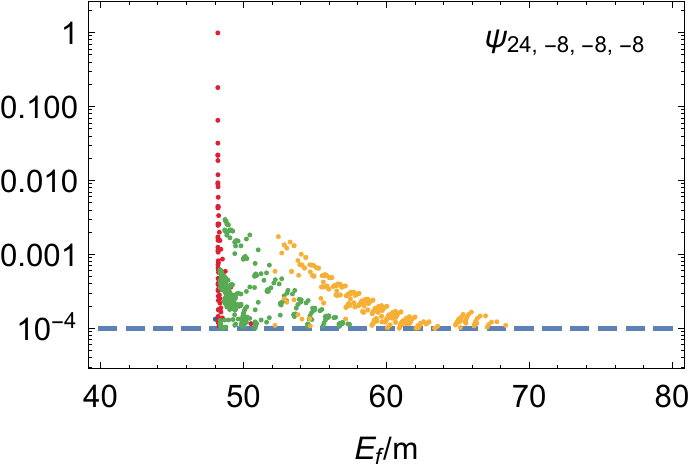}
\includegraphics[scale=0.83]{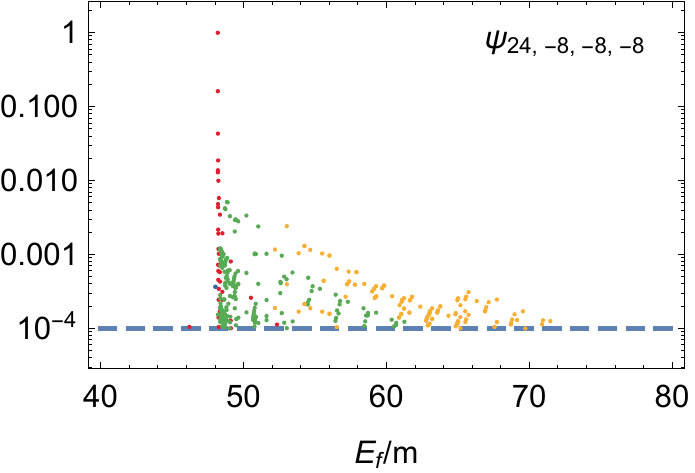}
\end{center}
\caption{\label{fig:dps}Plots of the eigenstates of the Hamiltonian for a range of values of $\Delta p$.  The left column is for $\Delta p=0.25m$, the middle column is for $\Delta p=0.5m$ and the right column is for $\Delta p=1m$.  Plots for $\Delta p=2m$ can be seen in the right column of Fig.~\ref{fig:Energy Eigenstates}.  The rows, the color coding, the dashed line, the vertical axis and the horizontal axis are the same as in Fig.~\ref{fig:Energy Eigenstates}.}
\end{figure*}
In this section, we turn to the dependence of our results on the size of the momentum spacing $\Delta p$, while keeping $\lambda=0.1m^2$.  A major long-term objective is to obtain useful information about the continuum limit $\Delta p\to0$.  In order to do this, we need to do the calculations in this paper at much smaller $\Delta p$ than we have already done.  In fact, we would like to do it for a range of small $\Delta p$ so that we can meaningfully extrapolate to $\Delta p\to0$.  This section will be an initial step in that direction.  We have calculated our eigenstates at a few values of $\Delta p$ and show plots for $\Delta p=0.25m, 0.5m$ and $1m$ in Fig.~\ref{fig:dps} while the results for $\Delta p=2m$ can be seen in the right column of Fig.~\ref{fig:Energy Eigenstates}.  For this section, we have kept $\lambda=0.1m^2$ and use the same cutoff on each state as we decreased $\Delta p$.  Because the cutoff has remained the same as we decrease $\Delta p$, the number of basis states above the cutoff has increased.  In particular, we have found that the vacuum $\Psi_v$ has 109, 360, 1211 and 4168 basis states above the cutoff, the two-particle eigenstate $\Psi_{24,-24}$ has 59, 173, 556 and 1796 basis states above the cutoff, and the four-particle eigenstate $\Psi^+_{24,-8,-8,-8}$ has 137, 393, 894 and 2486 basis states above the cutoff for $\Delta p=2m, 1m, 0.5m$ and $0.25m$, respectively.  In order to keep the ratio of the reduced Hilbert space size to the number of basis states above the cutoff roughly the same as we decreased $\Delta p$ (in order to make a fair comparison with perturbation theory and ensure the cyclic QSE code is equally effective at each $\Delta p$), we have tripled the reduced Hilbert space size each time we halfed $\Delta p$.  So, we used a reduced Hilbert space size of 700, 2100, 6300 and 18900 for $\Delta p=2m, 1m, 0.5m$ and $0.25m$, respectively.  This kept our code running at roughly the same effectiveness for each value of $\Delta p$.

As we look at the plots in Fig.~\ref{fig:dps}, we notice a few general features as we decrease $\Delta p$.  The first is that the general structure (the shape of the points on the plot) remains largely the same.  This is a good thing as it gives us some confidence that the results at larger $\Delta p$ are approximating those at small $\Delta p$.  The second is that the density of basis states (the density of points on the plot) increases as we decrease $\Delta p$.  This also makes sense as the free energy of the basis states (the horizontal axis of these plots) is a (nearly linear) function of the momentum spacing.  In particular, it takes the form $E_f=\sum_i\sqrt{n_i^2\Delta p^2+m^2}$, where $n_i$ is an integer unique to each free particle in the basis state and determines its momentum $p_i=n_i\Delta p$.  So, as $\Delta p$ decreases, a greater number of basis states fit into the same free energy region.  

Ideally, we would prefer to plot not the bare coefficient of each basis state (as we have done in Fig.~\ref{fig:dps} and throughout this paper), but rather the coefficient divided by the free-energy space between the basis states.  This would normalize the contribution of the basis states by their density and would allow for a more stable eigenstate as $\Delta p\to0$.  This was done in \cite{Christensen:2016naf} where only 0- and 2-particle basis states were included and the free-energy spacing between basis states was constant and, consequently, so was their density.  In the present work, however, we keep higher-multiplicity basis states and the free-energy spacing between basis states is not constant.  In fact, it is quite complicated.  We do not, presently, know the correct free-energy spacing to use for the normalization when higher-multiplicity basis states are included.  We feel that this is a very important topic for future research if this method is to succeed.   In the meantime, we will simply analyze the bare coefficient for the basis states and make a few comments about how normalizing by the density might affect our results.  All the plots in this section, and throughout this paper, use the bare coefficient.

\begin{figure}[!]
\begin{center}
\includegraphics[scale=0.89]{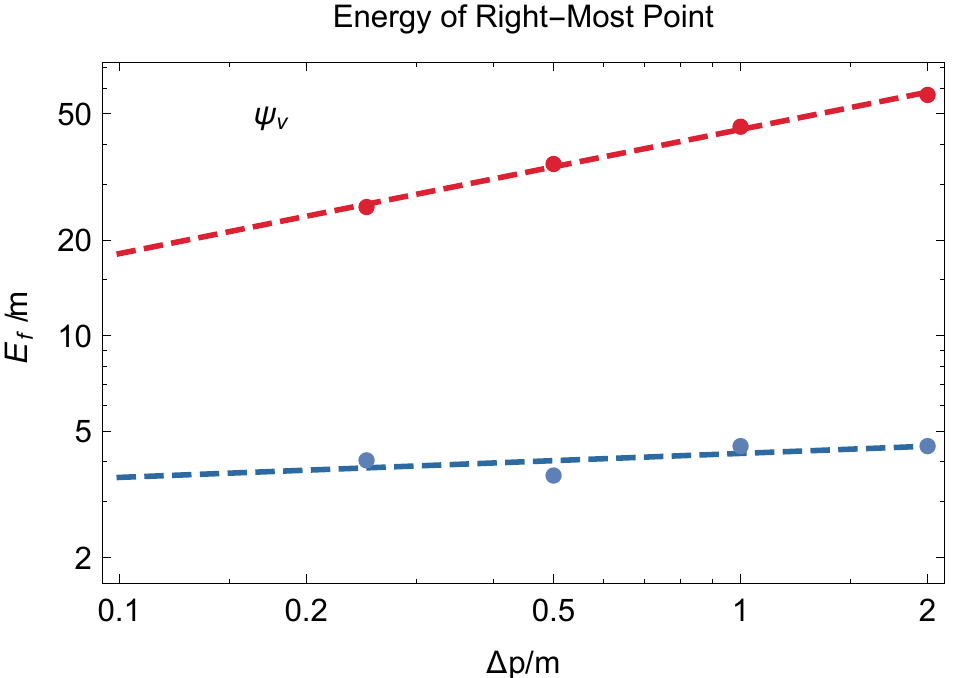}\\
\includegraphics[scale=0.89]{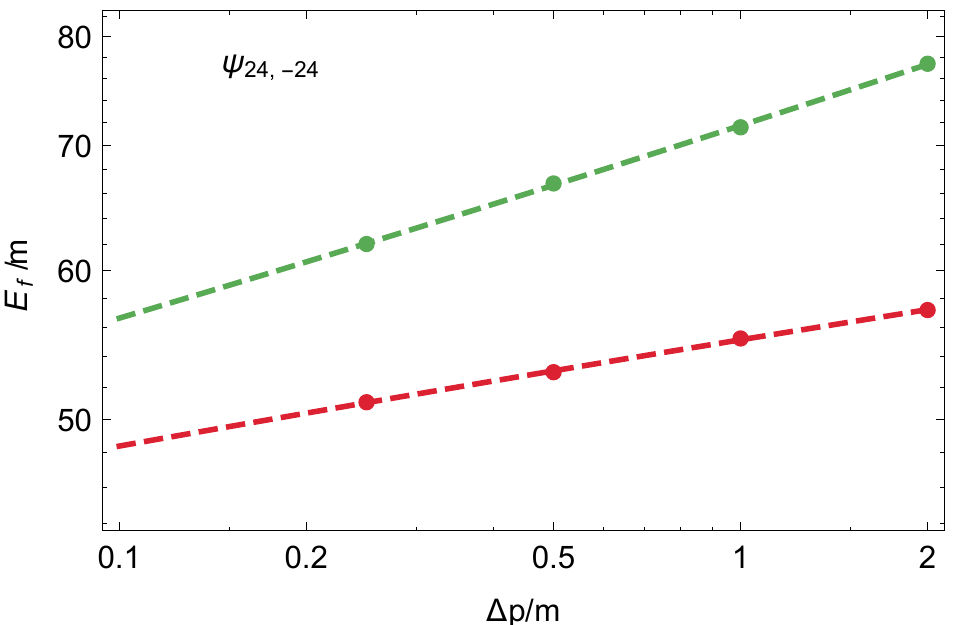}\\
\includegraphics[scale=0.89]{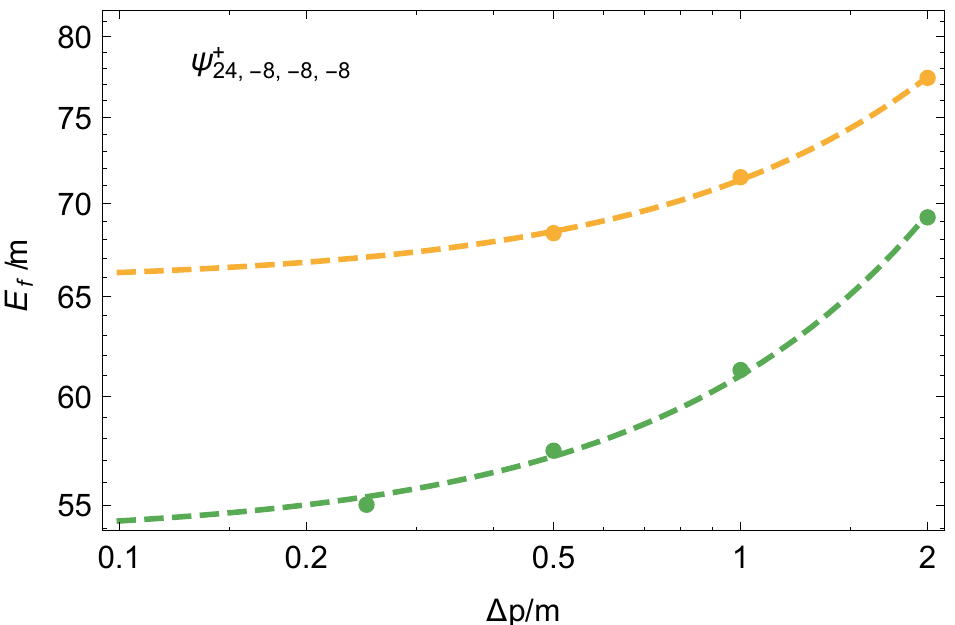}
\end{center}
\caption{\label{fig:endpoints Dp}Plots of the right-most point of each color from the plots in Fig.~\ref{fig:dps} as a function of $\Delta p$ along with their best-fit curves.  The horizontal axis is the momentum spacing divided by the mass while the vertical axis is the free energy of the right-most basis state divided by m. }
\end{figure}
The third feature that stands out about these plots is that, although the general shape remains the same, the fall off of the coefficients is more rapid as $\Delta p$ decreases.  For example, if we focus on the vacuum, we see that the 4-particle basis states extend out to approximately $57m$ before falling below the cutoff when $\Delta p=2m$, but only out to approximately $46m$ when $\Delta p=1m$, $35m$ when $\Delta p=0.5m$ and $26m$ when $\Delta p=0.25m$.  In fact, this can be well fit by a straight line on a log-log plot as seen in the top plot of Fig.~\ref{fig:endpoints Dp}, where the best-fit line is given by
\begin{equation}
\mbox{ln}\left(E_f\right) = 3.80+0.39\ \mbox{ln}\left(\Delta p\right)\ .
\end{equation}
The right-most 2-particle basis state (blue point) can also be fit by a straight line and is presented as the dashed blue line in the same plot.  
Moving to the two-particle eigenstate $\Psi_{24,-24}$, the right-most 6-particle basis state (green point) can be well fit by the straight line
\begin{equation}
\mbox{ln}\left(E_f\right) = 4.27+0.11\mbox{ln}\left(\Delta p\right)\ ,
\end{equation}
and is shown as the dashed green line in the middle plot of Fig.~\ref{fig:endpoints Dp} where we also present the right-most red point and its best-fit straight (dashed red) line.  We note that a straight-line best fit cannot continue to infinitesimal $\Delta p$ because that would imply that the right-most point in the plots of Fig.~\ref{fig:dps} goes to zero.  We believe that the reason for this is the lack of a proper density normalization as mentioned earlier in this section.  If we properly normalize these coefficients to the density of the basis states, we believe that the free energy of the right-most point would approach a constant nonzero value as $\Delta p\to0$.   That is, we believe with the proper normalization, they would not be well fit by a straight line, but rather something like an exponential curve (on a log-log plot).  

The four-particle eigenstate $\Psi^+_{24,-8,-8,-8}$, on the other hand, did have right-most points that appear to be gradually approaching a nonzero limit.  We have plotted the right most 8-particle basis state (yellow point) for this eigenstate in the bottom plot of Fig.~\ref{fig:endpoints Dp} along with its fit to an exponential curve (on a log-log plot).  We did not include the right-most 8-particle basis state at $\Delta p=0.25m$ for technical reasons, which we will explain in the next paragraph.  We also plotted the right-most 6-particle basis state (green point) in the same plot and also fit it with an exponential, given by
\begin{equation}
\mbox{ln}\left(E_f\right) = 3.98 + 0.13\Delta p\ .
\end{equation}
If we extrapolate this exponential curve all the way to $\Delta p\to0$, we estimate that the 6-particle basis states (green points) will end at approximately $E_f=53.5m$, in the continuum limit for the same cutoff.  Of course, properly normalizing by the density of basis states will probably make these curves shallower and affect this extrapolation to $\Delta p\to0$.

\begin{figure}[!]
\begin{center}
\includegraphics[scale=0.89]{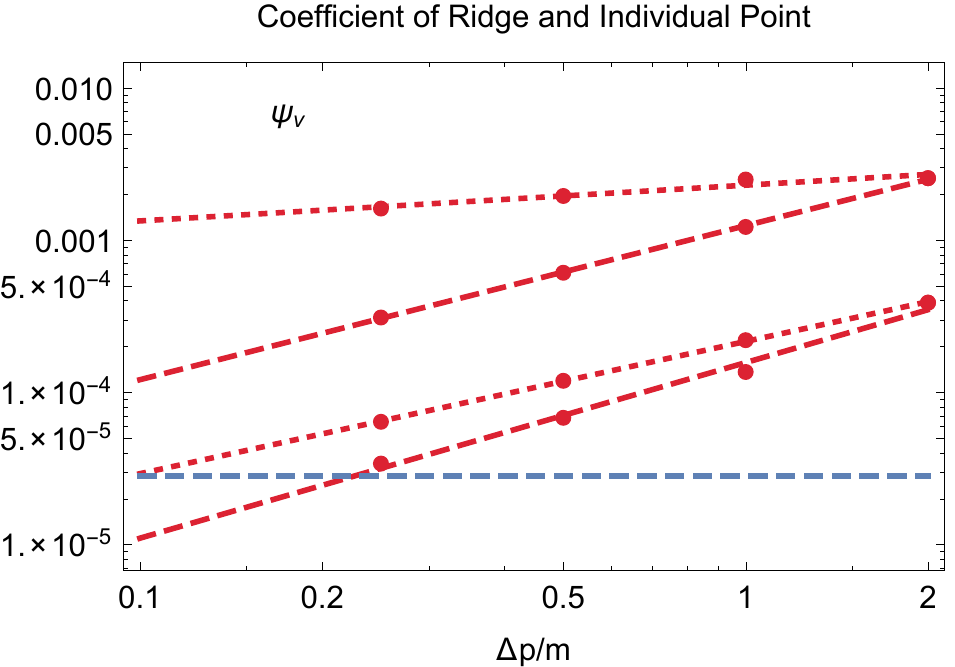}\\
\includegraphics[scale=0.89]{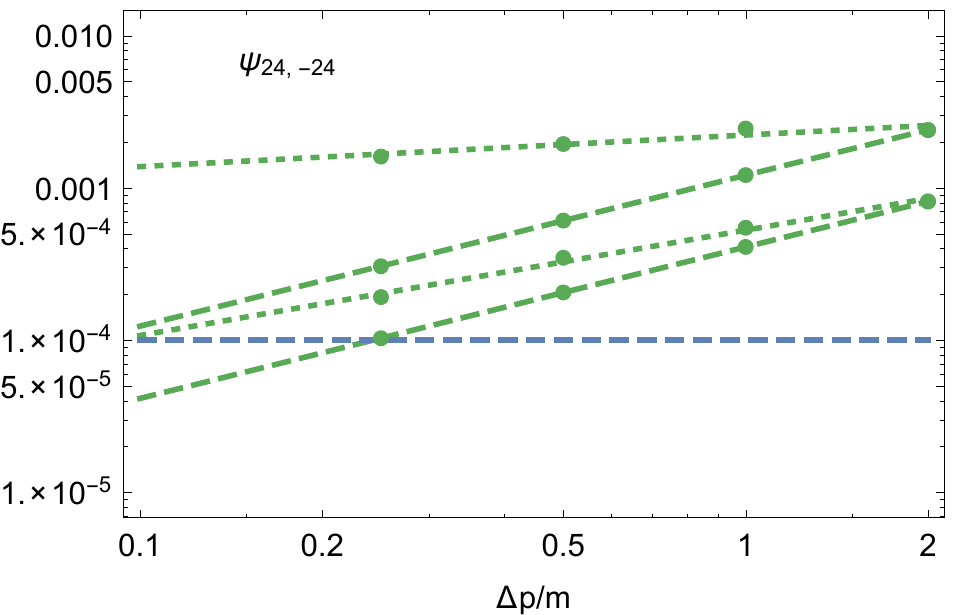}\\
\includegraphics[scale=0.89]{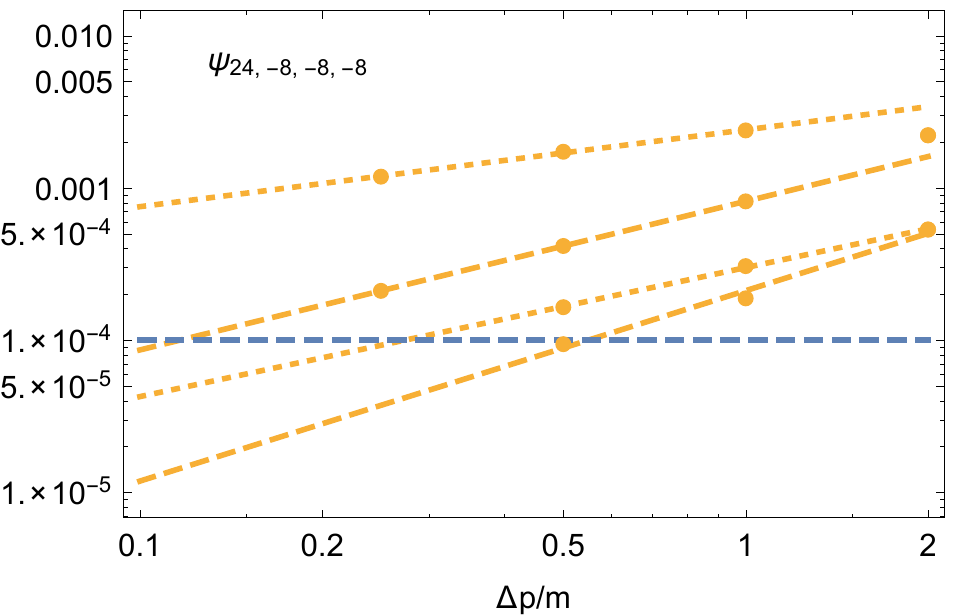}
\end{center}
\caption{\label{fig:Ridge Dp}Plots of the top ridge and for an individual point from the plots in Fig.~\ref{fig:dps} as a function of $\Delta p$ along with their best-fit curves.  The horizontal axis is the momentum spacing divided by the mass while the vertical axis is the absolute value of the coefficient for each point.  See the text for greater detail.}
\end{figure}
Although the right-most point is moving towards lower free energy as $\Delta p$ decreases, the basis states themselves do not move towards lower free energy.  Their free energy is fixed by their free-particle momenta.  Instead, what is happening is that their coefficients are diminishing as $\Delta p$ decreases.  The basis states that were just above the cutoff at $\Delta p=2m$ at the far right of the right plot in Fig.~\ref{fig:Energy Eigenstates} are being reduced to coefficient values below the cutoff so that they no longer appear in the plots for $\Delta p=1m$, $\Delta p=0.5m$ or $\Delta p=0.25m$ in Fig.~\ref{fig:dps}.  In particular, focusing on the four-particle eigenstate $\Psi^+_{24,-8,-8,-8}$ shown in the bottom row of Figs.~\ref{fig:Energy Eigenstates} and ~\ref{fig:dps}, we can see a small hump in the 8-particle basis states (yellow points) at $E_f\sim66m$.  As $\Delta p$ decreases and the density of basis states increases, we see this hump resolved in greater detail, however, it remains at the same free-energy position.  It's contribution to the eigenstate, however, does decrease and we see this as the hump sinks lower and lower as $\Delta p\to0$.  In particular, the highest point of the hump has coefficients of 0.00053, 0.00031 and 0.00017 at $\Delta p=2m, 1m$ and $0.5m$, respectively.  By the time, $\Delta p=0.25m$, the hump is completely below the cutoff.  This is why we did not include a point at  $\Delta p=0.25m$ in the extrapolation.  This can be seen as the bottom dotted line in the bottom plot of Fig.~\ref{fig:Ridge Dp} passes below the dashed blue line.

In order to explore this behavior further, we have plotted a series of points and curves in Fig.~\ref{fig:Ridge Dp}, which we will now explore.  The dashed blue line is the cutoff as usual.  The color coding of these points is the same as in Fig.~\ref{fig:dps}.  For each plot, we have begun with two points at $\Delta p=2m$ as can be seen at the right edge of these plots and then followed what happens with these points as $\Delta p$ decreases.  To understand these plots in detail, we will begin by focussing on the top plot, which is for the vacuum.  For this plot, we have focused on 4-particle basis states.  The top point  at $\Delta p=2m$ (in the top plot of Fig.~\ref{fig:Ridge Dp}) is the basis state $|0,0,0,0\rangle$ which has a free energy of 4m and a coefficient 0.00255.  It is the highest red point in the top right plot of Fig.~\ref{fig:Energy Eigenstates}.  As we move towards smaller $\Delta p$, two things happen.  The first is that the coefficient of the highest red point decreases.  In order to see this, we have plotted the coefficient of the highest red point for the vacuum at each value of $\Delta p$ as the top four red points of the top plot of Fig.~\ref{fig:Ridge Dp} and have fit a red dotted straight line to them.  This is the highest red line in the plot.  We can see that it has a very gentle slope downward as $\Delta p$ decreases.  However, the highest point is not the same basis state for all $\Delta p$.  In fact, at $\Delta p=1m$, it is the basis state $|-1m,0,0,1m\rangle$ with a coefficient 0.002499, at $\Delta p=0.5m$, it is the basis state $|-1m,0,0.5m,0.5m\rangle_+$ with a coefficient 0.00195, and at $\Delta p=0.25m$, it is the basis state $|-0.75m,0,0.25m,0.5m\rangle_+$ with a coefficient 0.00161.  The original basis state actually moves down at a faster rate, moving inside of the dense region of red points at the same energy.  In particular, the basis state $|0,0,0,0\rangle$ that was at the top at $\Delta p=2m$, has coefficients 0.00255, 0.00122, 0.00061 and 0.00031 at $\Delta p=2m, 1m, 0.5m$ and $0.25m$, respectively.  We plot this as the top red-dashed best-fit line and the points that it passes through.  This behavior is not unique to the highest point of a color.  Other points along the ridge of the color also do this.  The bottom red point (in the top plot of Fig. 12) at $\Delta p=2m$ is the basis state $|-8m,0,2m,6m\rangle_+$ and has free energy of 17.4m and coefficient 0.00039.  If we follow the top ridge of the red points at the same free energy, we find that it is the basis state $|-7m,-1m,0,8m\rangle_+$ at $\Delta p=1m$ with a coefficient 0.00022, it is the basis state $|-8m,0,0.5m,7.5m\rangle_+$ at $\Delta p=0.5m$ with a coefficient 0.00012, and it is the basis state $|-7.5,-0.25,0.5,7.25\rangle_+$ at $\Delta p=0.25m$ with a coefficient 0.00006.  We also plot these points along with their best fit line (bottom dotted red line) in the top plot of Fig.~\ref{fig:Ridge Dp}.  However, we also plot the same basis state $|-8m,0,2m,6m\rangle_+$ and follow its coefficient as the best-fit dashed red line below it.  It has coefficients 0.00014, 0.00007 and 0.00003 at $\Delta p=1m, 0.5m$ and $0.25m$.  We again see that the coefficient of the basis state is reduced faster than the height of the ridge showing that the basis state sinks into the red points at the free energy 17.4m.  All of these points are well fit by a straight line on a log-log plot.  However, as discussed earlier, we believe that this is due to using the bare coefficients of the basis states.  If, as we suspect we should, normalized the coefficients by the density of basis states, we believe these curves would flatten out as we approach smaller $\Delta p$ so that the density normalized coefficients would approach a constant nonzero value.

Focusing on the two-particle eigenstate $\Psi_{24,-24}$, we direct our attention to the middle plot of Fig.~\ref{fig:Ridge Dp}.  Similar to the previous case, we follow the highest green point of this eigenstate, which is at a free energy of 52.0m.  This is the basis state $|-24m,\allowbreak0,0,0,\allowbreak0,24m\rangle$, $|-24m,\allowbreak-1m,\allowbreak0,\allowbreak0,\allowbreak1m,\allowbreak24m\rangle$, $|-24m,-0.5m,\allowbreak-0.5m,0,\allowbreak1m,\allowbreak24m\rangle_+$ and $|-24m,-0.75m,\allowbreak0,0.25m,\allowbreak0.5m,\allowbreak24m\rangle_+$ with coefficient 0.00240, 0.00246, 0.00195 and 0.00161 at $\Delta p=2m, 1m, 0.5m$ and $0.25m$, respectively.  We have plotted this as the four green points along with their best-fit line at the top of the plot.  As before, the basis state at the top changes.  Each basis state sinks down faster than the height of the top green point.  In fact, the basis state $|-24m,0,0,0,0,24m\rangle$ has coefficients 0.00240, 0.00122, 0.00061 and 0.00031 at $\Delta p=2m, 1m, 0.5m$ and $0.25m$, respectively.  We have plotted these as the green points and top dashed green line in the plot.  Again, to show that this behavior is not special to the highest point, we consider a point at the higher free energy of 58.3m.  We find that the highest basis state at this free energy is $|-24m,-4m,0,0,4m,24m\rangle$, $|-24m,\allowbreak-5m,\allowbreak0,\allowbreak1m,\allowbreak4m,\allowbreak24m\rangle_+$, $|-24m,\allowbreak-4.5m,\allowbreak0,\allowbreak0.5m,\allowbreak4m,\allowbreak24m\rangle_+$ and $|-24m,\allowbreak-4.25m,\allowbreak0,\allowbreak0.25m,\allowbreak4m,\allowbreak24m\rangle_+$ with coefficients 0.00082, 0.00055, 0.00035 and 0.00020 at $\Delta p=2m, 1m, 0.5m$ and $0.25m$, respectively.  We have plotted this as the lower of the green points and best-fit straight dotted green line.  But, as before, the basis state itself falls off more quickly.  The basis state $|-24m,-4m,0,0,4m,24m\rangle$ has coefficients 0.00082, 0.00041, 0.00021 and 0.00010 at $\Delta p=2m, 1m, 0.5m$ and $0.25m$, respectively, and is plotted as the lower four green points and their associated dashed green best-fit line.  As before, we believe all of these curves would flatten out and approach a nonzero constant value if we normalized the coefficients with the density of basis states.

Turning to the four-particle eigenstate $\Psi^+_{24,-8,-8,-8}$, we focus on the bottom plot of Fig.~\ref{fig:Ridge Dp}.  We again choose two points at $\Delta p=2m$ to begin with.  The first is the highest yellow point from this eigenstate in Fig.~\ref{fig:Energy Eigenstates}.  It is the basis state $|-24m,0,0,0,0,8m,8m,8m\rangle_+$ at free energy 52.2m with a coefficient of 0.00222.  The top yellow point changes as $\Delta p$ decreases.  At $\Delta p=1m$, it is $|-8m,\allowbreak-8m,\allowbreak-8m,\allowbreak-1m,\allowbreak0,\allowbreak0,\allowbreak1m,\allowbreak24m\rangle_+$ with coefficient 0.00239, at $\Delta p=0.5m$, it is $|-8m,\allowbreak-8m,\allowbreak-8m,\allowbreak-0.5m,\allowbreak0,\allowbreak0,\allowbreak0.5m,\allowbreak24m\rangle_+$ with coefficient 0.00173, and at $\Delta p=0.25m$, it is $|-8m,\allowbreak-8m,\allowbreak-8m,\allowbreak-0.5m,\allowbreak-0.25m,\allowbreak0,\allowbreak0.75m,\allowbreak24m\rangle_+$ with coefficient 0.00119.  These four points are the top four yellow points of the bottom plot of Fig.~\ref{fig:Ridge Dp} along with their best-fit straight line in dotted yellow.  As in the previous eigenstates, the basis state $|-24m,0,0,0,0,8m,8m,8m\rangle_+$ itself, sinks down into the middle of the yellow region at a free energy of 52.2m.  It takes coefficients 0.00082, 0.00042 and 0.00021 at $\Delta p=1m, 0.5m$ and $0.25m$.  These points are plotted along with their best-fit dashed yellow line (the higher of the two).  We also show the behavior of the top of the hump located at a free energy of 65.6m.  At $\Delta p=2m$, this peak is held by the basis state $|-8m,\allowbreak-8m,\allowbreak-8m,\allowbreak-8m,\allowbreak0,\allowbreak2m,\allowbreak6m,\allowbreak24m\rangle_+$ with a coefficient of 0.00053.   As $\Delta p$ decreases, the coefficient of both this peak and this basis state decreases, that of the basis state is faster as before.  The peak is given by the basis states $|-8m,\allowbreak-8m,\allowbreak-8m,\allowbreak-8m,\allowbreak0,\allowbreak1m,\allowbreak7m,\allowbreak24m\rangle$ (coefficient 0.00031) at $\Delta p=1m$ and $|-8m,\allowbreak-8m,\allowbreak-8m,\allowbreak-8m,\allowbreak0,\allowbreak0.5m,\allowbreak7.5m,\allowbreak24m\rangle_+$ (coefficient 0.00017) at $\Delta p=0.5m$.  On the other hand, the basis state $|-8m,\allowbreak-8m,\allowbreak-8m,\allowbreak-8m,\allowbreak0,\allowbreak2m,\allowbreak6m,\allowbreak24m\rangle_+$ has coefficients 0.00019 and 0.00009 at $\Delta p=1m$ and $0.5m$, respectively.  They are plotted along with their best fit dotted and dashed lines, respectively at the bottom of the plot.  As we can see, both the peak of the hump as well as the basis state itself, sink below the cutoff by the time $\Delta p=0.25m$.  As in the previous paragraphs, we expect that normalizing the coefficients according to the density of the basis states would reduce the slope of these curves as $\Delta p\to0$ so that they approach constant nonzero values appropriate to the continuum limit.

\begin{figure}[!]
\begin{center}
\includegraphics[scale=0.89]{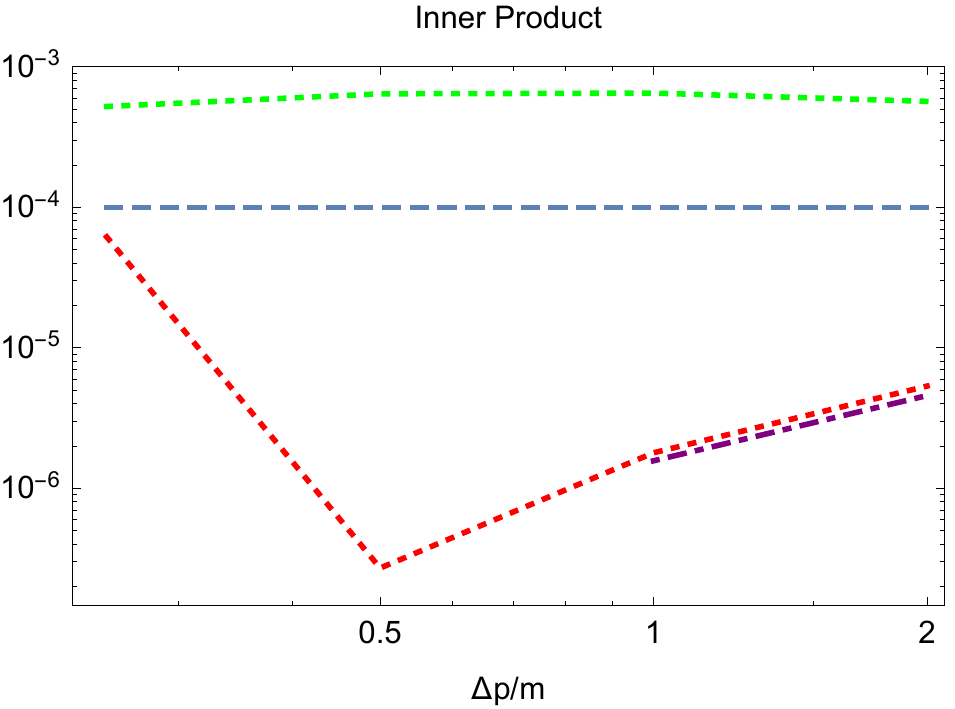}
\end{center}
\caption{\label{fig:IPvsDp}Plot of the inner product between the eigenstates $\Psi_{24,-24}$ and $\Psi^+_{24,-8,-8,-8}$ as a function of $\Delta p$.  The color coding is the same as in Fig.~\ref{fig:IPvsLambda}. }
\end{figure}
We have calculated the inner product between the eigenstates $\Psi_{24,-24}$ and $\Psi^+_{24,-8,-8,-8}$ and plot it in Fig.~\ref{fig:IPvsDp}.  The dashed blue line is the cutoff used, which is the approximate precision of this calculation.  The dotted green line at the top gives the first-order perturbation result which is above the cutoff and not consistent with zero.  As described in Sec.~\ref{sec:lambda}, this is due to not yet having sufficiently accurate coefficients and missing important basis states.  We can see that this problem persists at lower $\Delta p$.   The dot-dashed purple line gives the second-order perturbative result which, as we can see, is below the cutoff and therefore in agreement with a zero result.  In dotted red, we see the results of our cyclic QSE code which are also below the precision of the calculation and, therefore, also in agreement with zero.  We further see that the results of our QSE code are in perfect agreement with second-order perturbation theory where it was possible to calculate it.  The shape of the red dotted curve below the cutoff is unimportant as its shape depends on contributions below the precision of the calculation that were not included.  Therefore, we do not think that any trends, other than being in agreement with zero, can be extracted from this result.

\begin{figure}[!]
\begin{center}
\includegraphics[scale=0.89]{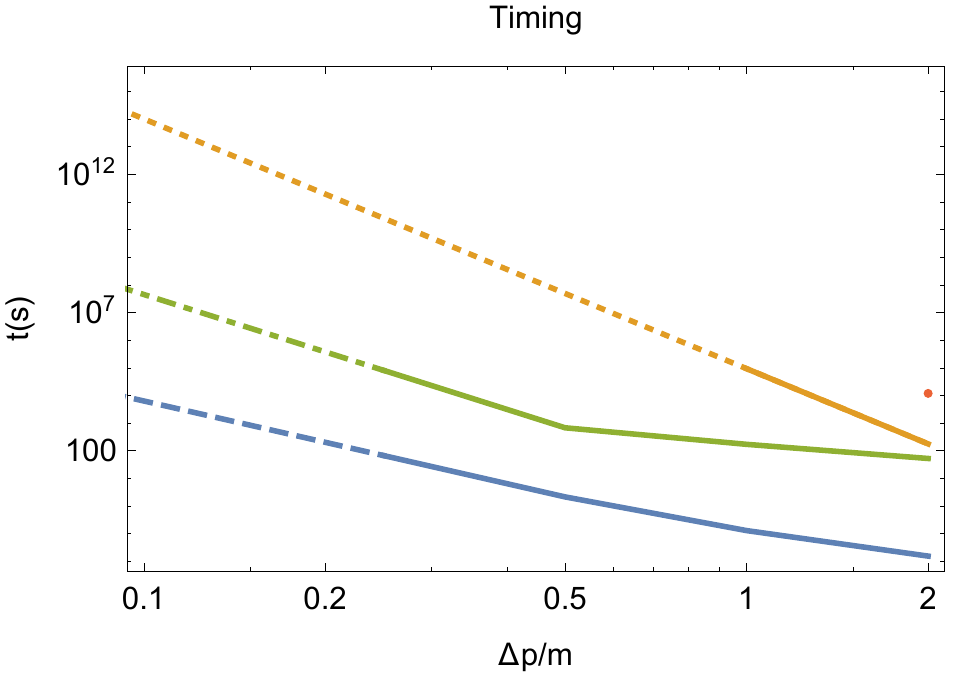}
\end{center}
\caption{\label{fig:time}Plot of the time the code requires for these calculations in seconds as a function of $\Delta p$.  The solid point and lines are measured while the dashed, dot-dashed and dotted lines are extrapolated based on the measured values at the smallest $\Delta p$.  The blue and yellow curves are for first- and second-order perturbation theory, respectively, while the red point is for third-order perturbation theory.  The green curve is the time it takes our cyclic QSE code per cycle.}
\end{figure}
As we have discussed, our goal is eventually to do this calculation at smaller $\Delta p$ and extrapolate to $\Delta p\to0$.  However, each decrease in $\Delta p$ takes an increase in computational time.  We have analyzed this time for the calculations we have done and present it in Fig.~\ref{fig:time} where the vertical axis is the time in seconds and the horizontal axis is the momentum spacing $\Delta p$ we used.  We base this on the time it takes to calculate the 2-particle eigenstate $\Psi_{24,-24}$.  We emphasize that all these calculations have been done on a single cpu on small servers and that improvements in time could certainly be achieved by parallelizing these calculations and potentially using supercomputer resources.  Moreover, since these times depend on the machine we used, we do not think the absolute times are the most meaningful aspect of this plot.  Rather, we are interested in the trends and how our cyclic QSE code compares to perturbation theory.

The solid curves come from the measured values using our calculations while the dashed, dot-dashed and dotted lines are extrapolations down to $\Delta p=0.1m$.  We have plotted first-order perturbation theory in blue.  It took 0.015s, 0.13s, 2.2s and 66s to complete first-order perturbation theory at $\Delta p=2m, 1m, 0.5m$ and $0.25m$, respectively.  We have plotted these points and joined them with a solid blue line that is very nearly linear on a log-log scale.  We have also plotted an extrapolation based on only the times at $\Delta p=0.25m$ and $0.5m$ to estimate how the time would grow as we further decreased $\Delta p$.  We estimate that it would take approximately $10^4s$ or approximately 3 hours at $\Delta p=0.1m$.  We were only able to achieve second-order perturbation theory at $\Delta p=2m$ and $1m$.  It took 170s and 84000s, respectively.  We plotted this data in orange and joined it with a solid orange line.  We also extrapolated this data with a dotted orange line to $\Delta p=0.1m$.  Looking at the extrapolation, we see that it would have taken approximately 3 years to complete second-order perturbation theory for $\Delta p=0.5m$ and much more for smaller $\Delta p$.  Furthermore, we see that the slope of the line for second order is greater than that of first order, so that as $\Delta p$ decreases, the time grows faster for second order than it does for first order.  Finally, we were only able to achieve third-order perturbation theory at $\Delta p=2m$, where it took 26000s.  Since we did not achieve third order at $\Delta p=1m$, we were not able to do an extrapolation.  However, based on first and second order, we expect that the slope would be even greater than that of second order.  

The reason that perturbation theory takes longer as $\Delta p$ decreases is that it has to calculate the contribution of every basis state connected by the Hamiltonian to the main basis state.  The number of these basis states grows exponentially with decreasing $\Delta p$.  At first order, it has to do this for every basis state connected by one application of the Hamiltonian.  However, at higher orders, the burden is greater because it has to calculate the contribution for every basis state connected to the main basis state by two applications of the Hamiltonian (at second order).  This requires a doubling of the number of sums performed in the code, because to determine if a test basis state is connected to the main basis state by two applications of the Hamiltonian, it has to check all intermediate basis state as in $\sum_i\langle b_m|V|b_i\rangle\langle b_i|V|b_t\rangle$, where $|b_m\rangle$ and $|b_t\rangle$ represent the main and test basis states, respectively.  At third order, the calculation has to be done for every basis state connected by three applications of the Hamiltonian, thus requiring a tripling of the sums in the code, as in $\sum_{i j}\langle b_m|V|b_i\rangle\langle b_i|V|b_j\rangle\langle b_j|V|b_t\rangle$.  This is the reason that second order has a greater slope than first order and third order is expected to have a greater slope than second order in Fig.~\ref{fig:time}.

Our cyclic QSE code, on the other hand, took 52s, 170s, 670s and 80000s per cycle for $\Delta p=2m, 1m, 0.5m$ and $0.25m$.  These times are plotted in green and are joined by a solid green line in Fig.~\ref{fig:time}.  Since we ran our code for ten cycles, the total time is one order of magnitude greater for each of these $\Delta p$.  Our code does significantly better than second-order perturbation theory.  Not only does it obtain just-as-good results (even better compared to naive second-order perturbation theory before the diagonalization step we added) but it is orders of magnitude more efficient as $\Delta p$ decreases.  We see that the time increase is nearly linear on a log-log plot between $\Delta p=2m$ and $0.5m$ and has a very shallow slope, shallower even than first-order perturbation theory.  But, between $\Delta p=0.5m$ and $0.25m$, the slope becomes steeper, greater than first-order, but still less steep than second-order perturbation theory.  The reason for this change in slope at $\Delta p=0.5m$ is that this is where the matrix size (that must be constructed and diagonalized) passes from a small memory footprint to a large one.  In particular, our matrix sizes were $700\times700$, $2100\times2100$, $6300\times6300$ and $18900\times18900$ for $\Delta p=2m, 1m, 0.5m$ and $0.25m$, respectively.   (The matrix size is equal to the reduced Hilbert space size squared.)  These matrix sizes are not significant compared to our memory resources until the last step at $\Delta p=0.25m$ where a computational bottleneck is encountered.  This bottleneck can likely be pushed down to lower $\Delta p$ with further clever computational techniques, however, it will eventually become impassable and is a critical and very relevant challenge for this technique.  On the other hand, we used a relatively large reduced Hilbert space size throughout this paper in order to have very high confidence in our results.  We will show in the next section that a much smaller reduced Hilbert space (and therefore a much smaller matrix size) is adequate for the vacuum and two-particle eigenstate, although not for the four-particle eigenstate.  However, as mentioned in previous sections, it is likely that the efficiency of the code at finding the important basis states for the four-particle eigenstate can be further improved.  Further research is necessary on this point.  In any case, it currently looks like there should not be any fundamental challenge to calculating two-particle eignestates using this method.

\section{\label{sec:Hsize}Dependence on the Size of the Reduced Hilbert Space and the Energy of the Eigenstate}
In this section, we explore the dependence on the size of the reduced Hilbert space and the energy of the eigenstate while keeping $\lambda=0.1m^2$ and $\Delta p=2m$.
\begin{figure}[!]
\begin{center}
\includegraphics[scale=0.89]{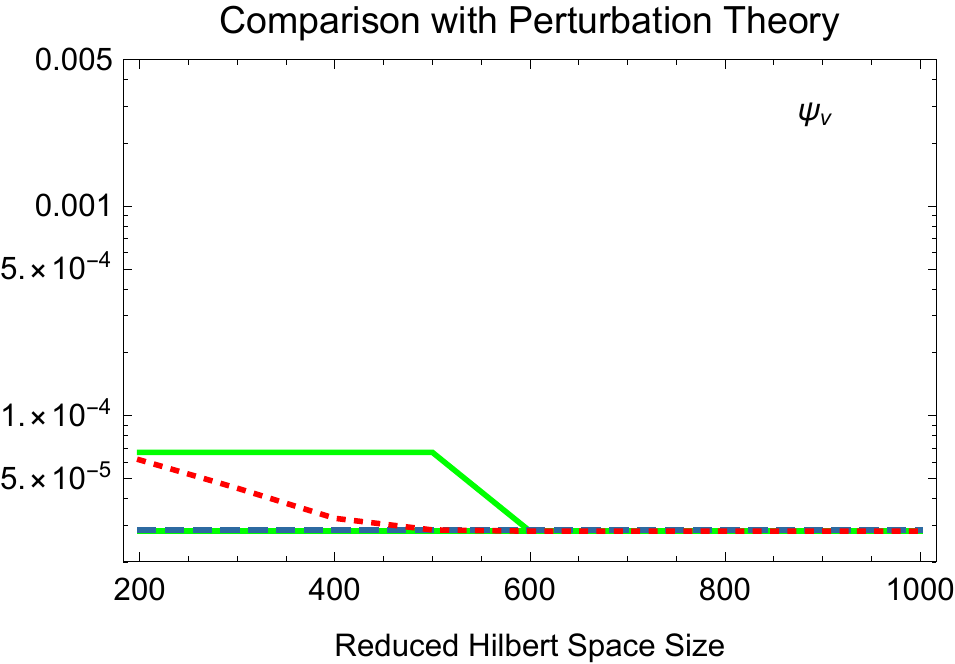}\\
\includegraphics[scale=0.89]{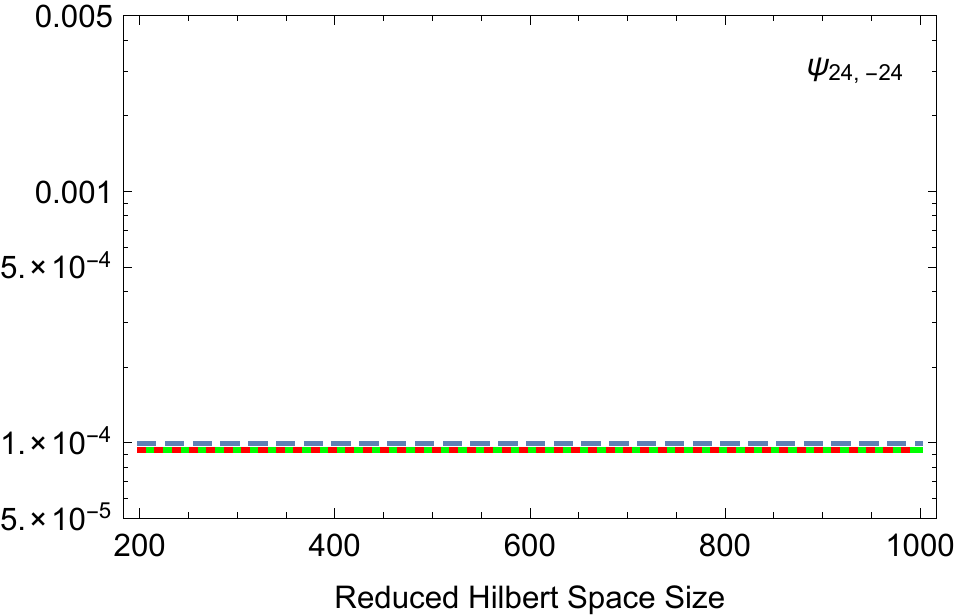}\\
\includegraphics[scale=0.89]{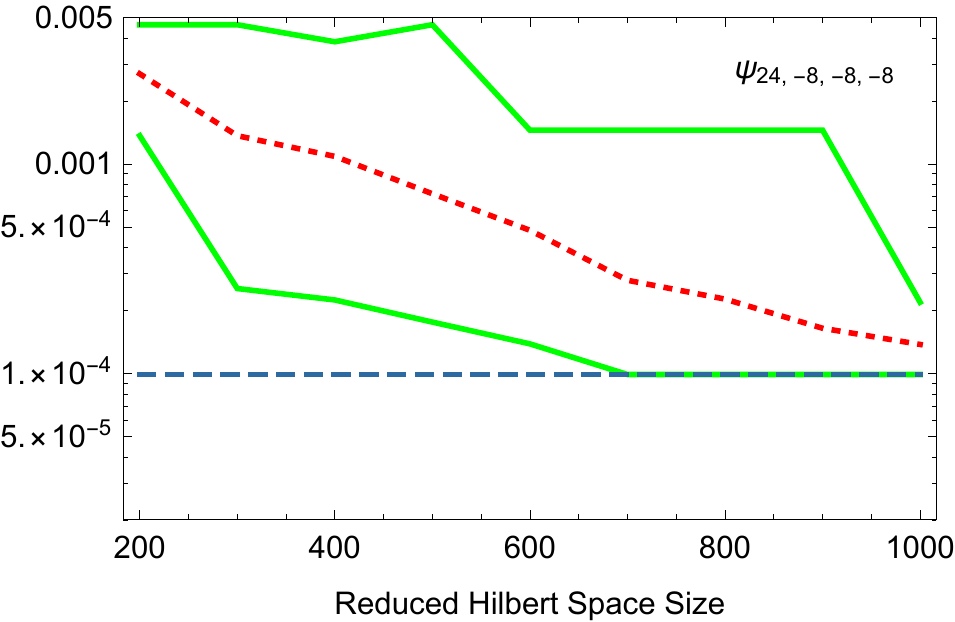}
\end{center}
\caption{\label{fig:DiffVsHsize}Plots of the distribution of largest difference between our cyclic QSE code and second-order perturbation theory for a range of sizes of the reduced Hilbert space.  The color coding is the same as in Fig.~\ref{fig:DiffVsLambda}.}
\end{figure}
We begin with the size of the reduced Hilbert space which is only a factor for our cyclic QSE code, not for perturbation theory.  During each iteration, after removing basis states below the cutoff (see App.~\ref{app:QSE method}), our cyclic QSE code randomly adds new basis states to the reduced Hilbert space until it reaches some predetermined size, that we call the reduced Hilbert space size.  The results are very sensitive to this size as we show in Fig.~\ref{fig:DiffVsHsize}, where the horizontal axis is the reduced Hilbert space size.  The plots in this figure show the largest absolute difference with second-order perturbation theory for different reduced Hilbert space sizes, with a range going from 200 to 1000 in increments of 100.  Since our cyclic QSE code is random, we ran it one-hundred independent times at each reduced Hilbert space size to build up a distribution of results.  For each plot in this figure, we show the cutoff we used in dashed blue, as usual. For each value of the reduced Hilbert space size, we plot the maximum largest differences encountered and join them with a solid green line.  Similarly, we plot the minimum largest differences with a solid green line.  We further plot the average largest difference with a dotted red curve.  

For the vacuum $\Psi_v$ (top plot of Fig.~\ref{fig:DiffVsHsize}), we see that once the reduced Hilbert space size is equal or greater than 600, our cyclic QSE code never misses any of the basis states above the cutoff.  Below this size, it does occasionally miss a basis state.  The solid green line begins at a value of approximately $7\times10^{-5}$ and remains there until a reduced size of 500.  This means that at least once in the one-hundred independent runs, a basis state near the cutoff was missed.  On the other hand, we can see by the red dotted line that this is not typical once the reduced size is 400 or greater.  Even at 300, our cyclic QSE code typically only misses a basis state  which is even closer to the cutoff, and therefore a less severe mistake.  However, even when the reduced Hilbert space size is on the smaller size, the difference with second-order perturbation theory is still not very severe and only slightly above the cutoff.  

The two particle eigenstate $\Psi_{24,-24}$, turns out to do even better.  The reason for this is, as we discussed in Secs.~\ref{sec:results} and \ref{sec:Higher Order}, that first-order perturbation theory finds all the basis states above the cutoff for this value of the coupling constant $\lambda$.  Their coefficients are not correct yet, but they are all there.  Therefore, all our cyclic QSE code needs to do is diagonalize the basis states found by first-order perturbation theory.  We can see this in the middle plot of Fig.~\ref{fig:DiffVsHsize} where all three lines are right at the precision of the calculation.  This is because all one-hundred independent trials gave exactly the same result, since only the diagonalization step was necessary.  The cyclic  part of the QSE code and the generation of random basis states to fill the reduced Hilbert space are both irrelevant, even overkill, for the two-particle eigenstate with our cutoff and value of $\lambda$.  If a much lower cutoff had been used or a larger $\lambda$, then basis states contributing at second and higher order would begin to contribute above the cutoff and our cyclic QSE code would find them cyclically as it does for the vacuum and four-particle eigenstate.  However, at this precision and $\lambda$, in one spatial dimension at least, it is unnecessary.  This suggests that if this precision is sufficient for a calculation of an elastic $2\to2$ scattering amplitude in two or three spatial dimensions, it may be sufficient to find the eigenstates at first order, diagonalize them and take the inner product.  Of course, we hasten to state that this suggestion must be tested.

The results for the four-particle eigenstate $\Psi^+_{24,-8,-8,-8}$ are plotted in the bottom of Fig.~\ref{fig:DiffVsHsize} where we see that the size of the reduced Hilbert space is extremely significant for this eigenstate.  This is due to the large number of important basis states missed at first order as well as their complex structures.  The good news, however, is that these basis states can be found by our cyclic QSE code if we increase the size of the reduced Hilbert space.  As we see in the plot, the general trend is towards smaller largest differences as the reduced space size increases.  By the time the reduced Hilbert space size is 1000, the maximum largest difference out of one-hundred independent trials is only $2\times10^{-4}$, where the cutoff was $1\times10^{-4}$ and the average was half of that.  On the other hand, when the reduced Hilbert space size is small even the minimum largest difference is significantly above the cutoff, even by an order of magnitude when the size is 200.

\begin{figure}[!]
\begin{center}
\includegraphics[scale=0.89]{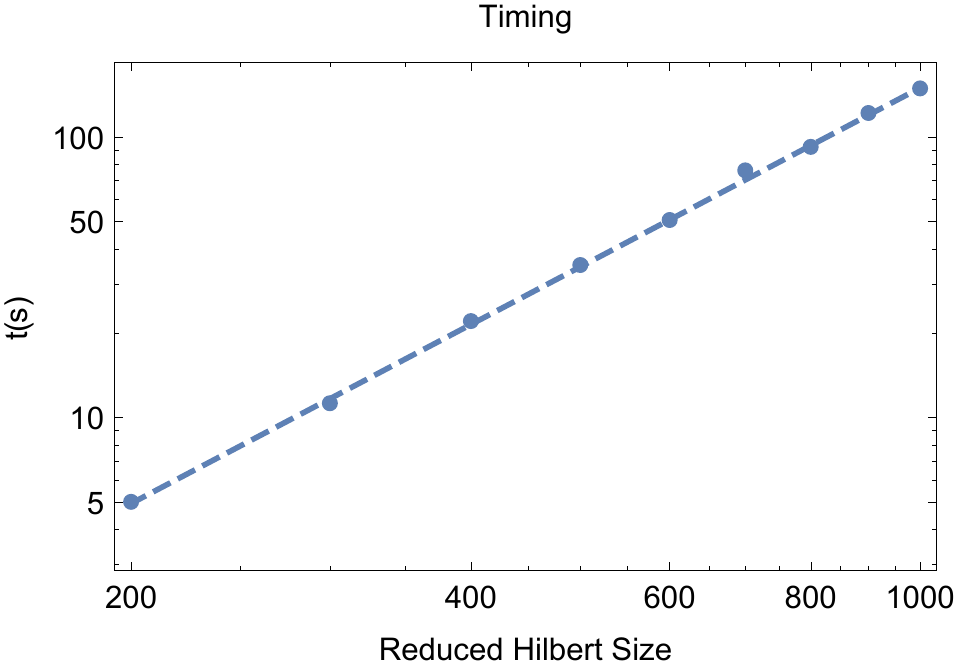}
\end{center}
\caption{\label{fig:TimeVsHsize}Plot of the time it takes our cyclic QSE code per cycle as a function of the reduced Hilbert space size.  The dots are data points and the dashed line is a best fit.}
\end{figure}
Although larger reduced Hilbert space sizes can always be used to increase confidence in the results, this also results in an increase in the time it takes to do the calculation.  The reason is that the reduced Hilbert space size directly determines the matrix size, as discussed in Sec.~\ref{sec:dp}.  In Fig.~\ref{fig:TimeVsHsize}, we show a plot of the time it takes our cyclic QSE code per cycle as a function of the size of the reduced Hilbert space.  We can see that it is very well fit by a straight line on a log-log scale.  This plot is complementary to the plot in Fig.~\ref{fig:time}.

As a final note, we would like to explore how the efficiency of our cyclic QSE method compares with perturbation theory when the eigenvalue energy increases and the eigenstate rises higher above the vacuum.  We will see that our cyclic QSE code is not very strongly affected by an increase in energy while perturbation theory is.  However, to make this comparison more striking, we will use hindsight to tune perturbation theory to be as efficient as possible.  To clarify, we will first describe two sets of parameters that have a large influence on the efficiency of perturbation theory but have very little effect on our cyclic QSE code.  Strictly speaking, in perturbation theory, we should calculate the contribution of all basis states connected to the main basis state by some number of applications of the Hamiltonian (1 at first order, 2 at second order and so on).  So, this means that we should consider all basis states with free energy all the way down to zero and all the way up to infinity.  Of course, this is impossible.  We have to cut this off for perturbation theory to finish in finite time.  Furthermore, only a tiny number of these basis states are important, as we have discussed.  So, we have a $E_{min}$ and $E_{max}$ in our perturbation theory code that restricts the range of basis states included in the perturbative calculations.  We also have these parameters for the cyclic QSE code, but we have found very little sensitivity to them (as long as they are low and high enough, respectively).  The reason is clear; the QSE code only deals with basis states connected to the basis states in the reduced Hilbert space by an application of the Hamiltonian.   So, if the basis states in the reduced Hilbert space are not very high in free energy, then it is not likely to choose a basis state far above them.  For the previous sections of this paper, we used an $E_{min}=0$ and an $E_{max}=80m$.  We could have raised $E_{min}$ for $\Psi_{24,-24}$ and $\Psi^+_{24,-8,-8,-8}$, but we wanted to allow the possibility that some basis states below the main basis state were above the cutoff.  Indeed, there were several just below the cutoff.  

There is another parameter important in the efficiency of the perturbative code.  It is $N_{max}$, the maximum number of free particles in a basis state.  Strictly speaking, perturbation theory should consider all basis states connected by the Hamiltonian once or twice (for first or second order).  This means that for $\Psi_{24,-24}$, we should keep up to 6 free particles in our basis states at first order and up to 10 free particles at second order.  For $\Psi^+_{24,-8,-8,-8}$, we should keep up to 8 and 12 free particles at first and second order, respectively.  In our calculations, up until this point, we have used an $N_{max}=10$ which includes all possible basis states for $\Psi_{24,-24}$ but falls slightly short for $\Psi^+_{24,-8,-8,-8}$.  We did this because choosing $N_{max}=12$ just took longer than we wanted to wait, especially since we knew with hindsight that 12-particle basis states were not important for these eigenstates.  However, we note that the timing plots in Figs.~\ref{fig:time} and \ref{fig:TimeVsHsize} and the times discussed in the text were for $\Psi_{24,-24}$ and included the full perturbative calculation.  On the other hand, with hindsight, we will reduce $N_{max}$ to 6 for $\Psi_{24,-24}$.  As we can see in the plots of these eigenstates, higher multiplicity basis states are never above the cutoff for $\lambda=0.1m^2$ and a cutoff of $1\times10^{-4}$.  This will reduce the time for perturbation theory significantly.  Although, it should be remembered that a slight increase in $\lambda$ brings 8-particle basis states above the cutoff.  So, the times we show for second-order perturbation theory are extremely generous to the perturbative time.  On the other hand, we have found our cyclic QSE code to be largely insensitive to this parameter.  Again, the reason is clear.  Our  QSE  code only randomly chooses basis states connected to the basis states already in the reduced Hilbert space.  And, it focuses more effort on those that are within $\pm2$ free particles of those already in the reduced Hilbert space.  So, as long as $N_{max}$ is not too low, our cyclic QSE code is unaffected by this parameter.  

\begin{figure}[!]
\begin{center}
\includegraphics[scale=0.89]{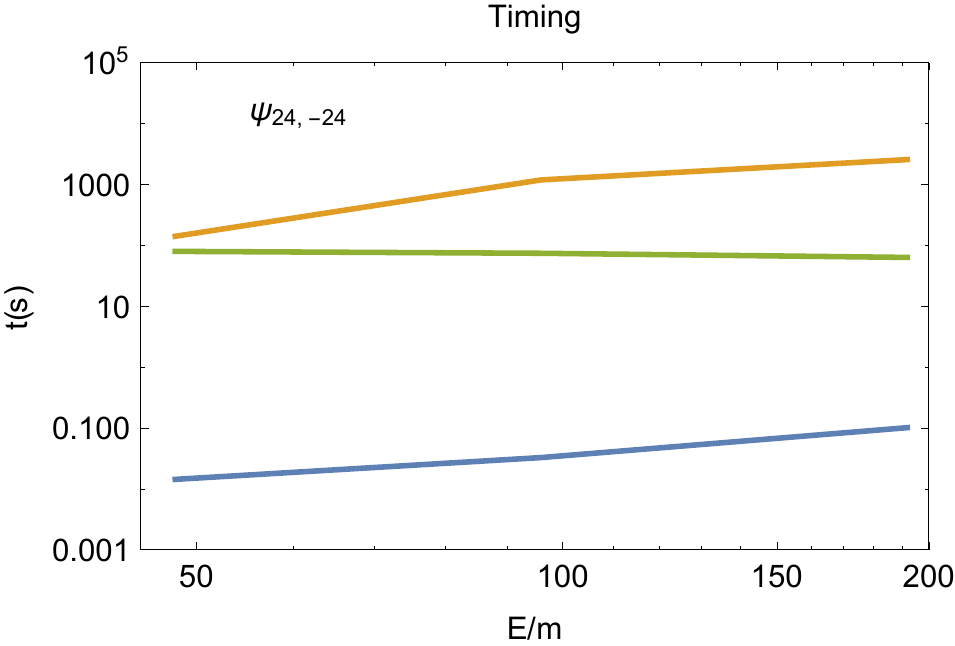}
\end{center}
\caption{\label{fig:TimeVsEnergy}Plots of the time it takes to calculate the eigenstate for different eigen-energies.  In blue and orange, we show the time it takes first- and second-order, respectively.  The time it takes our cyclic QSE code is shown in green. }
\end{figure}
In this comparison, then, we will push perturbation theory to near its efficiency limit.  We have recalculated $\Psi_{24,-24}$ with an $E_{min}=40m$, $E_{max}=80m$ and $N_{max}=6$.  We then calculated the analogous eigenstate at twice the energy, namely $\Psi_{48,-48}$.  We did this calculation with $E_{min}=90m$, $E_{max}=130m$ and $N_{max}=6$.  We also calculated the analogous eigenstate with four times the energy, namely $\Psi_{96,-96}$ with $E_{min}=190m$ and $E_{max}=230m$ and $N_{max}=6$.  Before we did these calculations, we first did the calculations with more generous parameters to make sure that we wouldn't miss any important basis states when we restricted to these more efficient parameters.  On the other hand, we tried our cyclic QSE code both with the original parameters and with these more restricted ones and saw no significant difference in time.  We have plotted the times for perturbation theory and our  QSE code in Fig.~\ref{fig:TimeVsEnergy}.  We have plotted the time it takes first- and second-order perturbation theory in blue and orange, respectively.  We see that first-order perturbation theory is much faster than our cyclic QSE code and second-order perturbation theory as before, but it increases in time as the eigenstate grows higher in energy.  With our very restrictive settings, the slope is not too steep and we do not expect it to cross the time it takes for our cyclic QSE code until quite high energies.  Second-order perturbation theory continues to take longer than our QSE code and also increases in time as the eigenstate increases in energy.  It also has a fairly shallow slope for these extreme parameter choices.  However, on the other hand, we see a remarkable feature of our QSE code plotted in green.  It is essentially flat!  Our cyclic QSE code is largely insensitive to the height of the eigenstate above the vacuum.  The reason for this is that the QSE code doesn't try to calculate everything.  It just searches the Hilbert space near the basis states it already has that are above the cutoff.

\section{\label{sec:conclusions}Summary and Conclusions}

In Sec.~\ref{sec:results}, we presented the results of our cyclic QSE code calculations of three eigenstates along with the first-order pertubative result.  One of the eigenstates was the vacuum $\Psi_v$, while two were ``high" above the vacuum.  The first was the eigenstate $\Psi_{24,-24}$ with two particles of momentum $24m$ and $-24m$, where $m$ is the mass parameter in the theory.  The second eigenstate was $\Psi^+_{24,-8,-8,-8}$, a parity-symmetric four-particle eigenstate with one particle of momentum $24m$ and the other three of momentum $-8m$.  We chose these two because they were potentially the type of eigenstates that could represent in and out states of a scattering S-Matrix element.  We plotted these eigenstates in Fig.~\ref{fig:Energy Eigenstates} along with the first-order perturbative approximation to them.  In the text, we described and compared and contrasted them.  We pointed out that although first-order perturbation theory failed to find many of the important basis states to the eigenstates (because they did not contribute until second order), our cyclic QSE code did find them.  We also pointed out that our cyclic QSE code obtained different coefficients for the basis states that were found by both it and first-order perturbation theory and claimed that the difference in coefficients was due to higher-order perturbative corrections.  

We ended this section with a calculation of the inner product between $\Psi_{24,-24}$ and $\Psi^+_{24,-8,-8,-8}$ (the S-Matrix between these eigenstates) and found that it was nonzero using the first-order perturbative results and was consistent with zero using the results of our cyclic QSE code.  Furthermore, we pointed out that the inner product should be zero because the energies of these eignestates (48.042m and 48.209m) were not the same and therefore, by energy conservation, it must be zero.  We noted that the reason first-order perturbation theory did not give a zero result is that we truncated the first-order perturbative state at a cutoff on the coefficients.  So, the first-order perturbative calculation of the inner product was not complete in this sense.  On the other hand, although our cyclic QSE code also only kept basis states above the cutoff, unlike first-order perturbation theory, it had an essentially complete set of basis states above the cutoff, including basis states that contribute at second order.  Moreover, it had more correct coefficients for those basis states.  Together, these properties allow it to achieve a more accurate result for the inner product.  Indeed, the inner product using the results of the cyclic QSE code was in agreement with what we expect, which is zero.

In Sec.~\ref{sec:Higher Order}, we compare the results of our cyclic QSE code with second- and third-order perturbation theory.  In Fig.~\ref{fig:Diffs}, we show the difference in coefficients for each basis state at first order on the left and with second order on the right.  We show that although the results of our QSE code disagrees with first-order perturbation theory, it is in full agreement with second and third order for the vacuum $\Psi_v$ and the two-particle eigenstate $\Psi_{24,-24}$.  For the four-particle eigenstate $\Psi^+_{24,-8,-8,-8}$, we show that the agreement with second-order perturbation theory is much better than with first order, but that there are four basis states that are still not in full agreement.  We point out that this is due to the random nature of our cyclic QSE code but that these missed basis states are very near the cutoff while basis states high above the cutoff are much more likely to be discovered by our QSE code.  We point out that two things are happening as our QSE code works.  The first is that it is discovering new basis states missed by first-order perturbation theory while the second is that it is diagonalizing the Hamiltonian with respect to whichever basis states it has found.  Although both of these are typically important, we find that for the two-particle eigenstate $\Psi_{24,-24}$, for these parameter values and cutoff, all the important basis states are found by first-order perturbation theory.  All that is necessary for this eigenstate is the diagonalization step.  For this reason, our QSE code shoots directly to agreement with second-order perturbation theory on the first iteration, as seen in the middle plot of Fig.~\ref{fig:DiffVsIt}.  On the other hand, for the vacuum and the four-particle eigenstate, our QSE code fills in most of the missing basis states in the first few iterations as seen in Fig.~\ref{fig:Iterations} and Fig.~\ref{fig:DiffVsIt}.  Since the agreement with second-order perturbation theory is better than that of frist order, we take this as a sign that our cyclic QSE code is working properly.

In Sec.~\ref{sec:lambda}, we scanned over the value of $\lambda$ and plotted the eigenstates for several values in Fig.~\ref{fig:lambdas}.  We described the dependence of several basis states for each eigenstate on $\lambda$ and how its growth related to first- or second-order perturbation theory.  In particular, we showed that basis states connected to the main basis state by one application of the Hamiltonian grew linearly with $\lambda$, when $\lambda$ was small and perturbative.  On the other hand, we showed that basis states connected to the main basis state by two applications of the Hamiltonian grew quadratically with $\lambda$, again when $\lambda$ was small and perturbative.  We also noted that the deviation from linear or quadratic growth at higher $\lambda$ was due to the increasing importance of higher-order corrections.  We also compared the results of our cyclic QSE code with second-order perturbation theory over the range of $\lambda$.  Since our code is random in nature, we ran our code 100 independent times for each value of $\lambda$ and compared each independent run with second-order perturbation theory.  We used these independent runs to build up a distribution of results.  In Fig.~\ref{fig:DiffVsLambda}, we plotted the range of largest differences between our QSE code and second-order perturbation theory.  We showed that the vacuum and two-particle eigenstates, $\Psi_v$ and $\Psi_{24,-24}$, are nearly always in agreement within the precision of the calculation.  On the other hand, we showed that the comparison for the four-particle eigenstate gets worse as $\lambda$ increases and is in bad agreement for $\lambda\gtrsim0.1m^2$.  We noted that this was due to increasingly missing important basis states as $\lambda$ increases and was a severe challenge for this code properly constructing four-particle eigenstates such as this one.  We suggested that it might be possible to further improve the QSE code for four-particle eigenstates and that that would be an important area of future research.  Finally, we also calculated the inner product of $\Psi_{24,-24}$ and $\Psi^+_{24,-8,-8,-8}$ for this range of $\lambda$ and plotted it in Fig.~\ref{fig:IPvsLambda}.  Since this result potentially depends on the random nature of our cyclic QSE code, we calculated it for the 100 independent trials of our QSE code and show the range of values in this plot.  We found the inner product to be in agreement with zero for the entire range of $\lambda$ and for all 100 trials at each value of $\lambda$.  We noted that this was true even though we missed important basis states in $\Psi^+_{24,-8,-8,-8}$ at large $\lambda$ because those missed basis states were below the cutoff in $\Psi_{24,-24}$, and thus, did not contribute above the cutoff.

In Sec.~\ref{sec:dp}, we analyzed the dependence of our results on the momentum spacing $\Delta p$.  We calculated the eigenstates at $\Delta p=2m, 1m, 0.5m$ and $0.25m$ and show plots for three of these $\Delta p$ in Fig.~\ref{fig:dps}.  We noted several important features of these eigenstates as $\Delta p$ decreased.  The first is that the shapes of the eigenstates are largely the same as $\Delta p$ decreases towards smaller values.  This gives us some confidence that the results with large $\Delta p$ approximate the results when $\Delta p$ is small.   Secondly, the density of basis states increases as $\Delta p$ decreases, filling in the gaps in the plots and resolving the structure with greater detail.  We accounted for this by understanding the dependence of the free energy on $\Delta p$ through $E_f=\sum_i\sqrt{n_i^2\Delta p^2+m^2}$.  Third, we found that the fall off of the basis states as their free energy increased was greater as $\Delta p$ became smaller.  We picked this apart and found that this was due to the coefficients of the basis states decreasing as $\Delta p$ diminished.  We claimed that this was mainly due to plotting the coefficients of the basis states directly rather than the coefficients normalized by the density of basis states.  If we had instead plotted the coefficients normalized by the density of basis states, we believe the fall of the basis states would stabilize and approach a constant value as $\Delta p\to0$.  But, we noted that we do not yet know the correct density of basis states to use for this normalization.  We believe this is an extremely important question to answer if we are to accurately calculate the limit of these eigenstates and the S-Matrix inner product between eigenstates as $\Delta p\to0$ and plan to research it further in the future.  We ended this section with a calculation of the inner product as a function of $\Delta p$ and plotted it in Fig.~\ref{fig:IPvsDp}.  We found that the first-order perturbative result continued to be greater than zero for the entire range that we analyzed while both second order and our cyclic QSE code were both in agreement with zero at the precision of our calculation.  Again, this was encouraging that our QSE-code result was in agreement with the higher orders of perturbation theory as well as with expectation based on physical arguments of energy conservation as $\Delta p\to0$.

In Sec.~\ref{sec:Hsize}, we explored the dependence of our code on the size of the reduced Hilbert space.  As we did with the dependence on $\lambda$, we calculated the eigenstates 100 independent times with our cyclic QSE code for each value of the reduced Hilbert space size.  We plotted the distribution of largest differences between the results of our QSE code and second-order perturbation theory in Fig.~\ref{fig:DiffVsHsize}.   We found that for the vacuum and two-particle eigenstate, $\Psi_v$ and $\Psi_{24,-24}$, our cyclic QSE code was in very good agreement with second-order perturbation theory for a large range of sizes of reduced Hilbert spaces from 200 up to 1000.  We even suggested this might be used to our advantage if we calculate elastic $2\to2$ scattering in two spatial dimensions.  On the other hand, we found that the four-particle eigenstate $\Psi^+_{24,-8,-8,-8}$ was extremely sensitive to the reduced Hilbert space size and that we only obtained good results for larger values on the order, or larger than, 1000.  On the one hand, increasing the size of the reduced Hilbert space is a great way to improve the accuracy of the results coming from our QSE code.  However, we note that, on the other hand, the time our QSE code takes to do the calculation is exponentially sensitive to the reduced space size.  We plotted the time it took versus the size of the reduced Hilbert space in Fig.~\ref{fig:TimeVsHsize}.

Also in Sec.~\ref{sec:Hsize}, we studied how the efficiency of our cyclic QSE code was affected as the eigenstate increased in energy above the vacuum.  We compared this with perturbation theory which took exponentially longer the higher the eigenstate was above the vacuum.  This was because the density of basis states that must be checked increases exponentially as the free energy increases.  On the other hand, we noted that the QSE code was essentially unaffected by an increase in the energy of the eigenstate.  We claimed that this was because the QSE code does not try to calculate the contribution of every basis state in the vicinity.  It simply searches nearby the basis states it already has and this algorithm seems to work well at a broad range of energies.

There are still several open questions that we feel are very important.  We have already mentioned that the search algorithm for new basis states to add to the reduced Hilbert space needs to be improved for four-particle eigenstates if this method is to be useful for their calculation.  We also mentioned that, in order to get the limit of these eigenstates and their inner products as $\Delta p\to0$, it is imperative that we understand the density of basis states better and use it to normalize the coefficients of the basis states.  However, there are others.  Although this QSE method appears to work well at the values of $\Delta p$ that we have calculated and it appears that it scales better than perturbation theory as $\Delta p\to0$ and as the eigenstate energy increases, it still grows too large for the very small values of $\Delta p$ that we are interested in.  We obtained results as low as $\Delta p=0.25m$ in this project but we believe we eventually need to get $\Delta p$ down to perhaps around $0.01m$ in order to get physically meaningful results in the continuum limit.  We think this may be possible with this method on a supercomputer, but we also think there may be significant ways to yet improve on the algorithm itself.  This will form a major line of our future research.

Beyond these points, it is our long-term goal to calculate non-trivial, non-zero S-matrix elements between scattering eigenstates.  We were unable to do this in the present project because no two eigenstates were degenerate in energy where we could achieve both our QSE-code calculation as well as second- and even, at one value of $\Delta p$, third-order perturbation theory for comparison.  In order to calculate non-zero inner products, we will either need to reduce $\Delta p$ to much smaller values where the eigenstates begin to overlap at the precision of the calculation or increase the spatial dimensions to two.  In the latter case, there would always be multiple two-particle eigenstates with the same magnitude of momenta, and therefore (due to the rotational symmetry) the same energy.  They would be degenerate.  On the other hand, the added complexity of increasing the spatial dimension will be a challenge.  However, we think that since we can control the granularity of the angle, we might make progress in this direction, and perhaps achieve a nonzero scattering matrix element.  In any case, achieving a nonzero scattering amplitude is a major objective of our future research.

Although it is not the purpose of this paper to give the technical details of how to extract the scattering amplitude from the non-zero inner product, perhaps it would be good to give some insight into how we hope to do this in the future.  Although our results will be effectively non-perturbative, our intended technique is based on the ideas of ``old-fashioned" perturbation theory.  Our discussion follows Weinberg \cite{Weinberg:1995mt}.  The S matrix can be split into a non-interacting piece and a scattering amplitude.  In Weinberg's very compact notation, this relation reads
\begin{equation}
S_{\beta\alpha}=\delta(\beta-\alpha)-2i\pi\delta(E_\beta-E_\alpha)T_{\beta\alpha}^+\ ,
\end{equation}
where the greek symbols $\alpha$ and $\beta$ represent the full list of quantum numbers for all the particles of the in and out states, respectively.  This includes the momenta, the spins, the charges and masses.  When $\alpha=\beta$, the in and out states are exactly the same with no interaction.  This is removed in the $\delta(\beta-\alpha)$ term.  $T_{\beta\alpha}^+$ is the scattering amplitude, up to a total-momentum preserving delta function.  It is given by the inner product
\begin{equation}
T_{\beta\alpha}^+=\left(\Phi_\beta,V\Psi_\alpha^+\right)\ ,
\label{eq:T matrix}
\end{equation}
where $\Phi_\beta$ is a free-particle state, analogous to our free-particle states $\langle p_1,\cdots|$, $\Psi_\alpha^+$ is a scattering eigenstate, and $V$ is the potential, the interacting part of the Hamiltonian.  In old-fashioned perturbation theory, the scattering eigenstate $\Psi_\alpha^+$ is expanded in a power-series in the coupling constant.  At leading order, $\Psi_\alpha^+=\Phi_\alpha$, and the leading order contribution is simply $\left(\Phi_\beta,V\Phi_\alpha\right)$, or using our discrete basis states, the leading order contribution to the scattering amplitude is given by
\begin{equation}
T_{\beta\alpha}^{+(0)} = \langle\beta|V|\alpha\rangle\ .
\end{equation}
Consider, for example, the elastic scattering of two particles in our $\lambda\phi^4$ theory.  The scattering eigenstate is dominated by the free two-particle basis state, say $|p_1,p_2\rangle$, therefore, the leading-order contribution to the elastic scattering amplitude is given by
\begin{equation}
T_{p_1,p_2;p_1,p_2}^{+(0)} = \langle p_1,p_2|V|p_1,p_2\rangle\ .
\end{equation}
If we look at the potential of our Hamiltonian given in Eq.~(\ref{eq:Discrete Hamiltonian}), we find a term that is $\lambda a_{p_1}^\dagger a_{p_2}^\dagger a_{p_1} a_{p_2}$, up to normalization factors.  This term will annihilate the two particles in the free-particle state on the right, then re-create them, leaving us with $\lambda\langle p_1,p_2|p_1,p_2\rangle=\lambda$, again up to normalization factors.  But, this is the well-known elastic $2\to2$ scattering amplitude at tree level in $\lambda\phi^4$ theory!  The normalization factors will have to be accounted for properly, of course, but they are the discrete-momentum analogs of the factors in the continuum theory, which has already been worked out in old-fashioned perturbation theory.  Naturally, we would like to go beyond tree level.  In old-fashioned perturbation theory, we find $\Psi_\alpha^+$ to higher order, as we have done in Appendix~\ref{app:perturbative solution}, showing the perturbative expansion in terms of the free-particle basis states in Eqs.~(\ref{eq:app:pert:cjn}) through (\ref{eq:app:pert:psi total}).  We then plug these higher-order contributions into Eq.~(\ref{eq:T matrix}).  As a result of these higher-order corrections to the scattering eigenstate, the scattering amplitude will pick up contributions from other basis states, and moreover, the contribution from each basis state will change slightly as higher orders in perturbation theory are included.  Additionally, because the contributions of the basis states at higher orders are energy dependent, it will lead to energy dependence in the scattering amplitude.  

So far, we are not stating anything new.  All of this existed long ago, before Feynman diagrams were even introduced \cite{Weinberg:1995mt}.  Our contribution is to suggest that these scattering eigenstates can be found more efficiently using the QSE algorithm and that the result will be effectively correct to all orders in perturbation theory, at least to the precision of the calculation, without doing the much more difficult perturbative calculations.  In this paper, we have shown that we can reproduce the scattering eigenstates with the QSE method and we have shown that it is more efficient and scales much better than perturbation theory as the momentum spacing $\Delta p$ decreases.  Once we have constructed the scattering eigenstate, we can simply plug it into Eq.~(\ref{eq:T matrix}) to determine the scattering amplitude.  We do this in the same way we would do it for perturbation theory, however,  we simply do it all at once rather than order by order in the coupling constant.  Of course, there will be challenges as we approach this result, both expected and unexpected.  Among the most important near-term expected challenges are that we will have to reduce the momentum spacing $\Delta p$ to be very small and determine the density of states so that our eigenstates asymptotically approach a stable continuum eigenstate.  We then hope to extrapolate our results to the continuum limit based on the small $\Delta p$ results.  

Our approach has some aspects in common with calculations of the S matrix on the lattice.  Our discretization of momentum spacing can be seen to come from a finite space with a periodic boundary condition.  However, we do not latticize space, we do not Euclideanize space and we do not deal directly with the fields.  In fact, it is partly our purpose to move away from the field formulation of particle physics.  Nevertheless, there is important research going into the lattice calculation of the S matrix, which complements and influences our own work.  An important breakthrough in this field was a method for calculating the elastic scattering amplitude on the lattice \cite{Luscher:1990ck}.  As part of that work, the authors state the importance of using finite spaces that are very large (the dual of our very small $\Delta p$) so that the ``wave function ... is accurately given by the free wave" near the boundaries and to suppress the effects of virtual particles going ``around the world".  Of course, we must achieve a small $\Delta p$ partly, at least, for the same reasons.   Lattice calculations of non-elastic scattering, on the other hand, have not been fully worked out yet, but a couple gateway references for the progress in this field are \cite{Briceno:2012rv,Romero-Lopez:2017gag}.

\appendix

\section{\label{sec:truncated}Truncated Hilbert Space Size}
\begin{figure}[!]
\begin{center}
\includegraphics[scale=0.89]{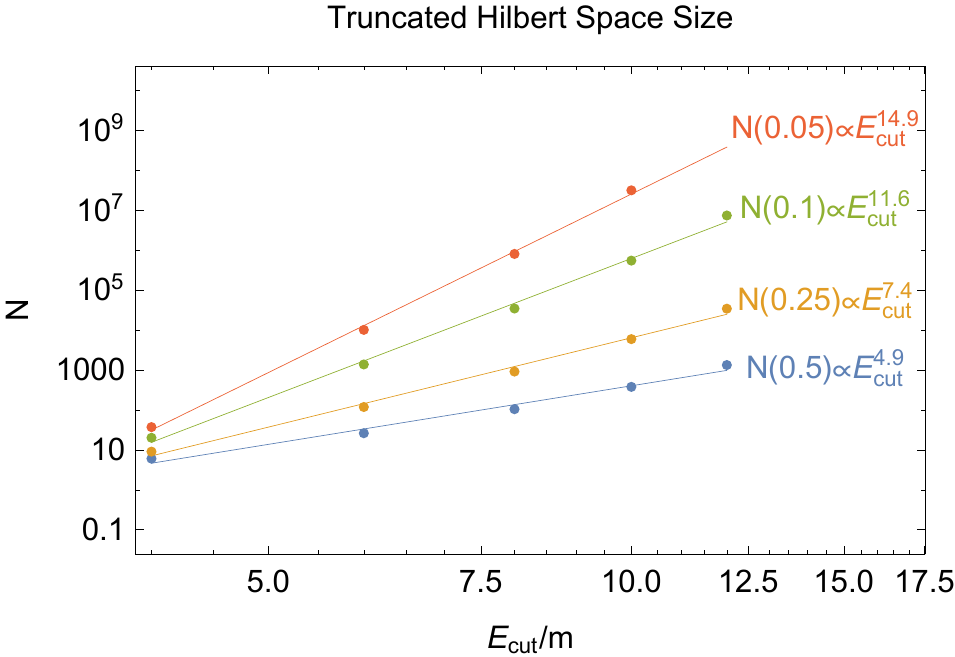}\\
%\vspace{0.05in}
\includegraphics[scale=0.89]{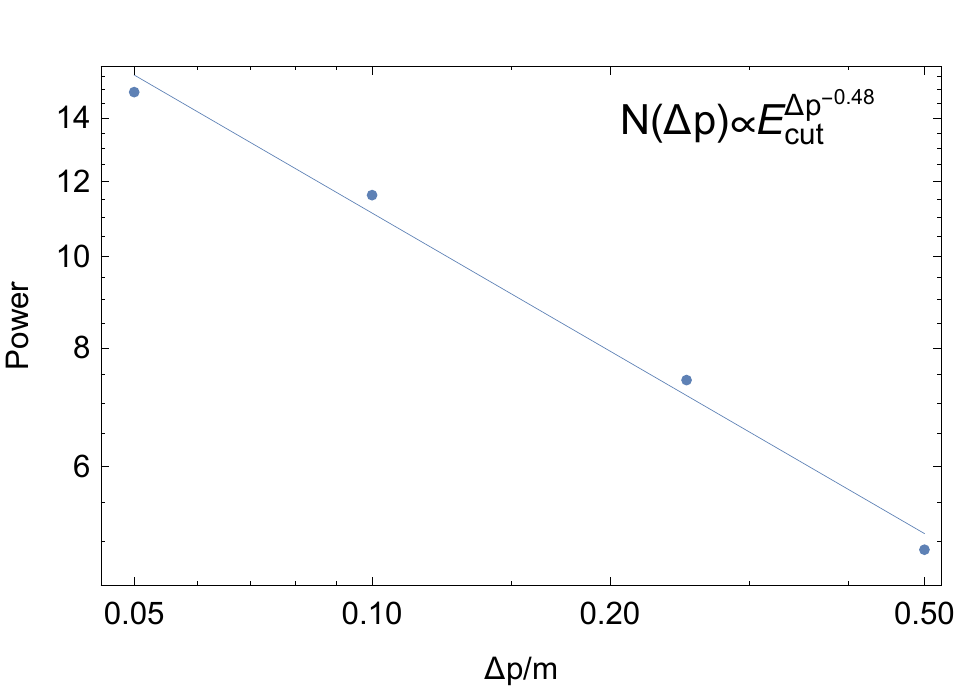}
\end{center}
\caption{\label{fig:Hilbert Space Size}The number of basis states in a Hilbert space truncated by an energy cutoff on the basis states in the top plot.  The dots are calculated and the lines are fit to the dots.  The slopes on a log-log plot are determined and plotted in the lower plot.  They are fit to a straight line, determining the slope as a function of $\Delta p$. }
\end{figure}

In \cite{Christensen:2016naf}, after discretizing momentum space, the Hilbert space was truncated by setting an upper limit on the free-particle energy of the basis states.  A limit on the number of particles to a maximum of two was also imposed.  The reason this was done is that, when all the basis states are kept, even with a cutoff on the free energy of the basis states, the Hilbert space grows too rapidly and quickly overcomes the ability of computers to diagonalize.  Indeed, it quickly overcomes the ability of computers to even store the Hamiltonian or the Hilbert space itself.   For illustration, we have plotted the size of the truncated Hilbert space as a function of the cutoff on the free-particle energies of the basis states in the top plot of Figure~\ref{fig:Hilbert Space Size}.  
%\begin{table}
%\begin{center}
%\begin{tabular}{|l|lllll|}
%\hline
%$E_{cut}$ & 2 & 4 & 6 & 8 & $\infty$\\
%\hline
%2 & 2 & 2 & 2 & 2 & 2\\
%4 & 36 & 37 & 37 & 37 & 37\\
%6 & 58 & 10,104 & 10,105 & 10,105 & 10,105\\
%8 & 79 & 39,737 & 806,519 & 806,520 & 806,520\\
%10 & 99 & 95,494 & 6,750,753 & 31,636,754 & 31,636,755\\
%\hline
%\end{tabular}
%\end{center}
%\caption{Test}
%\end{table}
We have done this for several momentum step sizes.  For example, if we set the momentum step to $\Delta p=0.05m$, the Hilbert space already contains over 100 million state with a cutoff of only 12m.  We have also fit straight lines to the results on a log-log plot.  As expected, we find that the slope of these lines increases as $\Delta p$ becomes smaller.  We have further plotted the slopes of these lines on the bottom plot of the same figure and fit a straight line to them on a log-log plot.  We find that the slope grows roughly as $1/\sqrt{\Delta p}$
\begin{equation}
N(\Delta p)\sim E_{cut}^{\Delta p^{-1/2}}\ ,
\end{equation}
which makes sense, since as $\Delta p\to0$, the slope should become infinite.  

On the other hand, in \cite{Christensen:2016naf}, we showed that for reasonable results, $\Delta p$ should be much smaller than $m$ while $E_{cut}$ should be much larger.  This means that, in order to achieve good results with a reasonably small $\Delta p$ and a reasonably high cutoff energy, while including multiple free particles in the basis states, we must find an alternative to diagonalizing the Hamiltonian while keeping all basis states below the cutoff.  Such an alternative was outlined in \cite{Lee:2000ac} and is called the Quasi-Sparse Eigenvector (QSE) method.  

\section{\label{app:QSE method}The Cyclic QSE Method}

Since the QSE method has been described in detail in \cite{Lee:2000ac}, we will review it here in the context of our own calculations and invite the reader to refer to \cite{Lee:2000ac} for more details on the method itself.  Since we are interested in perturbative coupling, each scattering eigenstate is dominated by one basis state (see Appendix~\ref{app:perturbative solution}).  For example, the vacuum approaches the 0-free-particle basis state in the limit $\lambda\to0$.  As the coupling constant turns on, other basis states begin to contribute.  Schematically, a perturbative eigenstate looks like
\begin{equation}
\Psi_m = c^{(0)}|m\rangle + \lambda\sum_{i=1}^\infty c^{(1)}_i|b^{(1)}_i\rangle + \lambda^2\sum_{i=1}^\infty c^{(2)}_i|b^{(2)}_i\rangle + \cdots\ ,
\label{eq:pert coeff}
\end{equation}
where $|m\rangle$ is the main basis state, $|b^{(1)}_i\rangle$ is a basis state contributing at first order, $|b^{(2)}_i\rangle$ is a basis state contributing at second order, and so on.  Although these sums are infinite, the contribution of the vast majority of the basis states is negligible, either because of the smallness of $\lambda^n$ or because of the smallness of $c^{(n)}_i$ which depends on the Hamiltonian matrix elements between basis states and inversely on their energy differences.  (See App.~\ref{app:perturbative solution} for details.)

Since calculations using these eigenstates will have working precisions, it does not make sense to keep or use basis states whose contributions are below that working precision.  So, if possible, we would like to truncate the full eigenstate by keeping only the basis states whose contributions to the eigenstate is greater than the precision of the calculation.  We will call this the cutoff on the basis states.   After throwing away any basis states whose contribution is smaller than the cutoff, we have something like
\begin{eqnarray}
\Psi_m &=& c^{(0)}|m\rangle + \lambda\sum_{i=1}^{N_1} c^{(1)}_i|b^{(1)}_i\rangle + \lambda^2\sum_{i=1}^{N_2} c^{(2)}_i|b^{(2)}_i\rangle\nonumber\\
&&+  \lambda^3\sum_{i=1}^{N_3} c^{(3)}_i|b^{(3)}_i\rangle\ ,
\end{eqnarray}
where we are assuming, for the sake of illuminating this technique, that all order-4 contributions are below the cutoff and only the first $N_i$ basis states at order $i$ are above the cutoff.  For some given precision and momentum spacing $\Delta p$, we might have $N_1+N_2+N_3$ is on the order of one-hundred to one-thousand, out of the infinite number of possible basis states in the full Hilbert space.  If possible, we would like to calculate only the contributions from these basis states in perturbation theory, however, unfortunately, we do not know, a priori, which basis states are above the cutoff.  So, we must calculate them all, or at least all of the basis states below some reasonable energy cutoff and throw away any below the precision cutoff.  Unfortunately, above first order, this is a very inefficient process and we would like to find something better.

What we would like is a search algorithm that explores the Hilbert space and extracts the most important basis states.  Once we have these, we can simply construct the Hamiltonian matrix in this reduced Hilbert space and directly diagonalize it.  This is what the QSE method does.  It accomplishes this cyclically, with each cycle getting closer to the complete set of most important basis states.  This algorithm works in the following five steps: 
\begin{enumerate}
\item Begin with a seed for the scattering state. This seed could be as simple as the main basis state, but since first-order perturbation theory is so efficient, we begin with the first-order perturbative result.  Of course, this is not yet sufficiently accurate for our purposes. 
\item Remove from the current scattering state all basis states whose contributions are below the desired precision cutoff. Call the remaining basis states the reduced Hilbert space. 
\item Randomly add new basis states to the reduced Hilbert space. Do this by choosing random basis states that are already in the reduced Hilbert space and act on them with a randomly chosen operator from the Hamiltonian. This results in a new basis state chosen randomly from the Hilbert space which is, however, ÒcloseÓ to the basis states already in the reduced Hilbert space. 
\item Construct the Hamiltonian matrix with the new reduced Hilbert space and diagonalize. 
\item Repeat steps 2-4 until the scattering state is achieved. 
\end{enumerate}
The original authors of \cite{Lee:2000ac} were not considering scattering states or scattering amplitudes.  In fact, their interest was in non-perturbative results.  Since every cycle of the QSE method potentially reaches a higher order of perturbation theory and it potentially constructs all the basis states above the precision cutoff, we say that it is ``effectively" non-perturbative.  However, if the coupling constant were truly above the perturbative range where each higher order in the perturbative series was more important, not less, it is not clear to us how this method can work.  Therefore, we do not believe it is truly non-perturbative in a strict sense.  It is only effectively non-perturbative in the perturbative regime.  As far as we can tell, the coupling can be large, but must still be perturbative for this method to work.  

In the rest of this appendix, we'll give details of how we choose a random basis state from the reduced Hilbert space and how we choose a random operator from the Hamiltonian.  Perhaps a reader will have a clever idea for improving it.  We will begin with our method for choosing a random basis state.  Each time we choose a basis state from the reduced Hilbert space, we begin by randomly choosing an integer.  If it is odd, then we choose a random basis state from the reduced Hilbert space with a flat distribution where they are all equally likely.  If the random integer is even, then we choose a basis state from the reduced Hilbert space weighted according to the log of their absolute coefficients.  We do this using a simple Monte-Carlo technique.  We first randomly choose a test basis state with a flat distribution.  We then choose a random number between the log of the precision cutoff (because the cutoff is less than one, this is a negative number) and zero.  If this random number is less than the log of the absolute value of the coefficient of the test basis state then we keep it.  If it is above it, we discard it and randomly choose a new test basis state.  We do this until we keep one.  

Once we have a basis state chosen randomly from the reduced Hilbert space, we need to randomly choose an operator from the Hamiltonian in Eq.~(\ref{eq:Discrete Hamiltonian}) to generate a new basis state that we add to the reduced Hilbert space.  Our first step is to choose a random integer between 0 and 13.  If it is 0, we use the operator $a_{-p_1}a_{-p_2}a_{-p_3}a_{-p_4}$ which annihilates four momenta.  We will come back to how we choose the momenta shortly.  First, we complete how we choose which operator we use.  If the integer is equal to 1, 2 or 3, we use the operator $a^\dagger_{p_1}a_{-p_2}a_{-p_3}a_{-p_4}$ which annihilates three momenta and creates one new momentum.  This reduces the number of free particles in the basis state by two and we have found that our code is more successful focusing on basis states with only two more, two fewer particles, or the same number of particles.  This is the reason we choose this operator three times more often than the one that annihilates four momenta.  In a similar vein, if the integer is equal to 4, 5 or 6, we use the operator $a^\dagger_{p_1}a^\dagger_{p_2}a_{-p_3}a_{-p_4}$ which annihilates two momenta and creates two new momenta.  This operator does not change the number of free particles in the basis state. It only changes one or two of the momenta.  If the integer is equal to 7, 8 or 9, we use the operator $a^\dagger_{p_1}a^\dagger_{p_2}a^\dagger_{p_3}a_{-p_4}$ which annihilates one momentum and creates three new momenta.  This operator increases the number of free particles in the basis state by two.  If the integer is equal to 10, we use the operator $a^\dagger_{p_1}a^\dagger_{p_2}a^\dagger_{p_3}a^\dagger_{p_4}$ which creates four new momenta and, as a result, increases the number of free particles in the basis state by four.  Again, since we find this is less important, we choose this operator less frequently.  Finally, if the integer is equal to 11, 12 or 13, we choose a special form of the operator $a^\dagger_{p_1}a^\dagger_{p_2}a_{-p_3}a_{-p_4}$ which simply shifts exactly two momenta by $\pm\Delta p$.  We will describe it in further detail below, but we find this to be one of the most important operators to fill in important basis states directly adjacent to the current basis states in the eigenstate.   If any of these operations are unsuccessful, we simply return to the beginning and choose a random new basis state from the reduced Hilbert space and a random new operator from the Hamiltonian.  We continue doing this until we fill the requested number of basis states.

When the operator $a_{p_1}a_{p_2}a_{p_3}a_{p_4}$ is chosen, we first make a list of all possible combinations of four momenta whose sum is zero that does not completely annihilate the basis state.  If no such combination can be found (for example if it is a 2-particle basis state), the function returns failure and the code chooses a new basis state and a new operator as discussed in the previous paragraph.  If it does find some momentum combinations that do not annihilate the basis state, then it randomly chooses one of them and annihilates the four corresponding free particles from the basis state.  It then returns this as a new basis state that is added to the reduced Hilbert space.  

If the operator $a^\dagger_{p_1}a_{p_2}a_{p_3}a_{p_4}$ is chosen, it make a list of all combinations of three momenta from the basis state.  It then randomly chooses one of these combinations and creates a free particle with a momentum equal to the sum of the three momenta.  This is followed by annihilations of the three momenta.  If this is successful, the new basis state is returned, else failure and the codes starts its search for a new basis state over.

If the operator $a^\dagger_{p_1}a^\dagger_{p_2}a_{p_3}a_{p_4}$ is chosen, first a list of combinations of two momenta is created from the basis state.  These two basis states will be annihilated.  We then randomly choose a momentum between $-E_{max}$ and $E_{max}$, which is a parameter that we can adjust.  We always set it to be higher than the highest basis state contributing above the precision cutoff.  For example, for the eigenstates shown in Fig.~\ref{fig:Energy Eigenstates}, we set $E_{max}$ to be $80m$.  This should be more than sufficient since the basis state energy is the sum of the energy of all the momenta in it.  One individual particle always has a much smaller momentum than this.  However, we keep this relatively high $E_{max}$ to be on the safe side.  This same $E_{max}$ is used for the rest of the operators we discuss in this appendix.  Once this random new momentum is chosen, the final momentum is given the value $p_1=p_3+p_4-p_2$, where $p_3$ and $p_4$ are the momenta chosen from those already existing and $p_2$ is the momentum randomly chosen between $\pm E_{max}$.  We then create two new free particles with momenta $p_1$ and $p_2$ followed by annihilation of free particles with momenta $p_3$ and $p_4$.  If this is successful, we return the new basis state, if not, we return failure.

When the operator $a^\dagger_{p_1}a^\dagger_{p_2}a^\dagger_{p_3}a_{p_4}$ is chosen, we begin by randomly choosing one of the momenta in the basis state.  We then choose two new momenta, each randomly between $-E_{max}$ and $E_{max}$.  The final momentum is taken as $p_1=p_4-p_2-p_3$.  Three new free particles are created with momenta $p_1, p_2$ and $p_3$ followed by one annihilation of a free particle with momentum $p_4$.  If success is found, the new basis state is returned, else failure is returned.

If the operator $a^\dagger_{p_1}a^\dagger_{p_2}a^\dagger_{p_3}a^\dagger_{p_4}$ is chosen, three new momenta are chosen, each randomly between $-E_{max}$ and $E_{max}$.  The final momentum is chosen as $p_1=-p_2-p_3-p_4$.  Four new free particles are created in the basis state with momenta $p_1, p_2, p_3$ and $p_4$.  If this is a success, the new basis state is returned, otherwise, failure.

Finally, if the special operator is chosen that simply shifts exactly two momenta by $\pm\Delta p$, then two of the momenta from the basis state are randomly chosen, say $p_1$ and $p_2$.  Free particles with momenta $p_1$ and $p_2$ are annihilated and two new free particles with momenta $p_1+\Delta p$ and $p_2-\Delta p$ are created.  If this is successful, the new basis state is returned.  If not, failure is returned.

After randomly creating a new basis state as described, we add either the P-even or the P-odd version of that basis state to the reduced Hilbert space.  For example, suppose the new basis state randomly generated is $|b\rangle$.  We then find the P reversed state $P|b\rangle$ by reversing all the momenta.  If these two states are the same ($P|b\rangle=|b\rangle$), we simply add the basis state $|b\rangle$.  If they are different, we choose a random integer.  If the random integer is even, we add the P-even basis state $(|b\rangle+P|b\rangle)/\sqrt{2}$.  If the random integer is odd, we add the P-odd basis state $(|b\rangle-P|b\rangle)/\sqrt{2}$.  We could have simply always added the P-even basis states as those were what we expected for the eignestates we studied, but we wanted to keep our algorithm more general and we wanted to be sure our algorithm was working correctly, so we included both P-even and P-odd basis states in our reduced Hilbert space at each cycle.   As expected, we found that the P-odd basis states were all found to be below the precision cutoff and removed from the reduced Hilbert space by our algorithm while all the basis states remaining above the cutoff were P even.

\section{\label{app:renormalization of the mass}The $\mathbf{\lambda\phi^4}$ Theory}

We already worked out the discrete Hamiltonian of our theory in two spacetime dimensions in Section I of \cite{Christensen:2016naf}.  However, we make one modification in the present paper.  We renormalize the mass in order to cancel the second term of Eq.~(10) in that paper.  We review the derivation and explain the modification here.

We begin with the Lagrangian of our theory
\begin{equation}
\mathcal{L} = \frac{1}{2}\partial_\mu\phi\partial^\mu\phi - \frac{1}{2}m^2\phi^2 - \frac{\lambda}{4!}\phi^4\ ,
\label{eq:Lagrangian}
\end{equation}
which, after Legendre transformation, gives the Hamiltonian
\begin{equation}
H = \int dx\left[\frac{1}{2}\left(\frac{\partial\phi}{\partial t}\right)^2+\frac{1}{2}\left(\frac{\partial\phi}{\partial x}\right)^2+\frac{1}{2}m^2\phi^2+\frac{\lambda}{24}\phi^4\right]\ .
\end{equation}
As before we replace the field $\phi$ with a linear combination of creation and annihilation operators
\begin{equation}
\phi(x) = \int\frac{dp}{2\pi}\frac{1}{\sqrt{2\omega}}\left[a(p)e^{i \left(\omega t-p x\right)}+a^\dagger(p)e^{-i \left(\omega t-p x\right)}\right]\ .
\end{equation}
However, unlike before, we define the free-particle energy as
\begin{equation}
\omega = \sqrt{p^2+\tilde{m}^2}\ ,
\end{equation}
where $\tilde{m}\neq m$.  After inserting this definition of the fields, expanding and normal ordering, we get the following contribution from the bare-mass term
\begin{eqnarray}
\int dx\frac{1}{2}m^2\phi^2 = \frac{1}{2}\int\frac{dp}{\left(2\pi\right)\left(2\omega\right)}m^2\Big[&a_pa_{-p}e^{i2\omega t}+2a_pa^\dagger_p&\nonumber\\
&+a^\dagger_pa^\dagger_{-p}e^{-i2\omega t}\Big]\ ,&\nonumber\\
\label{eq:app:m2 term}
\end{eqnarray}
where we have dropped a non-dynamical constant term.  On the other hand, the $\lambda\phi^4$ term, after normal ordering, also gives terms quadratic in the creation and annihilation operators, namely
\begin{eqnarray}
\int dx\frac{\lambda}{24}\phi^4 = \frac{1}{2}\int\frac{dp}{\left(2\pi\right)\left(2\omega\right)}\Delta m^2\Big[&a_pa_{-p}e^{i2\omega t}+2a_pa^\dagger_p&\nonumber\\
&+a^\dagger_pa^\dagger_{-p}e^{-i2\omega t}\Big]&\nonumber\\
&+\cdots\ ,
\label{eq:app:phi4 term}
\end{eqnarray}
where the dots represent the terms with four creation and annihilation operators that will not contribute to the renormalization of mass.  We have again dropped a non-dynamical constant term and defined
\begin{equation}
\Delta m^2 = \frac{\lambda}{4}\int\frac{dp'}{\left(2\pi\right)\omega'}\ .
\end{equation}
We see that the contribution to the Hamiltonian in Eq.~(\ref{eq:app:phi4 term}) has exactly the same form as the contribution from the bare-mass term in Eq.~(\ref{eq:app:m2 term}), therefore, it has the effect of renormalizing the mass.  We now take 
\begin{equation}
\tilde{m}^2 = m^2+\Delta m^2\ .
\end{equation}
We continue here with the rest of the textbook derivation of the free part of the Hamiltonian for the convenience of the reader.  The contribution from the spatial derivative is given by
\begin{eqnarray}
\int dx\frac{1}{2}\left(\frac{\partial\phi}{\partial x}\right)^2 = \frac{1}{2}\int\frac{dp}{\left(2\pi\right)\left(2\omega\right)}p^2\Big[&a_pa_{-p}e^{i2\omega t}+2a_pa^\dagger_p&\nonumber\\
&+a^\dagger_pa^\dagger_{-p}e^{-i2\omega t}\Big]\ ,&\nonumber\\
\end{eqnarray}
where an overall non-dynamical constant has been dropped.  When this is combined with the $m^2$ contribution from Eq.~(\ref{eq:app:m2 term}) and the $\phi^4$ contribution from Eq.~(\ref{eq:app:phi4 term}), we can use $\omega^2=p^2+\tilde{m}^2$ to obtain
\begin{displaymath}
\int dx\left[\frac{1}{2}\left(\frac{\partial\phi}{\partial x}\right)^2+\frac{1}{2}m^2\phi^2+\frac{\lambda}{24}\phi^4\right] =\hspace{1in}
\end{displaymath}
\vspace{-0.5cm}
\begin{equation}
\frac{1}{2}\int\frac{dp}{\left(2\pi\right)\left(2\omega\right)}\omega^2\Big[a_pa_{-p}e^{i2\omega t}+2a_pa^\dagger_p+a^\dagger_pa^\dagger_{-p}e^{-i2\omega t}\Big]+\cdots\ .
\end{equation}
Finally, the time derivative term gives
\begin{eqnarray}
\int dx\frac{1}{2}\left(\frac{\partial\phi}{\partial t}\right)^2 = \frac{1}{2}\int&\frac{dp}{\left(2\pi\right)\left(2\omega\right)}\omega^2\Big[-a_pa_{-p}e^{i2\omega t}&\nonumber\\
&+2a_pa^\dagger_p-a^\dagger_pa^\dagger_{-p}e^{-i2\omega t}\Big]\ ,&\nonumber\\
\end{eqnarray}
where, as usual, we drop a non-dynamical constant.  We see that the first and third terms cancel between the time derivative contribution and the other contributions while the second term adds.  Our final result is
\begin{equation}
H = \int\frac{dp}{\left(2\pi\right)}\omega a_p a^\dagger_p + \cdots\ ,
\end{equation}
where the dots, again, represent the terms with four creation and annihilation operators coming from the interaction.  

The final step, for our purposes, is to discretize momentum space.  We do this by taking $\int dp/\left(2\pi\right)\to\sum_p\Delta p$ and $2\pi\delta(p)\to\delta_p/\Delta p$, where $\Delta p$ is the momentum step size and $\delta_p$ is the Kronecker delta (equal to 1 if $p=0$ and 0, otherwise).  Consequently, we take $a(p)\to a_p/\sqrt{\Delta p}$ and $a^\dagger(p)\to a^\dagger_p/\sqrt{\Delta p}$ for the discrete creation and annihilation operators and find $\left[a_p,a^\dagger_{p'}\right] = \delta_{p,p'}$.  With this, the final discrete Hamiltonian is
%\begin{eqnarray}
%H &=&
%\sum_p \omega a^\dagger_p a_p 
%+\frac{\lambda\Delta p}{96}\sum_{p_1p_2p_3p_4}\frac{1}{\sqrt{\omega_1\omega_2\omega_3\omega_4}}\Big[\nonumber\\
%&&a_{p_1}a_{p_2}a_{p_3}a_{p_4} e^{i\left(\omega_1+\omega_2+\omega_3+\omega_4\right)t}\delta_{p_1+p_2+p_3+p_4}\nonumber\\
%&&+4a^\dagger_{p_1}a_{p_2}a_{p_3}a_{p_4} e^{i\left(-\omega_1+\omega_2+\omega_3+\omega_4\right)t}\delta_{-p_1+p_2+p_3+p_4}\nonumber\\
%&&+6a^\dagger_{p_1}a^\dagger_{p_2}a_{p_3}a_{p_4} e^{i\left(-\omega_1-\omega_2+\omega_3+\omega_4\right)t}\delta_{-p_1-p_2+p_3+p_4}\nonumber\\
%&&+4a^\dagger_{p_1}a^\dagger_{p_2}a^\dagger_{p_3}a_{p_4} e^{i\left(-\omega_1-\omega_2-\omega_3+\omega_4\right)t}\delta_{-p_1-p_2-p_3+p_4}\nonumber\\
%&&+a^\dagger_{p_1}a^\dagger_{p_2}a^\dagger_{p_3}a^\dagger_{p_4} e^{i\left(-\omega_1-\omega_2-\omega_3-\omega_4\right)t}\delta_{-p_1-p_2-p_3-p_4}\Big]\ ,\nonumber\\
%\label{eq:Discrete Hamiltonian}
%\end{eqnarray}
\begin{eqnarray}
H &=&
\sum_p \omega a^\dagger_p a_p 
+\frac{\lambda\Delta p}{96}\sum_{p_1+p_2+p_3+p_4=0}\frac{1}{\sqrt{\omega_1\omega_2\omega_3\omega_4}}\Big[\nonumber\\
&&a_{-p_1}a_{-p_2}a_{-p_3}a_{-p_4} e^{i\left(\omega_1+\omega_2+\omega_3+\omega_4\right)t}\nonumber\\
&&+4a^\dagger_{p_1}a_{-p_2}a_{-p_3}a_{-p_4} e^{i\left(-\omega_1+\omega_2+\omega_3+\omega_4\right)t}\nonumber\\
&&+6a^\dagger_{p_1}a^\dagger_{p_2}a_{-p_3}a_{-p_4} e^{i\left(-\omega_1-\omega_2+\omega_3+\omega_4\right)t}\nonumber\\
&&+4a^\dagger_{p_1}a^\dagger_{p_2}a^\dagger_{p_3}a_{-p_4} e^{i\left(-\omega_1-\omega_2-\omega_3+\omega_4\right)t}\nonumber\\
&&+a^\dagger_{p_1}a^\dagger_{p_2}a^\dagger_{p_3}a^\dagger_{p_4} e^{i\left(-\omega_1-\omega_2-\omega_3-\omega_4\right)t}\Big]\ ,\nonumber\\
\label{eq:Discrete Hamiltonian}
\end{eqnarray}
which is the same as Eq.~(10) of \cite{Christensen:2016naf}, except that the second term (quadratic in creation and annihilation operators) has been dropped due to the present mass renormalization. 

The reason we did not do this renormalization in \cite{Christensen:2016naf} is that we only kept basis states with up to two free particles in \cite{Christensen:2016naf}.  A truncated Hilbert space with only zero-particle and two-particle basis states completely decouple after this mass renormalization.  We kept the extra term in \cite{Christensen:2016naf} in order to achieve more interesting results with this highly truncated space.  However, now that we are including a large set of basis states, including basis states with greater numbers of free particles, it makes sense to use the renormalized mass from the beginning.

Before ending this appendix, we briefly describe our basis states.  They are the same as in \cite{Christensen:2016naf}, however, in this paper, we will write our basis states as $|p_1,p_2,\cdots,p_n\rangle$ where $p_1, p_2,\cdots,p_n$ is the list of the momenta of the free particles, where we have dropped the vector symbols for convenience since we are working in 1 spatial dimension.  Moreover, because we are dealing with bosons, the order does not matter.  In the $|p_1,p_2,\cdots,p_n\rangle$ example, there are $n$ free particles.  Some of the momenta could be the same.  If this is the case, that momentum is listed multiple times.  If there are no particles in the basis state, it is written $|\rangle$ and called the free vacuum.  These basis states are eigenstates of the free part of the Hamiltonian, namely $\sum_p\omega_pa^\dagger_p a_p$.  We call their free-Hamiltonian eigenvalue the free energy of the basis state and write it as $\tilde{\omega}$.  It is simply the sum of the free energy $\omega_p$ for each particle.   

The creation and annihilation operators acting on these states give
\begin{eqnarray}
a^\dagger_p|\cdots,p,\cdots\rangle &=& \sqrt{n_p+1}|p\cdots,p,\cdots\rangle\ ,\\
a_p|\cdots,p,\cdots\rangle &=& \sqrt{n_p}|\cdots,\cdots\rangle\ .
\end{eqnarray}
where $n_p$ is the number of times the momentum $p$ appears in the state.  (We have only showed it explicitly once, but there may be others in the $\cdots$.)  If it does not appear at all, then $n_p=0$.  The Hamiltonian matrix is then constructed by sandwiching the Hamiltonian operator in Eq.~(\ref{eq:Discrete Hamiltonian}) between all pairs of basis states in the (truncated or reduced) Hilbert space.  The inner product between two states is equal to 1 if they have exactly the same list of momenta (although, again, the order does not matter) and 0 otherwise.

\section{\label{app:perturbative solution}Perturbative Solution}
In this appendix, for completeness, we review time-independent perturbation theory, which can be found in many textbooks (e.g. \cite{Merzbacher}).  The Schrodinger equation can be written 
\begin{equation}
H|\psi\rangle = E|\psi\rangle\ ,
\end{equation}
where $H$ is the Hamiltonian, $|\psi\rangle$ is an eigenstate and $E$ is the associated eigenenergy.  We assume that the Hamiltonian can be written
\begin{equation}
H = H_o+\lambda V\ ,
\end{equation}
where $H_o$ is the free Hamiltonian and is assumed exactly solvable, $\lambda$ is a small coupling constant and $\lambda V$ is the interaction part of the Hamiltonian.  Because $\lambda$ is small, we can expand the eigenstate and the energy as a power series in $\lambda$ as in
\begin{eqnarray}
|\psi\rangle &=& |\psi^0\rangle + \lambda|\psi^1\rangle + \lambda^2|\psi^2\rangle + \cdots\\
E &=& E^0 + \lambda E^1 + \lambda^2 E^2 + \cdots\ ,
\end{eqnarray}
where the superscript on $\psi$ and $E$ are labels and not powers.  Plugging these expansions into the Schrodinger Equation, we obtain
\begin{displaymath}
\left(H_o+\lambda V\right)\left(|\psi^0\rangle + \lambda|\psi^1\rangle + \lambda^2|\psi^2\rangle + \cdots\right) =\hspace{1in}
\end{displaymath}
\begin{equation}
 \left(E^0 + \lambda E^1 + \lambda^2 E^2 + \cdots\right)\left(|\psi^0\rangle + \lambda|\psi^1\rangle + \lambda^2|\psi^2\rangle + \cdots\right).
\end{equation}
Since this equation must be satisfied for any value of the coupling constant $\lambda$, it must be satisfied order by order in $\lambda$.  Therefore, we get the system of equations
\begin{eqnarray}
H_o|\psi^0\rangle &=& E^0|\psi^0\rangle\nonumber\\
H_o|\psi^1\rangle + V|\psi^0\rangle &=& E^0|\psi^1\rangle + E^1|\psi^0\rangle\nonumber\\
H_o|\psi^2\rangle + V|\psi^1\rangle &=& E^0|\psi^2\rangle + E^1|\psi^1\rangle + E^2|\psi^0\rangle\nonumber\\
&\vdots&\ .
\label{eq:app:pert:system}
\end{eqnarray}
It turns out that this system of equations can be solved recursively.  We do this by introducing the eigenstates of the free Hamiltonian as a complete set of basis states.  We will call these basis states $|b_0\rangle, |b_1\rangle, |b_2\rangle,\cdots$ and their associated free Hamiltonian eigenvalues (their free energies) $\omega_0, \omega_1, \omega_2,\cdots$ so that $H_o|b_j\rangle = \omega_j|b_j\rangle$.  

We have seen above in Eq.~(\ref{eq:app:pert:system}) that $|\psi^0\rangle$ is an eigenstate of the free Hamiltonian, therefore, for notational convenience, we take it to be $|b_0\rangle$.  This tells us that 
\begin{equation}
E^0 = \omega_0\ .
\end{equation}
We next expand $|\psi^1\rangle$ in terms of the basis states so that
\begin{equation}
|\psi^1\rangle = \sum_j c_j^1 |b_j\rangle\ ,
\label{eq:app:pert:psi1sum}
\end{equation}
where $c_j^1$ are the coefficients of the basis states.  We plug this into the second line of Eq.~(\ref{eq:app:pert:system}) to obtain, after moving $|\psi^1\rangle$ to the left and $|\psi^0\rangle$ to the right,
\begin{equation}
\sum_j c_j^1\left(\omega_j-\omega_0\right)|b_j\rangle = \left(E^1-V\right)|b_0\rangle\ .
\end{equation}
We can see that this equation does not determine the coefficient of $|b_o\rangle$, $c_0^1$, because that term drops out of the left side ($\omega_0-\omega_0=0$).  In fact, if we take the inner product of this equation with $\langle b_0|$, we obtain the equation
\begin{equation}
E^1 = \langle b_0|V|b_0\rangle\ ,
\end{equation}
which determines the first-order contribution to the energy.  However, we would get this even if the sum over $j$ in Eq.~(\ref{eq:app:pert:psi1sum}) did not include $j=0$.  In fact, there is no way to determine $c_0^1$ from the perturbation series.  It is not unique.  This is a result of the perturbation series itself not being unique.  In fact, we can multiply the eigenstate $|\psi\rangle$ by any constant function (that passes through the Hamiltonian) and it will still be an eigenstate.  For example, we consider 
\begin{equation}
|\tilde{\psi}\rangle = \left(a_0+\lambda a_1 + \lambda^2 a_2 + \cdots\right)|\psi\rangle\ .
\end{equation}
We can see that this is also an eigenstate of $H$ with the same eigenvalue
\begin{equation}
H|\tilde{\psi}\rangle = E|\tilde{\psi}\rangle\ .
\end{equation}
We can use this freedom to completely remove the basis state $|b_0\rangle$ from all but the leading order solution.  To see this, suppose $|\psi^1\rangle$ has a nonzero $c_0^1$.   Now, consider the expansion of $|\tilde{\psi}\rangle$ in a power series in $\lambda$.  It is given by
\begin{equation}
|\tilde{\psi}\rangle = a_0|\psi^0\rangle + \lambda\left(a_1|\psi^0\rangle + a_0|\psi^1\rangle\right) + \cdots\ ,
\end{equation}
or, after plugging in the expansion in the basis states and focusing on $|\tilde{\psi}^1\rangle$, we have
\begin{equation}
|\tilde{\psi}^1\rangle = a_1|b_0\rangle + a_0\sum_j c_j^1|b_j\rangle\ .
\end{equation}
Therefore, we can completely remove $|b_0\rangle$ from $|\tilde{\psi}^1\rangle$ by taking 
\begin{equation}
a_1 = -a_0 c_0^1\ .
\end{equation}
We can do this, order by order, so that the only place $|b_0\rangle$ appears is in $|\tilde{\psi}^0\rangle$.  For the rest of this section, we will assume that this has been done and drop the tilde.  

Returning to Eq.~(\ref{eq:app:pert:psi1sum}), we now write 
\begin{equation}
|\psi^1\rangle = \sum_{j\neq0} c_j^1 |b_j\rangle\ .
\label{eq:app:pert:psi1tsum}
\end{equation}
We again plug this into the second line of Eq.~(\ref{eq:app:pert:system}) to obtain, 
\begin{equation}
\sum_{j\neq0} c_j^1\left(\omega_j-\omega_0\right)|b_j\rangle = \left(E^1-V\right)|b_0\rangle\ .
\label{eq:app:pert:1stO}
\end{equation}
If we take the inner product of this equation with $\langle b_0|$, we obtain the first-order contribution to the energy,
\begin{equation}
E^1 = \langle b_0|V|\psi^0\rangle\ ,
\end{equation}
where we have replaced $|b_0\rangle$ with $|\psi^0\rangle$, on the right.  If we, on the other hand, take the inner product of Eq.~(\ref{eq:app:pert:1stO}) with $\langle b_j|$, where $j\neq0$, we obtain
\begin{equation}
c_j^1\left(\omega_j-\omega_0\right) = -\langle b_j|V|b_0\rangle\ .
\end{equation}
Solving for $c_j^1$ gives us,
\begin{equation}
c_j^1 = \frac{\langle b_j|V|\psi^0\rangle}{\omega_0-\omega_j}\ ,
\label{eq:app:pert:cj1}
\end{equation}
where we have again replaced $|b_0\rangle$ with $|\psi^0\rangle$ (and we are assuming that $|b_j\rangle$ is not degenerate with $|b_0\rangle$).

We will also do second order explicitly since an important new term appears in the coefficient.  Taking
\begin{equation}
|\psi^2\rangle = \sum_{j\neq0} c_j^2 |b_j\rangle\ .
\label{eq:app:pert:psi2tsum}
\end{equation}
We plug this into the third line of Eq.~(\ref{eq:app:pert:system}) to obtain, 
\begin{equation}
\sum_{j\neq0} c_j^2\left(\omega_j-\omega_0\right)|b_j\rangle = \left(E^1-V\right)|\psi^1\rangle + E^2|\psi^0\rangle\ .
\label{eq:app:pert:2ndO}
\end{equation}
If we take the inner product of this equation with $\langle b_0|$, we obtain the second-order contribution to the energy,
\begin{equation}
E^2 = \langle b_0|V|\psi^1\rangle\ .
\end{equation}
If we take the inner product with $\langle b_j|$, where $j\neq0$, on the other hand, we obtain
\begin{equation}
c_j^2\left(\omega_j-\omega_0\right) = -\langle b_j|V|\psi^1\rangle + c_j^1 E^1\ ,
\end{equation}
so that
\begin{equation}
c_j^2 = \frac{1}{\omega_0-\omega_j}\left(\langle b_j|V|\psi^1\rangle - c_j^1E^1\right)\ .
\label{eq:app:pert:cj2}
\end{equation}

Continuing in this way, we find the general nth-order energy
\begin{equation}
E^n = \langle b_0|V|\psi^{n-1}\rangle\ ,
\end{equation}
where we remind the reader that
\begin{equation}
E = \sum_n \lambda^n E^n\ .
\end{equation}
We also find the general nth-order eigenstate
\begin{equation}
c_j^n = \frac{1}{\omega_0-\omega_j}\left(\langle b_j|V|\psi^{n-1}\rangle - \sum_{k=1}^{n-1}c_j^{k}E^{n-k}\right)\ ,
\label{eq:app:pert:cjn}
\end{equation}
where
\begin{equation}
|\psi^n \rangle = \sum_{j\neq0}c_j^n|b_j\rangle
\end{equation}
and
\begin{equation}
|\psi\rangle = \sum_n \lambda^n |\psi^n\rangle\ .
\label{eq:app:pert:psi total}
\end{equation}

So far, we have only described non-degenerate perturbation theory, which is appropriate when none of the basis states are degenerate, or even nearly degenerate.  However, this technique, as so far described, breaks down when a basis state is encountered whose free energy is close to that of the main basis state.  The reason is that the difference in free energy $(\omega_0-\omega_j)$ appears in the denominator of Eq.~(\ref{eq:app:pert:cjn}) causing the coefficient to be inappropriately large.  This signals the breakdown of non-degenerate perturbation theory.  The textbook method for dealing with this (e.g. \cite{Merzbacher}) is to directly diagonalize the degenerate (or nearly degenerate) sector first and then apply the perturbative formulas to the eigenstates of that initial diagonalization along with the other non-degenerate basis states.  However, this becomes complicated at higher orders of perturbation theory because the basis states may not be connected by just one factor of the potential.  As a result, the diagonalization step may involve many other basis states that may not be degenerate or nearly degenerate.  So, in fact, we must diagonalize a much larger sector.  However, it is not clear, a priori, what basis states should be included in this diagonalization since it is no longer just the nearby basis states.  Our approach is to simply calculate the coefficients using the naive formulas described in this section, realizing that some of them will be incorrectly large due to the breakdown of perturbation theory.  However, after doing this naive perturbative calculation, we fortunately have a clear set of basis states that can be directly diagonalized.  It may be larger than necessary, but it is certainly sufficient as it includes all the basis states connecting the main basis state and the degenerate and nearly degenerate basis states at the current order.   So, after doing the naive perturbative calculation, we simply construct the Hamiltonian using the set of basis states coming from perturbation theory and directly diagonalize it.  We do this for second- and third-order perturbation theory in this article.  We do not diagonalize the Hamiltonian for the first-order results because we do not run into a problem with degenerate basis states at first order.  We use the perturbative formulas as is at first order.


\begin{thebibliography}{99}



%%%%%%%%%%%%%%%%%%%%
% Perturbative
%%%%%%%%%%%%%%%%%%%%

  
 
 %\cite{Parke:1986gb}
\bibitem{Parke:1986gb} 
  S.~J.~Parke and T.~R.~Taylor,
  ``An Amplitude for $n$ Gluon Scattering,''
  Phys.\ Rev.\ Lett.\  {\bf 56}, 2459 (1986).
  %%CITATION = PRLTA,56,2459;%%
  %552 citations counted in INSPIRE as of 31 Aug 2015
 
  
  %\cite{Britto:2005fq}
\bibitem{Britto:2005fq} 
  R.~Britto, F.~Cachazo, B.~Feng and E.~Witten,
  ``Direct proof of tree-level recursion relation in Yang-Mills theory,''
  Phys.\ Rev.\ Lett.\  {\bf 94}, 181602 (2005)
  [hep-th/0501052].
  %%CITATION = HEP-TH/0501052;%%
  %679 citations counted in INSPIRE as of 31 Aug 2015
  
    %\cite{ArkaniHamed:2008gz}
\bibitem{ArkaniHamed:2008gz} 
  N.~Arkani-Hamed, F.~Cachazo and J.~Kaplan,
  ``What is the Simplest Quantum Field Theory?,''
  JHEP {\bf 1009}, 016 (2010)
  [arXiv:0808.1446 [hep-th]].
  %%CITATION = ARXIV:0808.1446;%%
  %293 citations counted in INSPIRE as of 31 Aug 2015

    %\cite{Feng:2011np}
\bibitem{Feng:2011np} 
  B.~Feng and M.~Luo,
  ``An Introduction to On-shell Recursion Relations,''
  Front.\ Phys.\  {\bf 7}, 533 (2012)
  [arXiv:1111.5759 [hep-th]].
  %%CITATION = ARXIV:1111.5759;%%
  %32 citations counted in INSPIRE as of 31 Aug 2015

  %\cite{Elvang:2013cua}
\bibitem{Elvang:2013cua} 
  H.~Elvang and Y.~t.~Huang,
  %``Scattering Amplitudes,''
  arXiv:1308.1697 [hep-th].
  %%CITATION = ARXIV:1308.1697;%%
  %207 citations counted in INSPIRE as of 31 Aug 2017


%\cite{Dixon:2013uaa}
\bibitem{Dixon:2013uaa} 
  L.~J.~Dixon,
  ``A brief introduction to modern amplitude methods,''
  arXiv:1310.5353 [hep-ph].
  %%CITATION = ARXIV:1310.5353;%%
  %18 citations counted in INSPIRE as of 31 Aug 2015
  
  %\cite{Benincasa:2013faa}
\bibitem{Benincasa:2013faa} 
  P.~Benincasa,
  ``New structures in scattering amplitudes: a review,''
  Int.\ J.\ Mod.\ Phys.\ A {\bf 29}, no. 5, 1430005 (2014)
  [arXiv:1312.5583 [hep-th]].
  %%CITATION = ARXIV:1312.5583;%%
  %3 citations counted in INSPIRE as of 31 Aug 2015
  
  
  
  

%\cite{Arkani-Hamed:2013jha}
\bibitem{Arkani-Hamed:2013jha} 
  N.~Arkani-Hamed and J.~Trnka,
  ``The Amplituhedron,''
  JHEP {\bf 1410}, 30 (2014)
  [arXiv:1312.2007 [hep-th]].
  %%CITATION = ARXIV:1312.2007;%%
  %57 citations counted in INSPIRE as of 09 sept. 2015
  



  %%%%%%%%%%%%%%%%%%%%
  %   QSE Method
  %%%%%%%%%%%%%%%%%%%%

%\cite{Lee:2000ac}
\bibitem{Lee:2000ac} 
  D.~Lee, N.~Salwen and D.~Lee,
  %``The Diagonalization of quantum field Hamiltonians,''
  Phys.\ Lett.\ B {\bf 503}, 223 (2001)
  doi:10.1016/S0370-2693(01)00197-6
  [hep-th/0002251].
  %%CITATION = doi:10.1016/S0370-2693(01)00197-6;%%
  %24 citations counted in INSPIRE as of 12 Sep 2016
  
  

  %%%%%%%%%%%%%%%%%%%%
  %   Our Previous Work
  %%%%%%%%%%%%%%%%%%%%


       %\cite{Christensen:2016naf}
\bibitem{Christensen:2016naf} 
  N.~Christensen,
  ``Diagonalizing the Hamiltonian of $\lambda \phi^4$ Theory in 2 Space-Time Dimensions,''
  arXiv:1603.01273 [hep-ph].
  %%CITATION = ARXIV:1603.01273;%%
  %1 citations counted in INSPIRE as of 07 Sep 2016


    %%%%%%%%%%%%%%%%%%%%
  %   More QSE Method
  %%%%%%%%%%%%%%%%%%%%

  
  
  %\cite{Lee:2000gm}
\bibitem{Lee:2000gm} 
  D.~Lee,
  %``Quasisparse eigenvector diagonalization and stochastic error correction,''
  Nucl.\ Phys.\ Proc.\ Suppl.\  {\bf 90}, 199 (2000)
  doi:10.1016/S0920-5632(00)00913-0
  [cond-mat/0008457].
  %%CITATION = doi:10.1016/S0920-5632(00)00913-0;%%
  %1 citations counted in INSPIRE as of 31 Aug 2017
  
  
    %\cite{Salwen:2000gn}
\bibitem{Salwen:2000gn} 
  N.~Salwen,
  %``EQSE diagonalization of the Hubbard model,''
  Nucl.\ Phys.\ Proc.\ Suppl.\  {\bf 90}, 202 (2000)
  doi:10.1016/S0920-5632(00)00897-5
  [cond-mat/0008458].
  %%CITATION = doi:10.1016/S0920-5632(00)00897-5;%%
  %1 citations counted in INSPIRE as of 31 Aug 2017

%\cite{Lee:2000xna}
\bibitem{Lee:2000xna} 
  D.~Lee, N.~Salwen and M.~Windoloski,
  %``Introduction to stochastic error correction methods,''
  Phys.\ Lett.\ B {\bf 502}, 329 (2001)
  doi:10.1016/S0370-2693(01)00198-8
  [hep-lat/0010039].
  %%CITATION = doi:10.1016/S0370-2693(01)00198-8;%%
  %14 citations counted in INSPIRE as of 31 Aug 2017

%\cite{Lee:2000yt}
\bibitem{Lee:2000yt} 
  D.~Lee,
  %``The Role of diagonalization within a diagonalization / Monte Carlo scheme,''
  Int.\ J.\ Mod.\ Phys.\ A {\bf 16S1C}, 1245 (2001)
  doi:10.1142/S0217751X01009430
  [hep-lat/0010095].
  %%CITATION = doi:10.1142/S0217751X01009430;%%

%\cite{Lee:2000yu}
\bibitem{Lee:2000yu} 
  D.~Lee,
  %``The Role of Monte Carlo within a diagonalization / Monte Carlo scheme,''
  Nucl.\ Phys.\ Proc.\ Suppl.\  {\bf 94}, 809 (2001)
  doi:10.1016/S0920-5632(01)01011-8
  [hep-lat/0010096].
  %%CITATION = doi:10.1016/S0920-5632(01)01011-8;%%

%\cite{Borasoy:2001pb}
\bibitem{Borasoy:2001pb} 
  B.~Borasoy and D.~Lee,
  %``Study of relativistic bound states in a scalar model using diagonalization / Monte Carlo methods,''
  Nucl.\ Phys.\ A {\bf 696}, 537 (2001)
  doi:10.1016/S0375-9474(01)01127-7
  [hep-ph/0101186].
  %%CITATION = doi:10.1016/S0375-9474(01)01127-7;%%
  
  
  
  %\cite{Salwen:2002dx}
\bibitem{Salwen:2002dx} 
  N.~Salwen,
  %``Non-perturbative methods in modal field theory,''
  hep-lat/0212035.
  %%CITATION = HEP-LAT/0212035;%%
  %4 citations counted in INSPIRE as of 31 Aug 2017
  
  
  
  
  
  
  
  
  
  %%%%%%%%%%%%%%%%%%%%
  %   Diagonalize the Hamiltonian
  %%%%%%%%%%%%%%%%%%%%
  
    %\cite{Salwen:1999pw}
\bibitem{Salwen:1999pw} 
  N.~Salwen and D.~Lee,
  %``A New approach to nonperturbative Minkowskian dynamics,''
  hep-th/9910103.
  %%CITATION = HEP-TH/9910103;%%
  %2 citations counted in INSPIRE as of 05 Jul 2016

  
  %\cite{Martinovic:2002bv}
\bibitem{Martinovic:2002bv} 
  L.~Martinovic,
  %``Spontaneous symmetry breaking in light front field theory,''
  Phys.\ Rev.\ D {\bf 78}, 105009 (2008)
  doi:10.1103/PhysRevD.78.105009
  [hep-th/0207137].
  %%CITATION = doi:10.1103/PhysRevD.78.105009;%%
  %6 citations counted in INSPIRE as of 31 Aug 2017
  
    %\cite{Wagner:1}
\bibitem{Wagner:1} 
  R.E.~Wagner, M.R.~Ware, A.M.~Vikartofsky, Q.~Su and R.~Grobe,
  ``Dynamics of Two- and Four-Boson Interactions in Dressed Vacuum States,"
  Int. J. Theor. Phys. \textbf{51}, 3787 (2012).
  
\bibitem{Wagner:2}
R.E.~Wagner, S.~Acosta, S.A.~Glasgow, Q.~Su and R.~Grobe,
``Quantum fluctuations in the dressed vacuum of a bosonic model system,"
J. Phys. A: Math. Theor. \textbf{45}, 275303 (2012).

  

%\cite{Hogervorst:2014rta}
\bibitem{Hogervorst:2014rta} 
  M.~Hogervorst, S.~Rychkov and B.~C.~van Rees,
  %``Truncated conformal space approach in d dimensions: A cheap alternative to lattice field theory?,''
  Phys.\ Rev.\ D {\bf 91}, 025005 (2015)
  doi:10.1103/PhysRevD.91.025005
  [arXiv:1409.1581 [hep-th]].
  %%CITATION = doi:10.1103/PhysRevD.91.025005;%%
  %43 citations counted in INSPIRE as of 31 Aug 2017
  
  
  %\cite{Rychkov:2014eea}
\bibitem{Rychkov:2014eea} 
  S.~Rychkov and L.~G.~Vitale,
  %``Hamiltonian truncation study of the $?^4$ theory in two dimensions,''
  Phys.\ Rev.\ D {\bf 91}, 085011 (2015)
  doi:10.1103/PhysRevD.91.085011
  [arXiv:1412.3460 [hep-th]].
  %%CITATION = doi:10.1103/PhysRevD.91.085011;%%
  %27 citations counted in INSPIRE as of 31 Aug 2017
  
  %\cite{Rychkov:2015vap}
\bibitem{Rychkov:2015vap} 
  S.~Rychkov and L.~G.~Vitale,
  %``Hamiltonian truncation study of the $\phi^4$ theory in two dimensions. II. The $\mathbb Z_2$ -broken phase and the Chang duality,''
  Phys.\ Rev.\ D {\bf 93}, no. 6, 065014 (2016)
  doi:10.1103/PhysRevD.93.065014
  [arXiv:1512.00493 [hep-th]].
  %%CITATION = doi:10.1103/PhysRevD.93.065014;%%
  %13 citations counted in INSPIRE as of 31 Aug 2017
  
  %\cite{Elias-Miro:2015bqk}
\bibitem{Elias-Miro:2015bqk} 
  J.~Elias-Miro, M.~Montull and M.~Riembau,
  %``The renormalized Hamiltonian truncation method in the large $E_T$ expansion,''
  JHEP {\bf 1604}, 144 (2016)
  doi:10.1007/JHEP04(2016)144
  [arXiv:1512.05746 [hep-th]].
  %%CITATION = doi:10.1007/JHEP04(2016)144;%%
  %8 citations counted in INSPIRE as of 31 Aug 2017
  
  %\cite{Bajnok:2015bgw}
\bibitem{Bajnok:2015bgw} 
  Z.~Bajnok and M.~Lajer,
  %``Truncated Hilbert space approach to the 2d $\phi^{4}$ theory,''
  arXiv:1512.06901 [hep-th].
  %%CITATION = ARXIV:1512.06901;%%
  %2 citations counted in INSPIRE as of 05 Jul 2016
  
  
  %\cite{Katz:2016hxp}
\bibitem{Katz:2016hxp} 
  E.~Katz, Z.~U.~Khandker and M.~T.~Walters,
  %``A Conformal Truncation Framework for Infinite-Volume Dynamics,''
  JHEP {\bf 1607}, 140 (2016)
  doi:10.1007/JHEP07(2016)140
  [arXiv:1604.01766 [hep-th]].
  %%CITATION = doi:10.1007/JHEP07(2016)140;%%
  %5 citations counted in INSPIRE as of 31 Aug 2017
  
  %\cite{Burkardt:2016ffk}
\bibitem{Burkardt:2016ffk} 
  M.~Burkardt, S.~S.~Chabysheva and J.~R.~Hiller,
  %``Two-dimensional light-front $\phi^4$ theory in a symmetric polynomial basis,''
  Phys.\ Rev.\ D {\bf 94}, no. 6, 065006 (2016)
  doi:10.1103/PhysRevD.94.065006
  [arXiv:1607.00026 [hep-th]].
  %%CITATION = doi:10.1103/PhysRevD.94.065006;%%
  %7 citations counted in INSPIRE as of 31 Aug 2017
  
  %\cite{Anand:2017yij}
\bibitem{Anand:2017yij} 
  N.~Anand, V.~X.~Genest, E.~Katz, Z.~U.~Khandker and M.~T.~Walters,
  %``RG flow from $\phi^4$ theory to the 2D Ising model,''
  JHEP {\bf 1708}, 056 (2017)
  doi:10.1007/JHEP08(2017)056
  [arXiv:1704.04500 [hep-th]].
  %%CITATION = doi:10.1007/JHEP08(2017)056;%%
  %2 citations counted in INSPIRE as of 31 Aug 2017


  
  
 
  
  %%%%%%%%%%%%%%%%%%%%%%%%
%   Weinberg
%%%%%%%%%%%%%%%%%%%%%%%%
 %\cite{Weinberg:1995mt}
\bibitem{Weinberg:1995mt} 
  S.~Weinberg,
  ``The Quantum theory of fields. Vol. 1: Foundations.''
  %%CITATION = INSPIRE-406190;%%
  %253 citations counted in INSPIRE as of 27 Feb 2018
  


%%%%%%%%%%%%%%%%%%%%%%%%
%   S-Matrix on Lattice
%%%%%%%%%%%%%%%%%%%%%%%%

%\cite{Luscher:1990ck}
\bibitem{Luscher:1990ck} 
  M.~Luscher and U.~Wolff,
  ``How to Calculate the Elastic Scattering Matrix in Two-dimensional Quantum Field Theories by Numerical Simulation,''
  Nucl.\ Phys.\ B {\bf 339}, 222 (1990).
  %doi:10.1016/0550-3213(90)90540-T
  %%CITATION = doi:10.1016/0550-3213(90)90540-T;%%
  %685 citations counted in INSPIRE as of 27 Feb 2018
  
  
  %\cite{Briceno:2012rv}
\bibitem{Briceno:2012rv} 
  R.~A.~Briceno and Z.~Davoudi,
  ``Three-particle scattering amplitudes from a finite volume formalism,''
  Phys.\ Rev.\ D {\bf 87}, no. 9, 094507 (2013)
  %doi:10.1103/PhysRevD.87.094507
  [arXiv:1212.3398 [hep-lat]].
  %%CITATION = doi:10.1103/PhysRevD.87.094507;%%
  %93 citations counted in INSPIRE as of 27 Feb 2018
  
  
  %\cite{Romero-Lopez:2017gag}
\bibitem{Romero-Lopez:2017gag} 
  F.~Romero-L—pez, C.~Urbach and A.~Rusetsky,
  ``Vector-Vector Scattering on the Lattice,''
  arXiv:1710.04524 [hep-lat].
  %%CITATION = ARXIV:1710.04524;%%
  %1 citations counted in INSPIRE as of 27 Feb 2018
  


%%%%%%%%%%%%%%%%%%%%%%%%
%   Others
%%%%%%%%%%%%%%%%%%%%%%%%


 
  
  \bibitem{Merzbacher}
  E.~Merzbacher,
  ``Quantum Mechanics,"
  1998, Wiley.

\end{thebibliography}
\end{document}